\documentclass[12pt, a4paper]{article}

\usepackage[a4paper, top=2.0cm, bottom=1.0cm, left=1.8cm, right=1.5cm, footskip = 1.0cm, includehead, includefoot]{geometry}
\usepackage{helvet} 			

\usepackage{csquotes}
\usepackage[utf8]{inputenc}
\usepackage[final]{graphicx}
\usepackage[margin=20mm, font=footnotesize, labelfont=bf, justification=justified]{caption}
\usepackage{dirtytalk}

\usepackage{graphicx}        	

\usepackage{times}
\usepackage[bottom]{footmisc}	
\usepackage{scrextend}
\deffootnote[1.8em]{1.8em}{1.8em}{\textsuperscript{\thefootnotemark}}
\usepackage{xcolor}			

\usepackage{newtxtext}
\usepackage{newtxmath}       	
\usepackage{braket}

\usepackage{mathtools}
\usepackage{amsmath}
\usepackage{subfigure}
\usepackage{tipa}
\usepackage{url}
\usepackage{authblk}
\usepackage{blindtext}
\usepackage[square, sort, comma, authoryear]{natbib}

\usepackage{MnSymbol}
\usepackage{hyperref}
\usepackage{tikzfig}
\usepackage{relsize}
\usepackage[outline]{contour}
\contourlength{.12em}

\tikzstyle{node}=[minimum size=0.3cm]
\tikzstyle{Z}=[fill={rgb,255:red,230; green,254; blue,230}, draw={rgb,255: red,61; green,77; blue,61}, shape=circle]
\tikzstyle{X}=[fill={rgb,255:red,255; green,135; blue,136}, draw={rgb,255: red,102; green,54; blue,54}, shape=circle]
\tikzstyle{Y}=[fill=zxblue, draw=zxdblue, shape=circle]
\tikzstyle{Z_big}=[fill={rgb,255:red,230; green,254; blue,230}, draw={rgb,255: red,61; green,77; blue,61}, shape=circle, minimum width=1.6em, font={\small}]
\tikzstyle{X_big}=[fill={rgb,255:red,255; green,135; blue,136}, draw={rgb,255: red,102; green,54; blue,54}, shape=circle, minimum width=1.6em, font={\small}]
\tikzstyle{Y_big}=[fill=zxblue, draw=zxdblue, shape=circle, minimum width=1.6em, font={\small}]
\tikzstyle{Z_tri}=[fill={rgb,255:red,230; green,254; blue,230}, draw={rgb,255: red,61; green,77; blue,61}, regular polygon, regular polygon sides=3, draw, shape border rotate=0, inner sep=0pt, minimum width=15pt, line width=0.75]
\tikzstyle{X_tri}=[fill={rgb,255:red,255; green,135; blue,136}, draw={rgb,255: red,102; green,54; blue,54}, regular polygon, regular polygon sides=3, draw, shape border rotate=0, inner sep=0pt, minimum width=15pt, line width=0.75]
\tikzstyle{Z_tri_inv}=[fill={rgb,255:red,230; green,254; blue,230}, draw={rgb,255: red,61; green,77; blue,61}, regular polygon, regular polygon sides=3, draw, shape border rotate=180, inner sep=0pt, minimum width=15pt, line width=0.75, font={\footnotesize}]
\tikzstyle{X_tri_inv}=[fill={rgb,255:red,255; green,135; blue,136}, draw={rgb,255: red,102; green,54; blue,54}, regular polygon, regular polygon sides=3, draw, shape border rotate=180, inner sep=0pt, minimum width=15pt, line width=0.75]
\tikzstyle{Z_tri_l}=[fill={rgb,255:red,230; green,254; blue,230}, draw={rgb,255: red,61; green,77; blue,61}, regular polygon, regular polygon sides=3, draw, shape border rotate=90, inner sep=0pt, minimum width=15pt, line width=0.75]
\tikzstyle{X_tri_l}=[fill={rgb,255:red,255; green,135; blue,136}, draw={rgb,255: red,102; green,54; blue,54}, regular polygon, regular polygon sides=3, draw, shape border rotate=90, inner sep=0pt, minimum width=15pt, line width=0.75]
\tikzstyle{H}=[fill=yellow, draw=black, shape=rectangle, minimum width=2mm, minimum height=2mm]
\tikzstyle{Y_box}=[draw=black, shape=rectangle, minimum width=2mm, minimum height=2mm, tikzit fill={rgb,255: red,255; green,128; blue,0}]
\tikzstyle{Z_med}=[fill={rgb,255:red,230; green,254; blue,230}, draw={rgb,255: red,61; green,77; blue,61}, shape=circle, minimum width=1.1em, font={\footnotesize}]
\tikzstyle{X_med}=[fill={rgb,255:red,255; green,135; blue,136}, draw={rgb,255: red,102; green,54; blue,54}, shape=circle, minimum width=1.1em, font={\footnotesize}]
\tikzstyle{Y_med}=[fill=zxblue, draw=zxdblue, font={\footnotesize}, minimum width=1.1em, font={\footnotesize}, shape=circle]
\tikzstyle{ZP}=[fill={rgb,255:red,230; green,254; blue,230}, draw={rgb,255: red,61; green,77; blue,61}, regular polygon, regular polygon sides=3, shape border rotate=30]
\tikzstyle{ZM}=[fill={rgb,255:red,230; green,254; blue,230}, draw={rgb,255: red,61; green,77; blue,61}, regular polygon, regular polygon sides=3, shape border rotate=-30]
\tikzstyle{XP}=[fill={rgb,255:red,255; green,135; blue,136}, draw={rgb,255: red,102; green,54; blue,54}, regular polygon, regular polygon sides=3, shape border rotate=30]
\tikzstyle{XM}=[fill={rgb,255:red,255; green,135; blue,136}, draw={rgb,255: red,102; green,54; blue,54}, regular polygon, regular polygon sides=3, shape border rotate=-30]
\tikzstyle{small_box}=[fill=white, draw=black, shape=rectangle, minimum width=1.5cm, minimum height=1.5cm, font={\footnotesize}]
\tikzstyle{med_rectangle}=[fill=white, draw=black, shape=rectangle, minimum width=2.8cm, minimum height=1.5cm, font={\footnotesize}]
\tikzstyle{big_rectangle}=[fill=white, draw=black, shape=rectangle, minimum width=4.5cm, minimum height=1.5cm, font={\footnotesize}]
\tikzstyle{smol_box}=[fill=white, draw=black, shape=rectangle, minimum width=1cm, minimum height=1cm]
\tikzstyle{poo}=[minimum height=0.7cm, minimum width=0.7cm, path picture={\node at (path picture bounding box.center) {\includegraphics[width=0.7cm] {figures/poo}};}]
\tikzstyle{Z_long}=[fill={rgb,255:red,230; green,254; blue,230}, draw={rgb,255: red,61; green,77; blue,61}, shape=rectangle, rounded corners=0.25cm, minimum height=0.5cm, inner sep=0.25em, font={\scriptsize}]
\tikzstyle{X_long}=[fill={rgb,255:red,255; green,135; blue,136}, draw={rgb,255: red,102; green,54; blue,54}, shape=rectangle, rounded corners=0.25cm, minimum height=0.5cm, inner sep=0.25em, font={\scriptsize}]
\tikzstyle{med_rectv}=[fill=white, draw=black, shape=rectangle, minimum width=1.1cm, minimum height=2.4cm, font={\scriptsize}, inner sep=0.2cm]
\tikzstyle{Z dot}=[fill={rgb,255: red,0; green,127; blue,0}, draw=black, shape=circle, minimum width=1.5mm]
\tikzstyle{X dot}=[fill={rgb,255:red,255; green,21; blue,0}, draw=black, shape=circle, minimum width=1.5mm]
\tikzstyle{Z phase dot}=[fill={rgb,255: red,0; green,127; blue,0}, draw=black, shape=circle, font={\footnotesize}]
\tikzstyle{X phase dot}=[fill={rgb,255:red,255; green,21; blue,0}, draw=black, shape=circle, font={\footnotesize}]
\tikzstyle{ZYa}=[draw=black, shape=rectangle, rectangle split, rectangle split parts=2, rectangle split horizontal, rectangle split part fill={zxgreen, zxblue}, rectangle split draw splits=false, minimum height=2mm, font={\tiny}]
\tikzstyle{YZ}=[draw=black, shape=rectangle, rectangle split, rectangle split parts=2, rectangle split horizontal, rectangle split part fill={zxblue, zxgreen}, rectangle split draw splits=false, minimum height=2mm, font={\tiny}]
\tikzstyle{XYa}=[draw=black, shape=rectangle, rectangle split, rectangle split parts=2, rectangle split horizontal, rectangle split part fill={zxred, zxblue}, rectangle split draw splits=false, minimum height=2mm, font={\tiny}]
\tikzstyle{YX}=[draw=black, shape=rectangle, rectangle split, rectangle split parts=2, rectangle split horizontal, rectangle split part fill={zxblue, zxred}, rectangle split draw splits=false, minimum height=2mm, font={\tiny}]
\tikzstyle{tiny_box}=[fill=white, draw=black, shape=rectangle, minimum width=1cm, minimum height=1cm, font={\footnotesize}]
\tikzstyle{scalar}=[fill=white, draw=black, shape=diamond, font={\scriptsize}]
\tikzstyle{black_dot}=[fill=black, draw=black, shape=circle, inner sep=0pt, minimum size=0.2cm, text height=2pt, text depth=0pt]
\tikzstyle{white_dot}=[fill=none, draw=black, shape=circle, inner sep=0pt, minimum size=0.4cm]

\tikzstyle{dashs}=[-, dashed, line width=0.15mm]
\tikzstyle{thick}=[-, line width=0.5mm]
\tikzstyle{arrow}=[->]
\tikzstyle{invisible}=[-, draw=none]
\tikzstyle{functor}=[-, fill={rgb,255: red,240; green,240; blue,240}]
\tikzstyle{boxedge}=[-, fill=white]
\tikzstyle{dashed_line}=[dashed, gray, line width=0.25mm]

\tikzstyle{black_dot}=[fill=black, draw=black, shape=circle, inner sep=0pt, minimum size=0.1cm, text height=2pt, text depth=0pt]
\tikzstyle{white_dot}=[fill=none, draw=black, shape=circle, inner sep=0pt, minimum size=0.2cm]

\tikzstyle{dashed_line}=[dashed, gray, line width=0.25mm]

\tikzstyle{color1}=[fill={rgb,255: red,155; green,155; blue,155}, line width=0.25mm]
\tikzstyle{color2}=[fill={rgb,255: red,180; green,180; blue,180}, line width=0.25mm]
\tikzstyle{color3}=[fill={rgb,255: red,205; green,205; blue,205}, line width=0.25mm]
\tikzstyle{color4}=[fill={rgb,255: red,230; green,230; blue,230}, line width=0.25mm]
\tikzstyle{color5}=[fill={rgb,255: red,255; green,255; blue,255}, line width=0.25mm]

\title{\huge A Quantum Natural Language Processing Approach to Musical Intelligence}
\author[1,2]{Eduardo Reck Miranda}
\author[2]{Richie Yeung}
\author[2]{Anna Pearson}
\author[2]{Konstantinos Meichanetzidis}
\author[2]{Bob Coecke}

\affil[1]{ICCMR, University of Plymouth, Plymouth, UK}
\affil[2]{Cambridge Quantum, Oxford, UK}
\affil[1]{\textit {eduardo.miranda@plymouth.ac.uk}}
\affil[2]{\textit {\{richie.yeung, anna.pearson, k.mei, bob.coecke\}@cambridgequantum.com}}

\begin{document}

\maketitle

\abstract{There has been tremendous progress in Artificial Intelligence (AI) for music, in particular for musical composition and access to large databases for commercialisation through the Internet. We are interested in further advancing this field, focusing on composition. In contrast to current `black-box' AI methods, we are championing an \emph{interpretable compositional} outlook on generative music systems. In particular, we are importing methods from the Distributional Compositional Categorical (DisCoCat) modelling framework for Natural Language Processing (NLP), motivated by musical grammars. Quantum computing is a nascent technology, which is very likely to impact the music industry in time to come. Thus, we are pioneering a Quantum Natural Language Processing (QNLP) approach to develop a new generation of intelligent musical systems. This work follows from previous experimental implementations of DisCoCat linguistic models on quantum hardware. In this chapter, we present \textit{Quanthoven}, the first proof-of-concept ever built, which (a) demonstrates that it is possible to program a quantum computer to learn to classify music that conveys different meanings and (b) illustrates how such a capability might be leveraged to develop a system to compose meaningful pieces of music. After a discussion about our current understanding of music as a communication medium and its relationship to natural language, the chapter focuses on  the techniques developed to (a) encode musical compositions as quantum circuits, and (b) design a quantum classifier. The chapter ends with demonstrations of compositions created with the system.}

\section{Introduction}
\label{subsec:intro}

When people say that John Coltrane's \textit{Alabama} is awesome or that Flow Composer's\footnote{Flow Composer is an AI lead sheet generator developed at Sony CSL Paris \citep{FlowComposer2021}. A lead sheet is a form of musical notation that specifies the essential elements (melody, lyrics, and harmony) of a song.} \textit{Daddy's car} is good, what do they mean by `awesome music' or `good music'? This is debatable. People have different tastes and opinions. And this is true whether the music is made by a human or a machine.

\medskip

The caveat here is not so much to do with the terms `awesome music' or `good music', or whether it is made by humans or machines. The caveat is with the word `mean'.

\medskip

In the last 20 years or so, there has been tremendous progress in Artificial Intelligence (AI) for music. But computers still cannot satisfactorily handle meaning in music in controlled ways that generalises between contexts. There is AI technology today to set up computers to compose a decent pastiche of, say, a Beethoven minuet; e.g., there are connectionist (a.k.a., `neural networks') systems for composition that have been trained on certain genres or styles. (For a comprehensive account of the state of the art of AI for music, please refer to \citep{Miranda2021}.) However, it is very hard to program a computer to compose a piece of music from a request to, say, `generate a piece for Alice's tea party'. How would it know how tea party music should sounds like, or who Alice is? And how would it relate such concepts with algorithms to compose? 

\medskip

A challenging task is to design generative systems with enough knowledge of musical structure, and ability to manipulate said structure, so that given requests for mood, purpose, style, and so on, are appropriately met. Systems that currently attempt to perform such task, specially for music recommendation systems, work in terms of finding and exploiting correlations in large amounts of human-annotated data; e.g., \citep{Recommend2019}. These for example would fill a preference matrix, encoding information such as `likes' for songs, which are categorised already in certain genres by listeners.

\medskip

Following an alternative route, which comes under the umbrella of \emph{Compositional Intelligence} \citep{coecke2021compositionality}, we are taking the first steps in addressing the aforememtioned challenge from a Natural Language Processing (NLP) perspective, which adopts a structure-informed approach. 

\medskip

Providing domain-specific structure to intelligent systems in a controlled and scalable way is a nontrivial challenge: we would need to ensure that the system does not significantly lose flexibility to adapt on different datasets. Nevertheless, having structure present in intelligent systems has potential benefits, such as increased interpretability. Another potential benefit is the need for fewer training parameters; in general, a learning system would need a set of free parameters so that it learns structure from scratch. But in our case, a structural approach is motivated by an analogy between grammar in language, and structure in musical compositions. Specifically to the present work, we are pioneering a Quantum Natural Language Processing (QNLP) approach to develop a new generation of intelligent musical systems. This in turn is motivated by our specific choice of mathematical grammar and a modelling framework for NLP, referred to as Distributional Compositional (DisCo). DisCo models are amenable to implementation on quantum computers due to a shared mathematical structure between grammar and quantum theory.

\medskip

Our aim in this work is to train parameterised quantum circuits to learn to classify music conveying different meanings. And then, use the classifier to compose music that conveys such meanings via rejection sampling. Here, the notion of `meaning' is defined as \textit{perceived properties of musical compositions, which might holistically characterise different types of music}. This is the first step towards the ambition of being able to ask an autonomous intelligent system to generate bespoke pieces of music with particular meanings, required purposes, and so on.

\medskip

Our classification system is trained on human-annotated examples. It scans given examples for musical elements, which we refer to as \emph{musical snippets}. A snippet can be as small as a single musical note. But it can also be a melody, a bar of music, or a group of bars. A main assumption made by the system - known also as the \emph{distributional hypothesis} - is that snippets that occur in similar contexts convey similar meanings. This is a hypothesis that regards meanings of words in text: words that appear in similar contexts convey similar meanings. Then, a combination of the meanings of all snippets determines the meaning of entire compositions, or sections thereof\footnote{For large pieces of music, it is assumed that different sections may convey different meanings, in the same way that sentences or paragraphs convey localised meanings within the overall narrative of a large text.}. These combinations follow \emph{grammatical} rules. Our aim is to input these rules into the machine along with meaning-encodings of the snippets, to allow it to identify different types of music. In the near future we hope it will be able to use this knowledge to compose pieces, or sections thereof, conveying specific meanings. Towards the end the chapter we provide a glimpse of how this might be achieved.

\medskip

Technically, the system assigns types to snippets, in analogy with a parser assigning grammatical part-of-speech tags to words, such as noun, adjective, verb, and so on. Types follow specific compositional rules, and in the specific grammar model that we employ - that is, \textit{pregroup grammar} - types obey algebraic relations.

\medskip

The rules induce a dependency structure onto a musical piece that is composed by the snippets. Then, \emph{meaning} enters the picture by sending types to vector spaces of specific dimensions. Snippet-meaning is encoded inside these vector spaces in terms of multilinear maps, i.e., \emph{tensors}. The compositional rules represented by the dependency structure induce a pattern of tensor-contractions between the tensors.

\medskip

The choice to embed meaning in vector spaces is motivated by a similar mathematical structure between the algebra of the types and the composition of tensors. Importantly, the tensors form a network, which reflects the dependencies as \emph{inherited} by the grammatical structure of a piece of music. However, as we are encoding meaning in tensors, we realise that the computational cost of manipulating and composing tensors is prohibitive for large numbers of snippets, even for mainframe supercomputers, let alone your average desktop. These would require high dimensionality of vector spaces along with a highly connected dependency network.

\medskip

Enter the quantum computer! Quantum processes constitute a powerful choice for a playground in which to encode meaning. After all, quantum theory can be formulated entirely in terms of complex-valued tensor networks. Specifically, quantum theory can be fully considered around one fundamental principle: the way quantum systems compose to form larger ones, and the mathematical operation that jointly describes multiple quantum systems is the tensor product.

\medskip

Emerging quantum computing technology promises formidable processing power for some tasks. As it turns out, and we will see this below, the mathematical tools that are being developed in QNLP research are quantum-native, by yet another analogy between mathematical structures. That is, they enable direct correspondences between language and quantum mechanics, and consequently music. In this case, (a) the meaning of words can be encoded as quantum states and (b) grammars would then correspond to quantum circuits \citep{CoeckeKissinger2017}.

\medskip

Despite the rapid growth in size and quality of quantum hardware in the last decade, quantum processors are still limited in terms of operating speed and noise-tolerance. However, they are adequate to implement proof-of-concept experiments of our envisaged system.

\medskip

In this chapter we present \textit{Quanthoven}, the first proof-of-concept ever built, which (a) demonstrates that it is possible to program a quantum computer to learn to classify music that conveys different meanings and (b) illustrates how such a capability might be leveraged to develop Artificial Intelligence systems able to compose meaningful pieces of music.

\medskip

The remainder of this chapter is structured as follows:
\begin{itemize}
    \item Section 2 explores our current understanding of music as a communication medium and its relationship to natural language, which motivates our research.
    \item Section 3 contextualises our work in relation to approaches to computational modelling of music and algorithmic composition.
    \item Section 4 introduces the very basics of quantum computing, just enough to follow the technical aspects of the quantum classifier (i.e., machine learning algorithm) discussed in this chapter.
    \item Section 5 introduces Distributional Compositional Categorical (DisCoCat) modelling, which uses Category Theory to unify natural language and quantum mechanics. It also shows how DisCoCat can model music.
    \item Section 6 describes the DisCoCat music model that we developed for this project, including the process by which musical pieces are generated using a bespoke grammar and converted into a quantum circuit. It delineates the machine learning algorithm and training methodology.
    \item Section 7 illustrates how the results from the classification can be leveraged to generate two classes of musical compositions.
\end{itemize}

\medskip

The chapter ends with final remarks. A number of appendices provide complementary information.

\section{Music and Meaning}

It is widely accepted that instrumental music\footnote{That is, music without singing. No lyrics. Only musical instruments.} is a non-verbal language. But what does music communicate? What is meaning in music?

\medskip

Some advocate that music communicates nothing meaningful because it cannot express ideas or tell stories in the same way that verbal languages can. Music has no `words' to express things like `Bob', `mussels and fries', `beer', `Bob loves mussels and fries with a beer', and so on. But this is a rather defeatist view.

\medskip

Conversely, others advocate that music can communicate messages of some sort. For example, the imitation of cuckoo birds in Ludwig van Beethoven's Pastoral Symphony is often thought to signify `springtime has arrived' \citep{Monelle1992}. But this is speculation. We do not know if Beethoven really wanted to communicate this. 

\medskip

There have been studies, however, suggesting that different types of chords, rhythms, or melodic shapes may convey specific emotions. For instance, the notion that a minor chord conveys sadness and a major one happiness has been demonstrated experimentally \citep{BakkerMartin2014}. And brain imaging studies have identified correlations between perceived musical signs and specific patterns of brain activity associated with feelings. These studies support the notion that music conveys, if anything, affective states \citep{Koelsch2014}  \citep{Daly2015}. But still, there is no treatise to date on how to communicate a specific feeling in a composition. There are no universal rules for doing this.

\medskip

Yet, it would be perfectly possible to have musical languages matching the expressive power of verbal languages. Creators of Esperanto-like verbal languages, such as Dothraki, made for the TV series \emph{The Game of Thrones}, have developed fully-fledged musical languages \citep{Peterson2015} \citep{Moore2019}. It is just that \emph{Homo sapiens} have not naturally evolved one as such. This is because we evolved musicality for purposes other than verbal communication. Thus, meaning in music is not quite the same as meaning in verbal languages. But what is it? To tackle this question, we ought to examine how the brain makes sense of language and music.

\subsection{Brain Resources Overlap}

Music and verbal languages do have a lot in common. There is scientific evidence that the brain resources that humans deploy to process music and verbal languages overlap to a great degree \citep{Jancke2012}. Stefan Koelsch proposed an ambitious neurocognitive model of music perception supported by extensive research into the neural substrates shared by music and language \citep{Koelsch2011}. Of course, there are differences too. For instance, whereas, the brain engages specialised areas for processing language (e.g., Broca's and Wernicke's areas), music processing tends to be distributed over a number of areas, which are not necessarily specialised for music.

\medskip

Let us consider, for instance, the notion of narrative, in text and music. We seem to deploy cognitive strategies to make sense of music, which are similar to those employed to read a text. This is because verbal languages and music share similar intrinsic structures \citep{PatelMorgan2016}. Or at least our brain thinks that this is the case. This is questionable. But it is a useful assumption, which is somewhat supported by experiments testing if musical cognitive functions can influence linguistic ones, and vice-versa.

\medskip

Notable experiments highlighted the benefits of musical proficiency for the acquisition of linguistic skills, in particular learning a second language \citep{MilovanovTervaniemi2011}  \citep{Patel2011}.  And Dawson et al. provided experimental evidence that native speakers of distinct languages process music differently \citep{Dawson2017}. This seems to be determined by the structural and phonological properties of their respective languages. For example, speakers of languages that has words whose meaning is determined by the duration of their vocalization tend to deal with rhythm more accurately than speakers of languages whose durations do not interfere with meaning.

\subsection{Meaning is Context}

In a similar fashion to narrative in text, musical narrative delineates relationships and changes of state of affairs from event to event \citep{JordanKafalenos1994}. Auditory events in a piece of music acquire significance within a structure that our mind's ear imposes on them. Different types of music define systems of relations among these elements, which induce us to assign them structural impressions; or create categories. The more we listen to the music of certain styles, the more familiar the categories that define such styles become. And the higher the chances that those assigned impressions might be elicited when we listen to such kinds of pieces of music again and again.

\medskip

Thus, we parse and interpret auditory streams according to our own made up mental schemes. We might as well refer to such schemes as \textit{musical grammars}. And the labels we give to impressions as \textit{meanings}.

\medskip

A piece of music may as well make one feel sad or happy, or remember Bob, or think of mussels and fries. And one might even hear Beethoven shouting from his grave that springtime has arrived. Thus, for instance, music with fast repetitive patterns might be categorizsed as `rhythmic', `energetic', or `exciting'. In contrast,  music composed of irregular successions of notes forming melodies and exquisite harmonies might be categorised as `melodic', `relaxing' or `impressionistic'. But this is not as simple as it appears to be. What happens when a piece is prominently rhythmic but is also melodic; e.g., ballroom dance music? We could think of a new category. But what if yet another piece is only slightly more melodic than rhythmic? The boundaries of such categories are not trivial to establish.

\medskip

Anyway, one thing is certain: the more culturally related a group of listeners are, the higher the chances that they might develop shared repertoires of mental schemes and impressions.

\medskip

Incidentally, we learn verbal languages in a similar way to what we just described. We make up the meaning of words and learn how to assemble them in phrases from the context in which they are used. We do this as we grow up and continue doing so throughout our lifetime \citep{Eliot1999}. Perhaps the main difference is that meaning in verbal languages needs to be more precise than meaning in music. That is, meaning is music tends to be more fluid than in language. As stated above, however, one can construct musical languages if one attempts to.

\section{Computational Modelling and Algorithmic Composition}

Scholars from both camps (i.e., linguistics and musicology) have developed comparable models to study linguistic and musical structures, and respective mental schemes \citep{Chomsky2006} \citep{LerdhalJackendoff1996}   \citep{Baroni1999}  \citep{Frank2012}. 

\medskip

Not surprisingly, computational models of linguistic processes have been successfully used to model musical processes and vice-versa. David Cope harnessed transition networks grammars \citep{Cope1991}, which have been used for the analysis and synthesis of natural language \citep{Woods1970}, to develop systems to analyse and synthesise music. A system to generate music using stochastic context-free grammars was proposed by \citep{PerchySarria2016}. Indeed, Noam Chomsky's context-free grammars \citep{HopcroftUllman1979} proved to be relevant for music in many ways. In particular transformative grammars \citep{Chomsky1975}, which suit well the widely adopted method of composing music referred to as variations on a theme. And \citep{Miranda2008} created a system whereby a group of interactive autonomous software agents evolved a common repertoire of intonations (or prosodies) to verbalise words. The agents evolved this by singing to each other. Moreover, evolutionary models informed by neo-Darwinism \citep{Gouyon2002} have been developed to study the origins of language and music almost interchangeably \citep{Miranda2010} \citep{Boer1999}.

\medskip

Generally, there have been two approaches to designing computer systems to generate music, which we refer to as the Artificial Intelligence (or AI) and the algorithmic approaches, respectively \citep{Miranda2014} \citep{Miranda2001}.

\medskip

The AI approach is concerned with embedding the system with musical knowledge to guide the generative process. For instance, computers have been programmed with rules of common practice for counterpoint and voicing in order to generate polyphonic music. And machine-learning technology has enabled computers to learn musical rules automatically from given scores, which are subsequently used to generate music \citep{Cope2000}. The linguistic-informed modelling mentioned above falls in this category.

\medskip

Conversely, the algorithmic approach is concerned with translating data generated from seemingly non-musical phenomena onto music. Examples of this approach abound. Computers have been programmed to generate music with fractals \citep{HsuHsu1990} and chaotic systems \citep{Harley1995}. And also with data from physical phenomena, such as particle collision \citep{Cherston2016}, and DNA sequences \citep{Miranda2020}.

\medskip

Aesthetically, the algorithmic approach tends to generate highly novel and rather unusual music; some may not even consider them as music. The AI approach tends to generate imitations of existing music; that is, imitations in the style of the music that was used to teach the AI system.  Neither approach can, however, satisfactorily 
respond to sophisticated requests such as `generate something for Alice's tea party' or 'compose a tune to make Bob feel energetic'. The QNLP outlook that we are developing is aimed at improving this scenario. In a way, it falls into `AI approach' category, but we are aiming beyond parodying existing styles.

\section{Brief Introduction to Quantum Computing}

A detailed introduction to quantum computing is beyond the scope of this chapter. This can be found in \citep{NielsenChuang_2011_TextBook}, \citep{CoeckeKissinger2017} and \citep{Sutor2019}. Nevertheless, in this section we briefly introduce a few basic concepts deemed necessary to follow the discussions in the following sections.

\medskip

Our starting point is a quantum bit, known as a \emph{qubit}, which is the basic unit of information carrier in a quantum computer. Physically, it is associated with a property of a physical system; e.g., the spin of an electron up or down along some axis. A qubit has a state $\ket{\psi}$, which lives in a 2-dimensional complex vector space, referred to as Hilbert space. The orthonormal basis vectors $\ket{0}$ and $\ket{1}$, related to measurement outcomes 0 and 1,  respectively, allow us to write the most general state of a qubit as a linear combination of the basis vectors known as \textit{superposition}: $\ket{\psi} = \alpha\ket{0} + \beta\ket{1}$ with $\alpha, \beta \in \mathbb{C}$ and $|\alpha|^2 + |\beta|^2 = 1$.

\begin{figure}[h]
\begin{center}\vspace{0.3cm}
\includegraphics[width=0.5\linewidth]{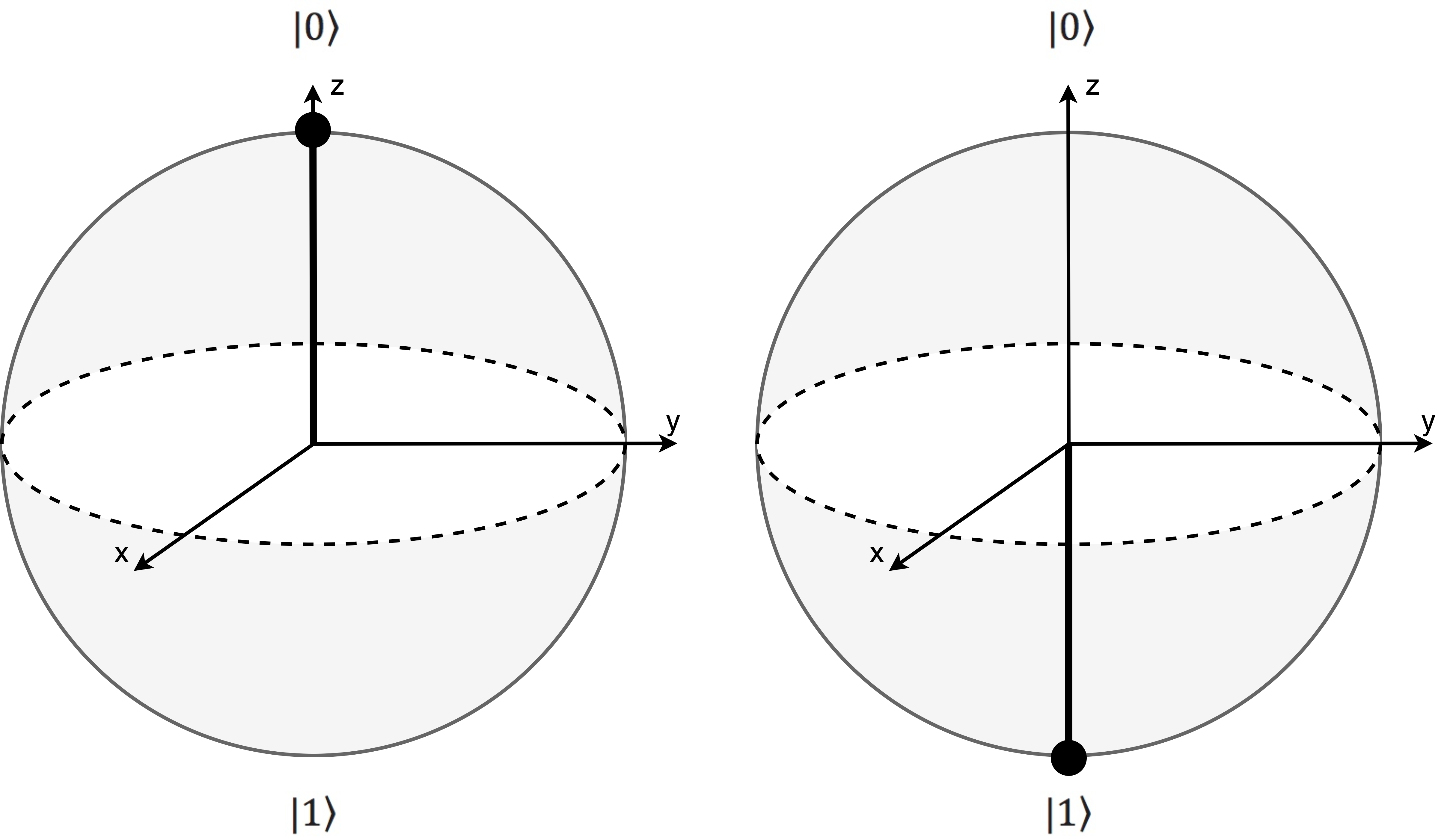}
\caption{The Pauli $\mathrm{\mathbf{X}}$ gate rotates the state vector (pointing upwards on the figure on the left side) by 180 degrees around the x-axis (pointing downwards on the figure on the right).}
\label{fig:BlochX}
\end{center}
\end{figure}

\medskip

An important aspect of quantum computers is that they are fundamentally probabilistic. This leads to the situation where even if one knows that a qubit is in a state $\ket{\psi}$, one can only obtain measurement outcomes $i= \{0,1\}$ with probability given by the Born rule $P(i) = | \braket{i | \psi} |^2$, which gives the square of the norm of the so-called amplitude $\braket{i | \psi}\in\mathbb{C}$. So, $\bra{i}$ is a quantum effect, also known as a `bra' in Dirac notation, and is the dual vector of the state, or `ket', $\ket{i}$. For the general single-qubit superposition in the paragraph above, $P(0) = |\alpha|^2$ and  $ P(1) = |\beta|^2$.

\medskip

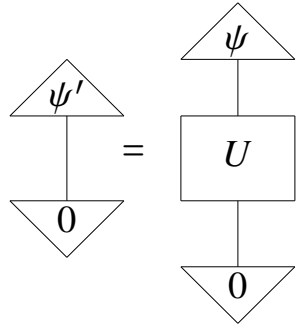
\begin{figure}[h]
\begin{center}
\scalebox{0.5}{\begin{tikzpicture}[{every node/.style}={scale=2}]
	\begin{pgfonlayer}{nodelayer}
		\node [style=none] (0) at (-9.5, 2.75) {};
		\node [style=none] (1) at (-8, 4.25) {};
		\node [style=none] (2) at (-6.5, 2.75) {};
		\node [style=none] (3) at (-9.5, 0.5) {};
		\node [style=none] (4) at (-6.5, 0.5) {};
		\node [style=none] (5) at (-8, -1) {};
		\node [style=none] (6) at (-8, 2.75) {};
		\node [style=none] (7) at (-8, 0.5) {};
		\node [style=none] (8) at (-5, 4.25) {};
		\node [style=none] (9) at (-3.5, 5.75) {};
		\node [style=none] (10) at (-2, 4.25) {};
		\node [style=none] (11) at (-3.5, 4.25) {};
		\node [style=none] (12) at (-2, -1.25) {};
		\node [style=none] (13) at (-3.5, -2.75) {};
		\node [style=none] (14) at (-5, -1.25) {};
		\node [style=none] (15) at (-3.5, -1.25) {};
		\node [style=none] (16) at (-5, 2.75) {};
		\node [style=none] (17) at (-2, 2.75) {};
		\node [style=none] (18) at (-5, 0.5) {};
		\node [style=none] (19) at (-2, 0.5) {};
		\node [style=none] (20) at (-3.5, 2.75) {};
		\node [style=none] (21) at (-3.5, 0.5) {};
		\node [style=none] (22) at (-6.25, 1.75) {\Large{$=$}};
		\node [style=none] (23) at (-8, 3.25) {\large{$\psi'$}};
		\node [style=none] (24) at (-8, 0) {\large{$0$}};
		\node [style=none] (25) at (-3.5, 1.75) {\large{$U$}};
		\node [style=none] (26) at (-3.5, 4.75) {\large{$\psi$}};
		\node [style=none] (27) at (-3.5, -1.75) {\large{$0$}};
	\end{pgfonlayer}
	\begin{pgfonlayer}{edgelayer}
		\draw (0.center) to (1.center);
		\draw (1.center) to (2.center);
		\draw (0.center) to (2.center);
		\draw (3.center) to (4.center);
		\draw (4.center) to (5.center);
		\draw (3.center) to (5.center);
		\draw (6.center) to (7.center);
		\draw (8.center) to (9.center);
		\draw (9.center) to (10.center);
		\draw (8.center) to (10.center);
		\draw (12.center) to (13.center);
		\draw (13.center) to (14.center);
		\draw (12.center) to (14.center);
		\draw (16.center) to (17.center);
		\draw (16.center) to (18.center);
		\draw (18.center) to (19.center);
		\draw (17.center) to (19.center);
		\draw (11.center) to (20.center);
		\draw (21.center) to (15.center);
	\end{pgfonlayer}
\end{tikzpicture}}
\caption{Diagram showing the evolution of an isolated qubit in initial state $\ket{\psi}$ with unitary map U composed with effect $\bra{0}$. (Taken from \citep{Lorenz2021} and used with permission.) }
\label{fig:robin1}
\end{center}
\end{figure}

In order to picture a single qubit, imagine a transparent sphere with opposite poles. From its centre, a vector whose length is equal to the radius of the sphere can point to anywhere on the surface. This sphere is called the \textit{Bloch sphere} and the vector is referred to as a \textit{state vector} (Fig.  \ref{fig:BlochX}). Before measurement, the evolution of a single qubit (that is, a qubit that is not interacting with any other qubits) is described by the transformation of its state vector with a unitary linear map $U$, so that $\ket{\psi'} = U\ket{\psi}$ (Fig. \ref{fig:robin1}). However if a qubit interacts with other qubits, then things become a little more complicated; more about this below.

In simple terms, quantum computers are programmed by applying sequences of unitary linear maps to qubits. Programming languages for quantum computing provide a number of such linear maps, referred to as \emph{gates}, which act on qubits. For instance, the `not gate', rotates the state vector by 180 degrees around the x-axis of the Bloch sphere (Fig. \ref{fig:BlochX}); that is, if the qubit vector is pointing to $\ket{0}$, then this gate flips it to $\ket{1}$, or vice-versa. This gate is often referred to as the `Pauli $\mathrm{\mathbf{X}}$ gate'. A more generic rotation $\mathrm{\mathbf{Rx}}(\theta)$ gate is typically available for quantum programming, where the angle for the rotation around the x-axis is specified. Obviously, $\mathrm{\mathbf{Rx}}(\pi)$ applied to $\ket{0}$ or $\ket{1}$  is equivalent to applying $\mathrm{\mathbf{X}}$ to $\ket{0}$  or $\ket{1}$.  Similarly, there are $\mathrm{\mathbf{Rz}}(\varphi)$ and $\mathrm{\mathbf{Ry}}(\theta)$ gates for rotations on the z-axis and y-axis of the Bloch sphere, respectively. An even more generic gate is typically available, which is a unitary rotation gate, with 3 Euler angles: $\mathrm{\mathbf{U(\theta, \varphi, \lambda)}}$. Essentially, all single-qubit quantum gates perform rotations, which change the amplitude distribution of the system. And in fact, any qubit rotation can be specified in terms of $\mathrm{\mathbf{U(\theta, \varphi, \lambda)}}$; for instance $\mathrm{\mathbf{Rx}}(\theta) = \mathrm{\mathbf{U}}(\theta, - \frac{\pi}{2}, \frac{\pi}{2}$).

\medskip

Quantum computation gets really interesting with gates that operate on multiple qubits, such as the controlled $\mathrm{\mathbf{X}}$ gate, or $\mathrm{\mathbf{CX}}$ gate; commonly referred to as the $\mathrm{\mathbf{CNOT}}$ gate. The $\mathrm{\mathbf{CNOT}}$ gate puts two qubits in \textit{entanglement}.

\medskip

Entanglement establishes a curious correlation between qubits. When considering its action on the computational states, the $\mathrm{\mathbf{CNOT}}$ gate applies an $\mathrm{\mathbf{X}}$ gate on a qubit only if the state of another qubit is $\ket{1}$. Thus, the $\mathrm{\mathbf{CNOT}}$  gate establishes a dependency of the state of one qubit with the value of another. In practice, any quantum gate can be made conditional and entanglement can take place between more than two qubits.

\medskip
An important gate for quantum computing is the Hadamard gate (referred to as the `$\mathrm{\mathbf{H}}$ gate'). It puts the qubit into a balanced superposition state consisting of an equal-weighted combination of two opposing states:$\left| \alpha \right|^2=0.5$ and   $\left| \beta \right|^2=0.5$.

\medskip

A combination of $\mathrm{\mathbf{H}}$ and $\mathrm{\mathbf{CNOT}}$ gates enables the implementation of the so-called Bell states; a form of maximally entangled qubits, which is explored later on in this chapter to represent grammatical structures.

\medskip

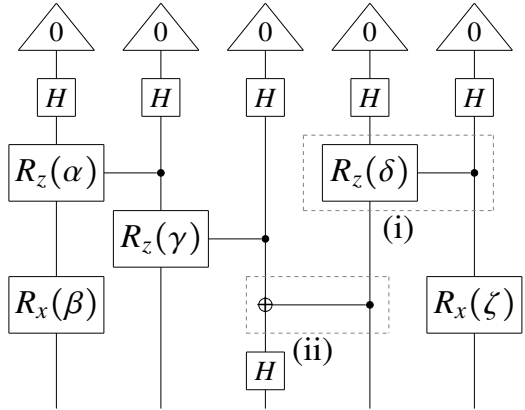
\begin{figure}[h]
\begin{center}
\scalebox{0.5}{\begin{tikzpicture}[{every node/.style}={scale=1.75}]
	\begin{pgfonlayer}{nodelayer}
		\node [style=none] (0) at (-9, 3.25) {};
		\node [style=none] (1) at (-8, 4.5) {};
		\node [style=none] (2) at (-7, 3.25) {};
		\node [style=none] (6) at (-8, 3.25) {};
		\node [style=none] (16) at (-8.5, 2.5) {};
		\node [style=none] (17) at (-7.5, 2.5) {};
		\node [style=none] (18) at (-8.5, 1.5) {};
		\node [style=none] (19) at (-7.5, 1.5) {};
		\node [style=none] (23) at (-8, 3.75) {{$0$}};
		\node [style=none] (28) at (-8, 2.5) {};
		\node [style=none] (30) at (-8, 2) {{$H$}};
		\node [style=none] (32) at (-9.25, 0.75) {};
		\node [style=none] (33) at (-6.75, 0.75) {};
		\node [style=none] (34) at (-9.25, -0.75) {};
		\node [style=none] (35) at (-6.75, -0.75) {};
		\node [style=none] (37) at (-8, 0) {\large{$R_z(\alpha)$}};
		\node [style=none] (38) at (-8, 0.75) {};
		\node [style=none] (39) at (-8, -0.75) {};
		\node [style=none] (40) at (-9.25, -2.75) {};
		\node [style=none] (41) at (-6.75, -2.75) {};
		\node [style=none] (42) at (-9.25, -4.25) {};
		\node [style=none] (43) at (-6.75, -4.25) {};
		\node [style=none] (45) at (-8, -3.5) {\large{$R_x(\beta)$}};
		\node [style=none] (46) at (-8, -2.75) {};
		\node [style=none] (47) at (-8, -4.25) {};
		\node [style=none] (48) at (-8, -6.25) {};
		\node [style=none] (49) at (-8, 1.5) {};
		\node [style=none] (50) at (-6.25, 3.25) {};
		\node [style=none] (51) at (-5.25, 4.5) {};
		\node [style=none] (52) at (-4.25, 3.25) {};
		\node [style=none] (53) at (-5.25, 3.25) {};
		\node [style=none] (54) at (-5.75, 2.5) {};
		\node [style=none] (55) at (-4.75, 2.5) {};
		\node [style=none] (56) at (-5.75, 1.5) {};
		\node [style=none] (57) at (-4.75, 1.5) {};
		\node [style=none] (58) at (-5.25, 3.75) {{$0$}};
		\node [style=none] (59) at (-5.25, 2.5) {};
		\node [style=none] (60) at (-5.25, 2) {{$H$}};
		\node [style=none] (61) at (-6.5, -1) {};
		\node [style=none] (62) at (-4, -1) {};
		\node [style=none] (63) at (-6.5, -2.5) {};
		\node [style=none] (64) at (-4, -2.5) {};
		\node [style=none] (65) at (-5.25, -1.75) {\large{$R_z(\gamma)$}};
		\node [style=none] (66) at (-5.25, -1) {};
		\node [style=none] (67) at (-5.25, -2.5) {};
		\node [style=none] (76) at (-5.25, 1.5) {};
		\node [style=none] (77) at (-5.25, -6.25) {};
		\node [style=none] (78) at (-6.75, 0) {};
		\node [style={black_dot}] (79) at (-5.25, 0) {};
		\node [style=none] (80) at (-3.5, 3.25) {};
		\node [style=none] (81) at (-2.5, 4.5) {};
		\node [style=none] (82) at (-1.5, 3.25) {};
		\node [style=none] (83) at (-2.5, 3.25) {};
		\node [style=none] (84) at (-3, 2.5) {};
		\node [style=none] (85) at (-2, 2.5) {};
		\node [style=none] (86) at (-3, 1.5) {};
		\node [style=none] (87) at (-2, 1.5) {};
		\node [style=none] (88) at (-2.5, 3.75) {{$0$}};
		\node [style=none] (89) at (-2.5, 2.5) {};
		\node [style=none] (90) at (-2.5, 2) {{$H$}};
		\node [style=none] (98) at (-2.5, 1.5) {};
		\node [style={black_dot}] (100) at (-2.5, -1.75) {};
		\node [style=none] (101) at (-2.5, -4.75) {};
		\node [style=none] (102) at (-4, -1.75) {};
		\node [style=none] (103) at (-0.75, 3.25) {};
		\node [style=none] (104) at (0.25, 4.5) {};
		\node [style=none] (105) at (1.25, 3.25) {};
		\node [style=none] (106) at (0.25, 3.25) {};
		\node [style=none] (107) at (-0.25, 2.5) {};
		\node [style=none] (108) at (0.75, 2.5) {};
		\node [style=none] (109) at (-0.25, 1.5) {};
		\node [style=none] (110) at (0.75, 1.5) {};
		\node [style=none] (111) at (0.25, 3.75) {{$0$}};
		\node [style=none] (112) at (0.25, 2.5) {};
		\node [style=none] (113) at (0.25, 2) {{$H$}};
		\node [style=none] (114) at (-1, 0.75) {};
		\node [style=none] (115) at (1.5, 0.75) {};
		\node [style=none] (116) at (-1, -0.75) {};
		\node [style=none] (117) at (1.5, -0.75) {};
		\node [style=none] (118) at (0.25, 0) {\large{$R_z(\delta)$}};
		\node [style=none] (119) at (0.25, 0.75) {};
		\node [style=none] (120) at (0.25, -0.75) {};
		\node [style=none] (121) at (0.25, 1.5) {};
		\node [style=none] (122) at (0.25, -6.25) {};
		\node [style=none] (124) at (1.5, 0) {};
		\node [style={white_dot}] (125) at (-2.5, -3.5) {};
		\node [style={black_dot}] (126) at (0.25, -3.5) {};
		\node [style=none] (127) at (2, 3.25) {};
		\node [style=none] (128) at (3, 4.5) {};
		\node [style=none] (129) at (4, 3.25) {};
		\node [style=none] (130) at (3, 3.25) {};
		\node [style=none] (131) at (2.5, 2.5) {};
		\node [style=none] (132) at (3.5, 2.5) {};
		\node [style=none] (133) at (2.5, 1.5) {};
		\node [style=none] (134) at (3.5, 1.5) {};
		\node [style=none] (135) at (3, 3.75) {{$0$}};
		\node [style=none] (136) at (3, 2.5) {};
		\node [style=none] (137) at (3, 2) {{$H$}};
		\node [style=none] (138) at (1.75, -2.75) {};
		\node [style=none] (139) at (4.25, -2.75) {};
		\node [style=none] (140) at (1.75, -4.25) {};
		\node [style=none] (141) at (4.25, -4.25) {};
		\node [style=none] (142) at (3, -3.5) {\large{$R_x(\zeta)$}};
		\node [style=none] (143) at (3, -2.75) {};
		\node [style=none] (144) at (3, -4.25) {};
		\node [style=none] (145) at (3, 1.5) {};
		\node [style=none] (146) at (3, -6.25) {};
		\node [style=none] (147) at (4.25, -3.5) {};
		\node [style={black_dot}] (148) at (3, 0) {};
		\node [style=none] (149) at (-3, -2.75) {};
		\node [style=none] (150) at (0.75, -2.75) {};
		\node [style=none] (151) at (-3, -4.25) {};
		\node [style=none] (152) at (0.75, -4.25) {};
		\node [style=none] (153) at (-1.5, 1) {};
		\node [style=none] (154) at (3.5, 1) {};
		\node [style=none] (155) at (-1.5, -1) {};
		\node [style=none] (156) at (3.5, -1) {};
		\node [style=none] (157) at (-1.25, -4.75) {\large{(ii)}};
		\node [style=none] (158) at (1, -1.5) {\large{(i)}};
		\node [style=none] (159) at (-2.5, -3.5) {\large{$+$}};
		\node [style=none] (160) at (-3, -4.75) {};
		\node [style=none] (161) at (-2, -4.75) {};
		\node [style=none] (162) at (-3, -5.75) {};
		\node [style=none] (163) at (-2, -5.75) {};
		\node [style=none] (165) at (-2.5, -5.25) {{$H$}};
		\node [style=none] (166) at (-2.5, -5.75) {};
		\node [style=none] (167) at (-2.5, -6.25) {};
	\end{pgfonlayer}
	\begin{pgfonlayer}{edgelayer}
		\draw (0.center) to (1.center);
		\draw (1.center) to (2.center);
		\draw (0.center) to (2.center);
		\draw (16.center) to (17.center);
		\draw (16.center) to (18.center);
		\draw (18.center) to (19.center);
		\draw (17.center) to (19.center);
		\draw (6.center) to (28.center);
		\draw (32.center) to (33.center);
		\draw (32.center) to (34.center);
		\draw (34.center) to (35.center);
		\draw (33.center) to (35.center);
		\draw (40.center) to (41.center);
		\draw (40.center) to (42.center);
		\draw (42.center) to (43.center);
		\draw (41.center) to (43.center);
		\draw (49.center) to (38.center);
		\draw (39.center) to (46.center);
		\draw (47.center) to (48.center);
		\draw (50.center) to (51.center);
		\draw (51.center) to (52.center);
		\draw (50.center) to (52.center);
		\draw (54.center) to (55.center);
		\draw (54.center) to (56.center);
		\draw (56.center) to (57.center);
		\draw (55.center) to (57.center);
		\draw (53.center) to (59.center);
		\draw (61.center) to (62.center);
		\draw (61.center) to (63.center);
		\draw (63.center) to (64.center);
		\draw (62.center) to (64.center);
		\draw (76.center) to (66.center);
		\draw (67.center) to (77.center);
		\draw (78.center) to (79);
		\draw (80.center) to (81.center);
		\draw (81.center) to (82.center);
		\draw (80.center) to (82.center);
		\draw (84.center) to (85.center);
		\draw (84.center) to (86.center);
		\draw (86.center) to (87.center);
		\draw (85.center) to (87.center);
		\draw (83.center) to (89.center);
		\draw (98.center) to (101.center);
		\draw (102.center) to (100);
		\draw (103.center) to (104.center);
		\draw (104.center) to (105.center);
		\draw (103.center) to (105.center);
		\draw (107.center) to (108.center);
		\draw (107.center) to (109.center);
		\draw (109.center) to (110.center);
		\draw (108.center) to (110.center);
		\draw (106.center) to (112.center);
		\draw (114.center) to (115.center);
		\draw (114.center) to (116.center);
		\draw (116.center) to (117.center);
		\draw (115.center) to (117.center);
		\draw (121.center) to (119.center);
		\draw (120.center) to (122.center);
		\draw (125) to (126);
		\draw (127.center) to (128.center);
		\draw (128.center) to (129.center);
		\draw (127.center) to (129.center);
		\draw (131.center) to (132.center);
		\draw (131.center) to (133.center);
		\draw (133.center) to (134.center);
		\draw (132.center) to (134.center);
		\draw (130.center) to (136.center);
		\draw (138.center) to (139.center);
		\draw (138.center) to (140.center);
		\draw (140.center) to (141.center);
		\draw (139.center) to (141.center);
		\draw (145.center) to (143.center);
		\draw (144.center) to (146.center);
		\draw (124.center) to (148);
		\draw [style={dashed_line}] (149.center) to (150.center);
		\draw [style={dashed_line}] (149.center) to (151.center);
		\draw [style={dashed_line}] (151.center) to (152.center);
		\draw [style={dashed_line}] (150.center) to (152.center);
		\draw [style={dashed_line}] (153.center) to (155.center);
		\draw [style={dashed_line}] (153.center) to (154.center);
		\draw [style={dashed_line}] (154.center) to (156.center);
		\draw [style={dashed_line}] (155.center) to (156.center);
		\draw (160.center) to (161.center);
		\draw (160.center) to (162.center);
		\draw (162.center) to (163.center);
		\draw (161.center) to (163.center);
		\draw (166.center) to (167.center);
	\end{pgfonlayer}
\end{tikzpicture}}
\caption{Example of a quantum circuit of the kind used in our system: the Hadamard gate $\mathrm{\mathbf{H}}$, the X-rotation gate $\mathrm{\mathbf{Rx(\beta)}}$ by angle $\beta$, the controlled Z-rotation gate (i), part of which is a Z-rotation gate $\mathrm{\mathbf{Rz(\delta)}}$ by angle $\delta$, and finally the $\mathrm{\mathbf{CNOT}}$ gate. (Taken from \citep{Lorenz2021} and used with permission.)}
\label{fig:robin2}
\end{center}
\end{figure}

A quantum program is often depicted as a circuit diagram of quantum gates, showing sequences of gate operations on the qubits (Figure \ref{fig:robin2}). So, if one has $n$ qubits, then their joint state space is given by the tensor product of their individual state spaces and has dimension $2^n$. The evolution of many qubits interacting with each other is given by a (global) exponentially large unitary map acting on the entire joint exponentially large state space.

\medskip

There are two ways of thinking of a quantum circuit. On an abstract level, it can be viewed as the application of linear map to a vector, which computes the entire state of the system at a later time. As stated above, these complex-valued matrices and vectors are exponentially large in terms of the amount of qubits required to encode them. Therefore, simulations of quantum systems are believed to be a hard task for classical computers. And on a physical, or operational, level, a quantum circuit is a set of instructions for a quantum computer to execute.

\medskip

In the circuit model of quantum computation, qubits typically start at easily preparable states, $\ket{0}$, and then a sequence of gates are applied. Next, the qubits are read and the results are stored in standard digital memory, which are accessible for further handling. In practice, a quantum computer works alongside a classical computer, which in effect acts as the interface between the user and the quantum machine. The classical computer enables the user to handle the measurements for practical applications. As a quantum computer is probabilistic, a circuit must be run several times to give statistics to estimate outcome probabilities. The design of a circuit must be such that it encodes the problem in a way that the outcome probabilities obtained give the answer to what one wishes to solve. 

\medskip

Building and running a quantum computer is an engineering task on a completely different scale to that of building and running a classical computer. Not only must qubits be shielded from random errors picked up from the environment, one must also avoid unwanted interactions between multiple qubits within the same device. In order to avoid such problems, the number of successive operations on qubits (i.e., `circuit depth') are limited on current quantum hardware. It is expected that to give an advantage for large scale problems one must have many fault-tolerant qubits, which are obtained by error correction techniques. These involve encoding the state of a `logical' qubit in the state of many `physical' qubits. As this is a difficult engineering task, quantum computers with fault-tolerant qubits are not available at the time of writing. Current machines are known as Noisy Intermediate-Scale Quantum (NISQ) Computers.  At the time of writing, they have circa 100 physical qubits. But this number is increasing fast.

\section{DisCoCat Modelling}

Development in the growing new field of QNLP is greatly facilitated by the Distributional Compositional Categorical (DisCoCat) modelling of natural language semantics. In DisCoCat, grammar dictates the composition of word-meanings to derive the meaning of a whole sentence \citep{Coecke2010}.

\medskip

DisCoCat is a natural framework to develop natural language-like generative music systems with quantum computing.  At the core of DisCoCat, as the model was originally formulated, is Joachim Lambek's, algebraic model of pregroup grammar; the curious reader may refer to \citep{Coecke2013} for a more rigorous mathematical treatment.

\medskip

In DisCoCat, the compositional structure of language is captured by (symmetric) monoidal categories, a.k.a. \emph{process theories}. This mathematical formalism comes equipped with a formal graphical notation in the form of \emph{string diagrams}. Meaning is encoded by a mapping, which endows semantics to the diagram representing the grammatical structure of a text. This is in fact how DisCoCat was introduced in the first instance, by sending words to vectors, or in general, higher-order tensors, which are connected according to grammatical dependencies to form a tensor network. Contracting the tensors along the connections results in the derivation of the meaning of a text.

\medskip

Categorical quantum mechanics (CQM) is a framework that reformulates quantum theory in terms of process theories. In this context, string diagrams describe quantum protocols involving quantum states, transformations, and measurements \citep{AbramskyCoecke2008},  \citep{CoeckeKissinger2017}. Low-level and fine-grained diagrammatic calculi, which build on the philosophy of CQM, such as ZX-calculus \citep{CoeckeDuncan2011}, are promising complementary --- or even alternative --- approaches to reasoning about quantum systems and processes. We see then that string diagrams provide a common formalism, via which we can associate language with quantum theory. After all, Hilbert spaces, which is where quantum states are encoded, are vector spaces. By analogy, many-body quantum states encode word-meanings. And grammatical reductions correspond to processes such as quantum maps, quantum effects, and measurements. To take advantage of such analogy, we employ DisCoPy \citep{Felice2020} to perform grammar-based musical experiments on quantum processors.  DisCoPy is an open-source toolbox for manipulating string diagrams and implementing mappings to underlying semantics of our choosing. 

\medskip

Among its features, DisCoPy includes tools for defining Context-Free Grammars (CFGs), Lambek's pregroup grammars \citep{Lambek1999}, tensor networks \citep{Orus2014}, ZX-calculus, and other diagram-based reasoning processes, under the same compositional paradigm. In essence, and particular to this work, DisCoPy provides the means for mapping representations of musical systems into quantum circuits, encoding quantum processes, to be implemented on a quantum computer. 

\medskip

Prior to the work preseted here, \citep{Meichanetzidis2020} and \citep{Meichanetzidis_2021} used DisCoPy and Cambridge Quantum's compiler $\textrm{t}|\textrm{ket}\rangle$ \citep{Sivarajah_2020} to develop and deploy a pioneering QNLP experiment on quantum hardware. Here the sentences were represented as parameterised quantum circuits and the parameters were trained to perform a sentence-classification task. A larger-scale experiment of the same nature was reported by \citep{Lorenz2021}. The models built for those experiments have been further developed into a Python library for QNLP known as \texttt{lambeq} \citep{kartsaklis2021lambeq}, which is named in homage to Joachim Lambek. Indeed, part of the work reported in this chapter was conducted in tandem with the development of \texttt{lambeq}.

\subsection{A Musical DisCoCat Model}

In pregroup grammars, the syntax of a sentence is captured by a finite product of words of different pregroup types. In our case, musical snippets will play the role of words, and musical compositions the role of sentences.

\medskip

Let us define a musical composition $\sigma$ as a finite product of musical snippets $w$ as follows: $\sigma = \prod w_i$, where the product operation captures the sequential nature of the string representing a musical sequence, in analogy with words being placed side by side in a sentence. Each snippet is assigned a pregroup type $t_w = \prod b_i^{k_i}$ comprising a product of basic types $b_i$ from a set of types $B$. Basic types also have adjoints: ${k_i} \in\{\dots, ll, l, \_, r, rr, \dots\}$. Then, the type of a musical composition $\sigma$ is simply the product of the types of its snippets.

\medskip

 A composition is deemed valid, that is, grammatically correct --- or musical --- if and only if its type reduces to a special type $s \in B$. These reductions take place by means of an algebraic process of pair-wise annihilations of basic types and their adjoints according to the following rules: $b^l \cdot b \to 1$, $b \cdot b^r \to 1$, $b^{ll} \cdot b^l \to 1$, $b^r \cdot b^{rr} \to 1$, $\dots$. To visualise this, let's go back to natural language and consider a sentence with a transitive verb: $\emph{Bob plays guitar}$. Here, the transitive verb is a process that expects two input nouns of, say, type $n$, on each side, in order to output a sentence type $s$. The type of the transitive verb is therefore denoted as $n^r \cdot s \cdot n^l$.  There exists a grammatical reduction following the algebraic rules of pregroup types, as shown in Eq. \ref{eq:pregroup_types}.

\begin{equation}
n \cdot (n^r \cdot s \cdot n^l) \cdot n \to (n \cdot n^r) \cdot s \cdot (n^l \cdot n) \to 1 \cdot s \cdot 1 \to s
\label{eq:pregroup_types}
\end{equation}

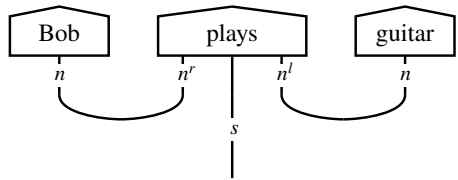
\begin{figure}[h]
\begin{center}
\scalebox{1.3}{\begin{tikzpicture}[baseline={(current bounding box).east}, node distance=5mm, text height=5pt, text depth=0pt, {every node/.style}={scale=0.6}]
	\begin{pgfonlayer}{nodelayer}
		\node [style=none] (0) at (2.75, 0) {};
		\node [style=none, fill=white, anchor=west, scale=0.85] (1) at (3.125, -0.2) {$n$};
		\node [style=none] (2) at (3.25, 0) {};
		\node [style=none] (3) at (3.25, -0.4) {};
		\node [style=none] (4) at (2.75, 0) {};
		\node [style=none] (5) at (3.75, 0) {};
		\node [style=none] (6) at (3.75, 0.4) {};
		\node [style=none] (7) at (3.25, 0.5) {};
		\node [style=none] (8) at (2.75, 0.4) {};
		\node [style=none, fill=white, scale=1] (9) at (3.25, 0.2) {Bob};
		\node [style=none, fill=white, anchor=west, scale=0.85] (10) at (4.375, -0.2) {$n^{r}$};
		\node [style=none] (11) at (4.5, 0) {};
		\node [style=none] (12) at (4.5, -0.4) {};
		\node [style=none] (14) at (5, 0) {};
		\node [style=none] (15) at (5, -0.4) {};
		\node [style=none, fill=white, anchor=west, scale=0.85] (16) at (5.375, -0.2) {$n^{l}$};
		\node [style=none] (17) at (5.5, 0) {};
		\node [style=none] (18) at (5.5, -0.4) {};
		\node [style=none] (19) at (4.25, 0) {};
		\node [style=none] (20) at (5.75, 0) {};
		\node [style=none] (21) at (5.75, 0.4) {};
		\node [style=none] (22) at (5, 0.5) {};
		\node [style=none] (23) at (4.25, 0.4) {};
		\node [style=none, fill=white, scale=1] (24) at (5, 0.2) {plays};
		\node [style=none, fill=white, anchor=west, scale=0.85] (25) at (6.625, -0.2) {$n$};
		\node [style=none] (26) at (6.75, 0) {};
		\node [style=none] (27) at (6.75, -0.4) {};
		\node [style=none] (28) at (6.25, 0) {};
		\node [style=none] (29) at (7.25, 0) {};
		\node [style=none] (30) at (7.25, 0.4) {};
		\node [style=none] (31) at (6.75, 0.5) {};
		\node [style=none] (32) at (6.25, 0.4) {};
		\node [style=none, fill=white, scale=1] (33) at (6.75, 0.2) {guitar};
		\node [style=none] (34) at (3.875, -0.65) {};
		\node [style=none] (35) at (6.125, -0.65) {};
		\node [style=none] (36) at (5, -1.25) {};
		\node [style=none, fill=white, anchor=west, scale=0.85] (37) at (4.9, -0.75) {$s$};
	\end{pgfonlayer}
	\begin{pgfonlayer}{edgelayer}
		\draw [in=90, out=-90] (2.center) to (3.center);
		\draw [-, fill=white] (4.center)
			 to (5.center)
			 to (6.center)
			 to (7.center)
			 to (8.center)
			 to cycle;
		\draw [in=90, out=-90] (11.center) to (12.center);
		\draw [in=90, out=-90] (14.center) to (15.center);
		\draw [in=90, out=-90] (17.center) to (18.center);
		\draw [-, fill=white] (19.center)
			 to (20.center)
			 to (21.center)
			 to (22.center)
			 to (23.center)
			 to cycle;
		\draw [in=90, out=-90] (26.center) to (27.center);
		\draw [-, fill=white] (28.center)
			 to (29.center)
			 to (30.center)
			 to (31.center)
			 to (32.center)
			 to cycle;
		\draw [in=180, out=-90, looseness=0.72] (3.center) to (34.center);
		\draw [in=0, out=-90, looseness=0.72] (12.center) to (34.center);
		\draw [in=180, out=-90, looseness=0.72] (18.center) to (35.center);
		\draw [in=0, out=-90, looseness=0.72] (27.center) to (35.center);
		\draw [in=90, out=-90] (15.center) to (36.center);
	\end{pgfonlayer}
\end{tikzpicture}}
\caption{Pregroup diagram for the transitive sentence `Bob plays guitar'.}
\label{fig:fig1}
\end{center}
\end{figure}

\medskip
The DisCoCat string diagram for Eq. \ref{eq:pregroup_types} is shown in Fig. \ref{fig:fig1}. The wires carry a type. They bent in a U-shape, or cups, represent the reductions. Complex networks of processes and relationships can be designed by connecting boxes with input and output wires, and observing that the types are respected and are reducible as required.

\medskip

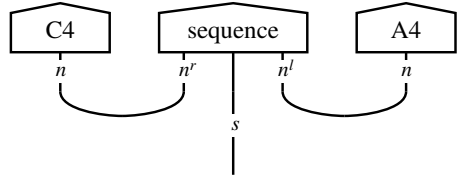
\begin{figure}[h]
\begin{center}
\scalebox{1.3}{\begin{tikzpicture}[baseline={(current bounding box).east}, node distance=5mm, text height=5pt, text depth=0pt, {every node/.style}={scale=0.6}]
	\begin{pgfonlayer}{nodelayer}
		\node [style=none] (0) at (2.75, 0) {};
		\node [style=none, fill=white, anchor=west, scale=0.85] (1) at (3.125, -0.2) {$n$};
		\node [style=none] (2) at (3.25, 0) {};
		\node [style=none] (3) at (3.25, -0.4) {};
		\node [style=none] (4) at (2.75, 0) {};
		\node [style=none] (5) at (3.75, 0) {};
		\node [style=none] (6) at (3.75, 0.4) {};
		\node [style=none] (7) at (3.25, 0.5) {};
		\node [style=none] (8) at (2.75, 0.4) {};
		\node [style=none, fill=white, scale=1] (9) at (3.25, 0.2) {C4};
		\node [style=none, fill=white, anchor=west, scale=0.85] (10) at (4.375, -0.2) {$n^{r}$};
		\node [style=none] (11) at (4.5, 0) {};
		\node [style=none] (12) at (4.5, -0.4) {};
		\node [style=none] (14) at (5, 0) {};
		\node [style=none] (15) at (5, -0.4) {};
		\node [style=none, fill=white, anchor=west, scale=0.85] (16) at (5.375, -0.2) {$n^{l}$};
		\node [style=none] (17) at (5.5, 0) {};
		\node [style=none] (18) at (5.5, -0.4) {};
		\node [style=none] (19) at (4.25, 0) {};
		\node [style=none] (20) at (5.75, 0) {};
		\node [style=none] (21) at (5.75, 0.4) {};
		\node [style=none] (22) at (5, 0.5) {};
		\node [style=none] (23) at (4.25, 0.4) {};
		\node [style=none, fill=white, scale=1] (24) at (5, 0.2) {sequence};
		\node [style=none, fill=white, anchor=west, scale=0.85] (25) at (6.625, -0.2) {$n$};
		\node [style=none] (26) at (6.75, 0) {};
		\node [style=none] (27) at (6.75, -0.4) {};
		\node [style=none] (28) at (6.25, 0) {};
		\node [style=none] (29) at (7.25, 0) {};
		\node [style=none] (30) at (7.25, 0.4) {};
		\node [style=none] (31) at (6.75, 0.5) {};
		\node [style=none] (32) at (6.25, 0.4) {};
		\node [style=none, fill=white, scale=1] (33) at (6.75, 0.2) {A4};
		\node [style=none] (34) at (3.875, -0.65) {};
		\node [style=none] (35) at (6.125, -0.65) {};
		\node [style=none] (36) at (5, -1.25) {};
		\node [style=none, fill=white, anchor=west, scale=0.85] (37) at (4.9, -0.75) {$s$};
	\end{pgfonlayer}
	\begin{pgfonlayer}{edgelayer}
		\draw [in=90, out=-90] (2.center) to (3.center);
		\draw [-, fill=white] (4.center)
			 to (5.center)
			 to (6.center)
			 to (7.center)
			 to (8.center)
			 to cycle;
		\draw [in=90, out=-90] (11.center) to (12.center);
		\draw [in=90, out=-90] (14.center) to (15.center);
		\draw [in=90, out=-90] (17.center) to (18.center);
		\draw [-, fill=white] (19.center)
			 to (20.center)
			 to (21.center)
			 to (22.center)
			 to (23.center)
			 to cycle;
		\draw [in=90, out=-90] (26.center) to (27.center);
		\draw [-, fill=white] (28.center)
			 to (29.center)
			 to (30.center)
			 to (31.center)
			 to (32.center)
			 to cycle;
		\draw [in=180, out=-90, looseness=0.72] (3.center) to (34.center);
		\draw [in=0, out=-90, looseness=0.72] (12.center) to (34.center);
		\draw [in=180, out=-90, looseness=0.72] (18.center) to (35.center);
		\draw [in=0, out=-90, looseness=0.72] (27.center) to (35.center);
		\draw [in=90, out=-90] (15.center) to (36.center);
	\end{pgfonlayer}
\end{tikzpicture}}
\caption{Pregroup diagram for musical form shown in Fig. \ref{fig:fig3}.}
\label{fig:fig2}
\end{center}
\end{figure}

\begin{figure}[h]
\begin{center}
\includegraphics[width=0.24\textwidth]{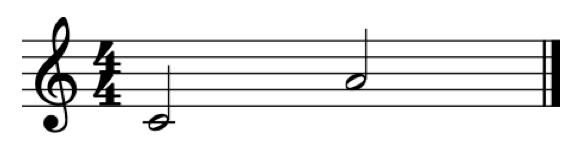}
\caption{A sequence of two notes: C4 and A4.}
\label{fig:fig3}
\end{center}
\end{figure}

Such pregroup modelling is directly applicable to music. For instance, consider the example shown in Fig. \ref{fig:fig2}. Here we defined a relationship $n^r \cdot s \cdot n^l$ between two musical notes (type $n$): C4 and A4.  The relationship states that note A4 follows note C4 (Fig. \ref{fig:fig3}). In this particular example, $\emph{sequence}$ is not a verb but an actual action to be carried out by some hypothetical generative music system.

\begin{figure}[h]
\begin{center}
\scalebox{1.3}{\begin{tikzpicture}[scale=0.8, baseline={(current bounding box).east}, node distance=5mm, text height=5pt, text depth=0pt, {every node/.style}={scale=0.6}]
	\begin{pgfonlayer}{nodelayer}
		\node [style=none] (0) at (2.75, 0) {};
		\node [style=none, fill=white, anchor=west, scale=0.85] (1) at (3.125, -0.2) {$n$};
		\node [style=none] (2) at (3.25, 0) {};
		\node [style=none] (3) at (3.25, -0.4) {};
		\node [style=none] (4) at (2.75, 0) {};
		\node [style=none] (5) at (3.75, 0) {};
		\node [style=none] (6) at (3.75, 0.4) {};
		\node [style=none] (7) at (3.25, 0.5) {};
		\node [style=none] (8) at (2.75, 0.4) {};
		\node [style=none, fill=white, scale=1] (9) at (3.25, 0.2) {p4};
		\node [style=none, fill=white, anchor=west, scale=0.85] (10) at (4.375, -0.2) {$n^{r}$};
		\node [style=none] (11) at (4.5, 0) {};
		\node [style=none] (12) at (4.5, -0.4) {};
		\node [style=none] (13) at (5, 0) {};
		\node [style=none] (14) at (5, -0.4) {};
		\node [style=none, fill=white, anchor=west, scale=0.85] (15) at (5.375, -0.2) {$n^{r}$};
		\node [style=none] (16) at (5.5, 0) {};
		\node [style=none] (17) at (5.5, -0.4) {};
		\node [style=none] (18) at (4.25, 0) {};
		\node [style=none] (19) at (6.75, 0) {};
		\node [style=none] (20) at (6.75, 0.4) {};
		\node [style=none] (21) at (5.5, 0.5) {};
		\node [style=none] (22) at (4.25, 0.4) {};
		\node [style=none, fill=white, scale=1] (23) at (5.5, 0.2) {s4};
		\node [style=none] (33) at (3.875, -0.65) {};
		\node [style=none, fill=white, anchor=west, scale=0.85] (37) at (5.875, -0.2) {$n^{r}$};
		\node [style=none] (38) at (6, 0) {};
		\node [style=none] (39) at (6, -0.4) {};
		\node [style=none] (40) at (6.5, 0) {};
		\node [style=none] (41) at (6.5, -0.4) {};
		\node [style=none] (42) at (5.75, 0) {};
		\node [style=none] (43) at (6.5, -2.25) {};
		\node [style=none, fill=white, anchor=west, scale=0.85] (44) at (6.4, -0.75) {$s$};
		\node [style=none] (45) at (1.25, 0) {};
		\node [style=none, fill=white, anchor=west, scale=0.85] (46) at (1.625, -0.2) {$n$};
		\node [style=none] (47) at (1.75, 0) {};
		\node [style=none] (48) at (1.75, -0.4) {};
		\node [style=none] (49) at (1.25, 0) {};
		\node [style=none] (50) at (2.25, 0) {};
		\node [style=none] (51) at (2.25, 0.4) {};
		\node [style=none] (52) at (1.75, 0.5) {};
		\node [style=none] (53) at (1.25, 0.4) {};
		\node [style=none, fill=white, scale=1] (54) at (1.75, 0.2) {p9};
		\node [style=none] (55) at (-0.25, 0) {};
		\node [style=none, fill=white, anchor=west, scale=0.85] (56) at (0.125, -0.2) {$n$};
		\node [style=none] (57) at (0.25, 0) {};
		\node [style=none] (58) at (0.25, -0.4) {};
		\node [style=none] (59) at (-0.25, 0) {};
		\node [style=none] (60) at (0.75, 0) {};
		\node [style=none] (61) at (0.75, 0.4) {};
		\node [style=none] (62) at (0.25, 0.5) {};
		\node [style=none] (63) at (-0.25, 0.4) {};
		\node [style=none, fill=white, scale=1] (64) at (0.25, 0.2) {p7};
		\node [style=none] (65) at (-1.75, 0) {};
		\node [style=none, fill=white, anchor=west, scale=0.85] (66) at (-1.375, -0.2) {$n$};
		\node [style=none] (67) at (-1.25, 0) {};
		\node [style=none] (68) at (-1.25, -0.4) {};
		\node [style=none] (69) at (-1.75, 0) {};
		\node [style=none] (70) at (-0.75, 0) {};
		\node [style=none] (71) at (-0.75, 0.4) {};
		\node [style=none] (72) at (-1.25, 0.5) {};
		\node [style=none] (73) at (-1.75, 0.4) {};
		\node [style=none, fill=white, scale=1] (74) at (-1.25, 0.2) {p5};
		\node [style=none] (75) at (3.5, -1) {};
		\node [style=none] (76) at (3, -1.5) {};
		\node [style=none] (77) at (2.5, -2) {};
		\node [style=none, fill=white, anchor=west, scale=0.85] (78) at (4.875, -0.2) {$n^{r}$};
	\end{pgfonlayer}
	\begin{pgfonlayer}{edgelayer}
		\draw [in=90, out=-90] (2.center) to (3.center);
		\draw [-, fill=white] (4.center)
			 to (5.center)
			 to (6.center)
			 to (7.center)
			 to (8.center)
			 to cycle;
		\draw [in=90, out=-90] (11.center) to (12.center);
		\draw [in=90, out=-90] (13.center) to (14.center);
		\draw [in=90, out=-90] (16.center) to (17.center);
		\draw [-, fill=white] (18.center)
			 to (19.center)
			 to (20.center)
			 to (21.center)
			 to (22.center)
			 to cycle;
		\draw [in=180, out=-90, looseness=0.72] (3.center) to (33.center);
		\draw [in=0, out=-90, looseness=0.72] (12.center) to (33.center);
		\draw [in=90, out=-90] (38.center) to (39.center);
		\draw [in=90, out=-90] (40.center) to (41.center);
		\draw [in=90, out=-90] (41.center) to (43.center);
		\draw [in=90, out=-90] (47.center) to (48.center);
		\draw [-, fill=white] (49.center)
			 to (50.center)
			 to (51.center)
			 to (52.center)
			 to (53.center)
			 to cycle;
		\draw [in=90, out=-90] (57.center) to (58.center);
		\draw [-, fill=white] (59.center)
			 to (60.center)
			 to (61.center)
			 to (62.center)
			 to (63.center)
			 to cycle;
		\draw [in=90, out=-90] (67.center) to (68.center);
		\draw [-, fill=white] (69.center)
			 to (70.center)
			 to (71.center)
			 to (72.center)
			 to (73.center)
			 to cycle;
		\draw [in=-180, out=-90, looseness=0.75] (48.center) to (75.center);
		\draw [in=-90, out=0, looseness=0.75] (75.center) to (14.center);
		\draw [in=-180, out=-90, looseness=0.75] (58.center) to (76.center);
		\draw [in=-90, out=0, looseness=0.75] (76.center) to (17.center);
		\draw [in=-180, out=-90, looseness=0.75] (68.center) to (77.center);
		\draw [in=-90, out=0, looseness=0.75] (77.center) to (39.center);
	\end{pgfonlayer}
\end{tikzpicture}}
\caption{Pregroup diagram for a musical composition form.}
\label{fig:fig4}
\end{center}
\end{figure}
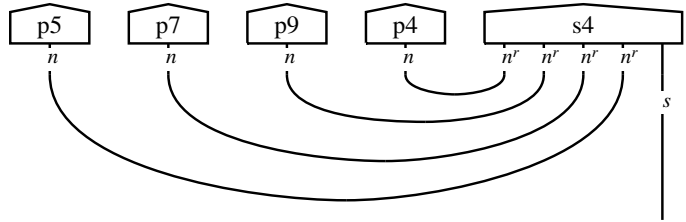

\medskip

\begin{figure}[h]
\begin{center}
\includegraphics[width=0.5\textwidth]{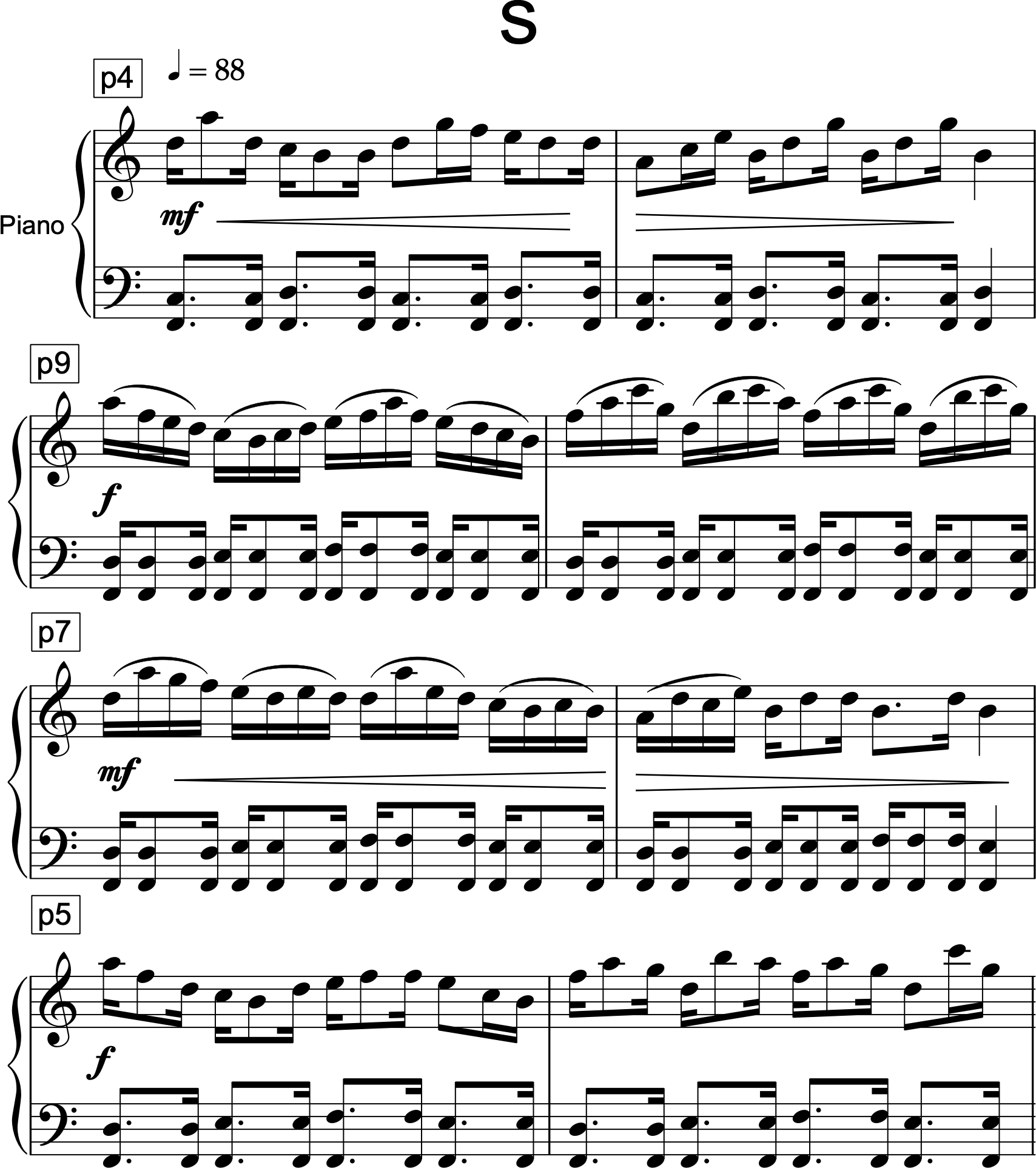}
\caption{A short musical composition corresponding to the pregroup diagram in Fig.\ref{fig:fig4}.}
\label{fig:pregroup_music}
\end{center}
\end{figure}

Fig. \ref{fig:fig4} depicts an example of a pregroup diagram representing a short musical composition, which is shown in Fig. \ref{fig:pregroup_music}. In this case there are five musical snippets (labelled as p4, p9, p7 and p5), each of which comprising several musical notes forming two entire musical bars each.

\medskip

The semantic map is a functor $\mathcal{F}$ that sends basic types ($n$, $s$) to vector spaces ($N$, $S$). Take the example in Fig. \ref{fig:fig1}, `Bob plays guitar'. The DisCoCat diagram can be decomposed as $\text{diagram} = (\text{Bob} \otimes \text{plays} \otimes \text{guitar}) \circ \text{cups}$.  The second word `plays' is a transitive verb and has the pregroup type $n^r \cdot s \cdot n^l$, whilst `Bob' and `guitar' have pregroup type $n$. The functor $\mathcal{F}$ sends the diagram to vector space semantics by dropping the adjoints ($l, r$) and sending the pregroup type to the corresponding vector spaces. Based on their pregroup types, the words of the sentence will be sent to tensors of the corresponding vector space (Eq. \ref{eq:tensors}).

\begin{equation}
\begin{split}
\mathcal{F}(\text{Bob}), \mathcal{F}(\text{guitar}) \in \mathcal{F}(n) = N  \\
 \mathcal{F}(\text{plays}) \in \mathcal{F}(n^r \cdot s \cdot n^l) = N \otimes S \otimes N
\label{eq:tensors}
\end{split}
\end{equation}

\medskip

Moreover, $\mathcal{F}$ translates all reductions $\text{cups}$ to tensor contractions. Conceptually, the functor can be thought of as something that performs tensor contraction between the words and the cups to return a vector for the whole sentence (Eq. \ref{eq:contraction}). In section \ref{subsec:preg2qcirc}, we will relay the details of the corresponding mapping to quantum processes.

\begin{equation}
\mathcal{F}(\text{diagram}) =
(\mathcal{F}(\text{Bob}) \otimes \mathcal{F}(\text{plays}) \otimes \mathcal{F}(\text{guitar}))
\circ \mathcal{F}(\text{cups}) \in S
\label{eq:contraction}
\end{equation}

\medskip

Exactly the same principles are applicable to music. For instance, assuming the example in Fig. \ref{fig:fig2}, and a set of musical notes $N$, the relationship $\emph{sequence}$ acquires the meaning  $s = C4 \cdot R \cdot A4$ with $\lbrace C4, A4 \rbrace \in N$ and  $R \in (N \otimes S \otimes N)$.

\section{Machine Learning of Music}

\begin{figure}[h]
\begin{center}
\includegraphics[width=0.6\textwidth]{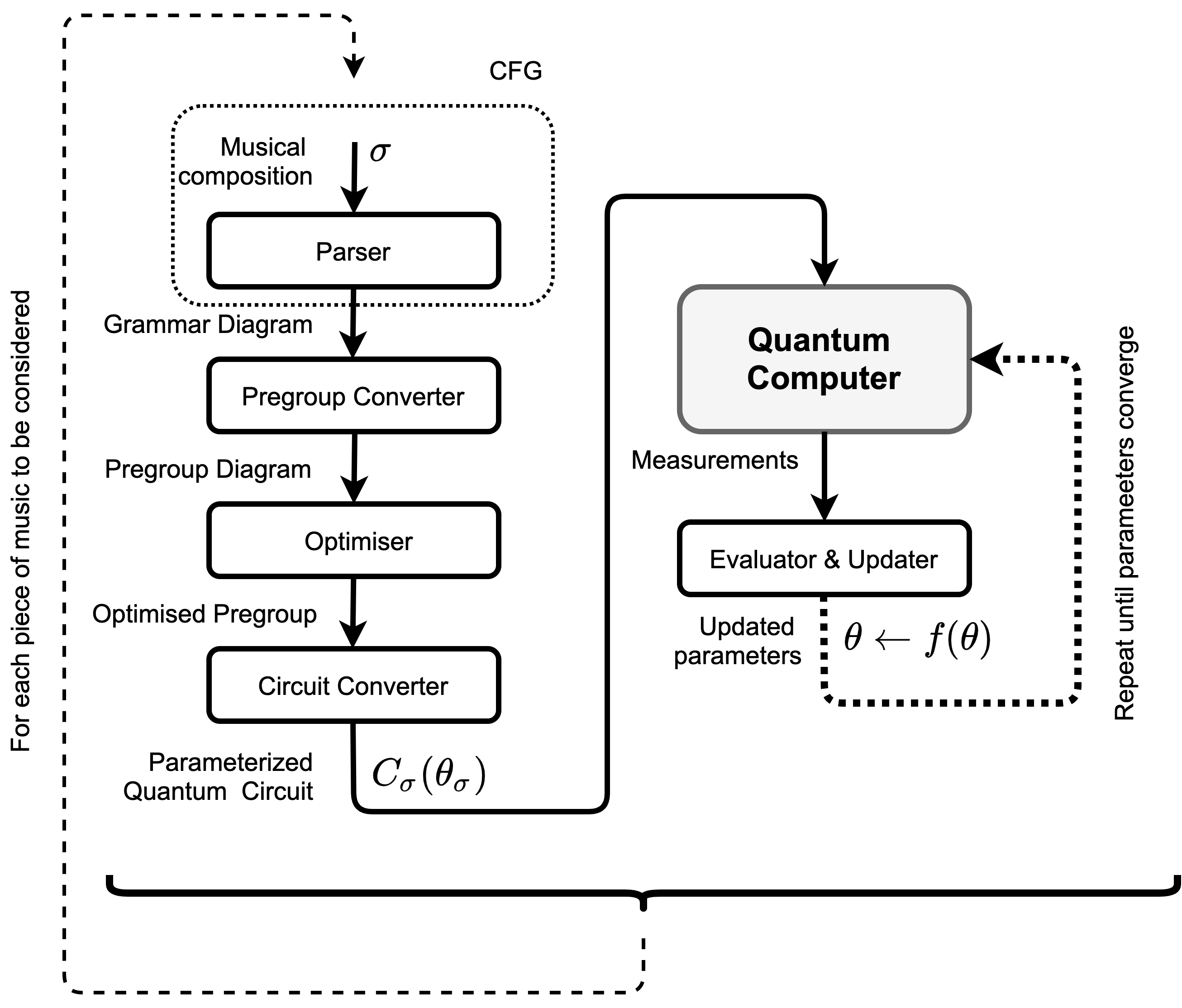}
\caption{Schematic overview of our quantum machine learning algorithm.}
\label{fig:fig6}
\end{center}
\end{figure}

In this section we introduce our DisCoCat music model and the quantum machine learning algorithm that we built to differentiate between two categories of music. Example code for the machine learning process can be found at [ \href{https://github.com/CQCL/Quanthoven}{\color{blue}{\underline{https://github.com/CQCL/Quanthoven}}} ]. A schematic overview of the algorithm is shown in Fig. \ref{fig:fig6}.

\medskip

In a nutshell, the algorithm learns from corpora generated by a bespoke context-free grammar (CFG). However, the quantum machine learning algorithm needs to see the structure of the compositions in the form of parameterised quantum circuits. Thus, the system needs to transform their underlying CFG structures into quantum circuits. This is done by converting CFGs into pregroups diagrams, which are then optimised before they are translated into quantum circuits. This optimisation is highly advisable because cups tend to produce more qubits than necessary when converting pregroup diagrams into a quantum circuit. The system then generates instructions for an appropriate quantum computer to run the circuit.

\subsection{Generating a Training Corpus with a Context-Free Grammar}
\label{sec:cfg}

Here we define a Context-Free Grammar (CFG) to generate musical compositions for piano, which will be used to teach the machine learning algorithm. And \textit{Quanthoven} will use this CFG to generate new pieces after it has learnt to classify them.

\medskip

The lexicon of our CFG contains four types of musical snippets. As it was mentioned already, a snippet is the equivalent of a word in natural language. Whereas a word is formed by letters from an alphabet, a snippet is formed by notes from a pitch framework\footnote{In the context of this work, a musical pitch framework defines the notes that characterise a specific musical style or culture.  An octave is defined by two notes, one having twice the pitch of another. Different pitch frameworks have their own way of subdividing the octave to create distinct scales of musical notes. Conventionally, there are 1,200 cents to the octave. Western European music normally subdivides the octave into 12 notes, equally-spaced at 100 cents. This forms the so-called chromatic scale.  However, some cultures subdivide the octave into notes with different spacings between one another. For example, Slendro Indonesian music subdivides the octave into five notes, spaced by 240 cents from each other. And most Middle Eastern music divides the octave into six notes spaced by 200 cents. Furthermore, no matter how an octave is divided, the notes can be played at different registers, or heights, depending on the instrument at hand. For instance, most pianos can produce 88 notes, accommodating seven octaves. That is, one can play seven different notes `E' on the piano (or `Mi' in traditional nomenclature). The snippets discussed in this article are made from the piano's 88 note-set.}. Likewise, whilst combinations of words form sentences, combinations of snippets form musical sequences; that is, musical compositions.

\medskip

As the second level of the Chomsky hierarchy, CFGs are expressive enough to model a vast portion of natural language. An informal, diagrammatic representation of a CFG is to draw words in the lexicon as boxes with one output wire. And production rules can be represented as boxes with multiple input wires and one output wire. Then, sequences can be built by freely composing these boxes. A more detailed explanation about combining CFGs with monoidal category theory can be found in Appendix \ref{app:appendix1}.

\medskip

\begin{figure}[h]
\begin{center}
\scalebox{1.2}{\begin{tikzpicture}
	\begin{pgfonlayer}{nodelayer}
		\node [style=none] (54) at (0, -0.5) {};
		\node [style=none] (55) at (0, -0.75) {};
		\node [style=none] (56) at (0, -1.125) {};
		\node [style=none, right] (57) at (0.1, -0.975) {\scriptsize e$_1$};
		\node [style=none] (58) at (-0.25, -0.75) {};
		\node [style=none] (59) at (0.25, -0.75) {};
		\node [style=none] (60) at (0.25, -0.25) {};
		\node [style=none] (61) at (-0.25, -0.25) {};
		\node [style=none, fill=white] (62) at (0, -0.5) {$p_i$};
		\node [style=none] (63) at (2, -0.5) {};
		\node [style=none] (64) at (2, -0.75) {};
		\node [style=none] (65) at (2, -1.125) {};
		\node [style=none, right] (66) at (2.1, -0.975) {\scriptsize e$_2$};
		\node [style=none] (67) at (1.75, -0.75) {};
		\node [style=none] (68) at (2.25, -0.75) {};
		\node [style=none] (69) at (2.25, -0.25) {};
		\node [style=none] (70) at (1.75, -0.25) {};
		\node [style=none, fill=white] (71) at (2, -0.5) {$s_i$};
		\node [style=none] (72) at (4, -0.5) {};
		\node [style=none] (73) at (4, -0.75) {};
		\node [style=none] (74) at (4, -1.125) {};
		\node [style=none, right] (75) at (4.1, -0.975) {\scriptsize e$_3$};
		\node [style=none] (76) at (3.75, -0.75) {};
		\node [style=none] (77) at (4.25, -0.75) {};
		\node [style=none] (78) at (4.25, -0.25) {};
		\node [style=none] (79) at (3.75, -0.25) {};
		\node [style=none, fill=white] (80) at (4, -0.5) {$t_i$};
		\node [style=none] (81) at (-2, -0.5) {};
		\node [style=none] (82) at (-2, -0.75) {};
		\node [style=none] (83) at (-2, -1.125) {};
		\node [style=none, right] (84) at (-1.9, -0.975) {\scriptsize s};
		\node [style=none] (85) at (-2.25, -0.75) {};
		\node [style=none] (86) at (-1.75, -0.75) {};
		\node [style=none] (87) at (-1.75, -0.25) {};
		\node [style=none] (88) at (-2.25, -0.25) {};
		\node [style=none, fill=white] (89) at (-2, -0.5) {$g_i$};
	\end{pgfonlayer}
	\begin{pgfonlayer}{edgelayer}
		\draw [in=90, out=-90] (55.center) to (56.center);
		\draw [-, fill=white] (58.center)
			 to (59.center)
			 to (60.center)
			 to (61.center)
			 to cycle;
		\draw [in=90, out=-90] (64.center) to (65.center);
		\draw [-, fill=white] (67.center) to (68.center);
		\draw [-, fill=white] (68.center) to (69.center);
		\draw [-, fill=white] (69.center) to (70.center);
		\draw [-, fill=white] (70.center) to (67.center);
		\draw [in=90, out=-90] (73.center) to (74.center);
		\draw [-, fill=white] (76.center) to (77.center);
		\draw [-, fill=white] (77.center) to (78.center);
		\draw [-, fill=white] (78.center) to (79.center);
		\draw [-, fill=white] (79.center) to (76.center);
		\draw [in=90, out=-90] (82.center) to (83.center);
		\draw [-, fill=white] (85.center) to (86.center);
		\draw [-, fill=white] (86.center) to (87.center);
		\draw [-, fill=white] (87.center) to (88.center);
		\draw [-, fill=white] (88.center) to (85.center);
	\end{pgfonlayer}
\end{tikzpicture}}
\caption{Diagrammatic representation of snippets.}
\label{fig:snippets}
\end{center}
\end{figure}
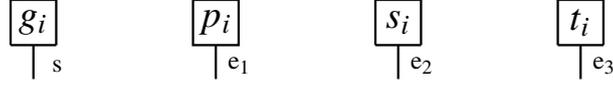

Verbal languages have different \textit{types} of words, such as nouns, verbs, adjectives, and so on. Grammatical rules then dictate how they are combined to form sentences. Similarly, our composition system contains four \textit{types} of snippets: ground (g), primary (p), secondary (s) and tertiary (t) snippets. Some examples are shown in Fig. \ref{fig:fig7}. These will be represented below as $s$, $e_1$, $e_2$ and $e_3$, respectively. We annotate the wires of the boxes with these types to enforce how the lexicon and production rules can be combined in a grammatical way to form musical compositions (Fig. \ref{fig:snippets}).

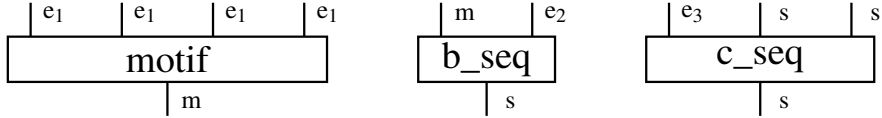
\begin{figure}[h]
\begin{center}
\scalebox{1.2}{\begin{tikzpicture}
	\begin{pgfonlayer}{nodelayer}
		\node [style=none] (0) at (0.75, 0.5) {};
		\node [style=none] (1) at (0.75, 1.125) {};
		\node [style=none] (2) at (0.75, 0.75) {};
		\node [style=none, right] (3) at (0.85, 1) {\scriptsize e$_1$};
		\node [style=none] (4) at (1.75, 1.125) {};
		\node [style=none] (5) at (1.75, 0.75) {};
		\node [style=none, right] (6) at (1.85, 1) {\scriptsize e$_1$};
		\node [style=none] (7) at (2.75, 1.125) {};
		\node [style=none] (8) at (2.75, 0.75) {};
		\node [style=none, right] (9) at (2.85, 1) {\scriptsize e$_1$};
		\node [style=none] (10) at (3.75, 1.125) {};
		\node [style=none] (11) at (3.75, 0.75) {};
		\node [style=none, right] (12) at (3.85, 1) {\scriptsize e$_1$};
		\node [style=none] (13) at (2.25, 0.25) {};
		\node [style=none] (14) at (2.25, -0.125) {};
		\node [style=none, right] (15) at (2.35, 0.025) {\scriptsize m};
		\node [style=none] (16) at (0.5, 0.25) {};
		\node [style=none] (17) at (4, 0.25) {};
		\node [style=none] (18) at (4, 0.75) {};
		\node [style=none] (19) at (0.5, 0.75) {};
		\node [style=none, fill=white] (20) at (2.25, 0.5) {motif};
		\node [style=none] (21) at (5.25, 0.5) {};
		\node [style=none] (22) at (5.25, 1.125) {};
		\node [style=none] (23) at (5.25, 0.75) {};
		\node [style=none, right] (24) at (5.35, 1) {\scriptsize m};
		\node [style=none] (25) at (6.25, 1.125) {};
		\node [style=none] (26) at (6.25, 0.75) {};
		\node [style=none, right] (27) at (6.35, 1) {\scriptsize e$_2$};
		\node [style=none] (28) at (5.75, 0.25) {};
		\node [style=none] (29) at (5.75, -0.125) {};
		\node [style=none, right] (30) at (5.85, 0.025) {\scriptsize s};
		\node [style=none] (31) at (5, 0.25) {};
		\node [style=none] (32) at (6.5, 0.25) {};
		\node [style=none] (33) at (6.5, 0.75) {};
		\node [style=none] (34) at (5, 0.75) {};
		\node [style=none, fill=white] (35) at (5.75, 0.5) {b\_seq};
		\node [style=none] (36) at (7.75, 0.5) {};
		\node [style=none] (37) at (7.75, 1.125) {};
		\node [style=none] (38) at (7.75, 0.75) {};
		\node [style=none, right] (39) at (7.85, 1) {\scriptsize e$_3$};
		\node [style=none] (40) at (8.75, 1.125) {};
		\node [style=none] (41) at (8.75, 0.75) {};
		\node [style=none, right] (42) at (8.85, 1) {\scriptsize s};
		\node [style=none] (43) at (9.75, 1.125) {};
		\node [style=none] (44) at (9.75, 0.75) {};
		\node [style=none, right] (45) at (9.85, 1) {\scriptsize s};
		\node [style=none] (46) at (8.75, 0.25) {};
		\node [style=none] (47) at (8.75, -0.125) {};
		\node [style=none, right] (48) at (8.85, 0.025) {\scriptsize s};
		\node [style=none] (49) at (7.5, 0.25) {};
		\node [style=none] (50) at (10, 0.25) {};
		\node [style=none] (51) at (10, 0.75) {};
		\node [style=none] (52) at (7.5, 0.75) {};
		\node [style=none] (53) at (8.75, 0.5) {c\_seq};
	\end{pgfonlayer}
	\begin{pgfonlayer}{edgelayer}
		\draw [in=90, out=-90] (1.center) to (2.center);
		\draw [in=90, out=-90] (4.center) to (5.center);
		\draw [in=90, out=-90] (7.center) to (8.center);
		\draw [in=90, out=-90] (10.center) to (11.center);
		\draw [in=90, out=-90] (13.center) to (14.center);
		\draw [-, fill=white] (16.center)
			 to (17.center)
			 to (18.center)
			 to (19.center)
			 to cycle;
		\draw [in=90, out=-90] (22.center) to (23.center);
		\draw [in=90, out=-90] (25.center) to (26.center);
		\draw [in=90, out=-90] (28.center) to (29.center);
		\draw [-, fill=white] (31.center)
			 to (32.center)
			 to (33.center)
			 to (34.center)
			 to cycle;
		\draw [in=90, out=-90] (37.center) to (38.center);
		\draw [in=90, out=-90] (40.center) to (41.center);
		\draw [in=90, out=-90] (43.center) to (44.center);
		\draw [in=90, out=-90] (46.center) to (47.center);
		\draw [-, fill=white] (49.center)
			 to (50.center)
			 to (51.center)
			 to (52.center)
			 to cycle;
	\end{pgfonlayer}
\end{tikzpicture}}
\caption{Diagrammatic representation of musical sequences.}
\label{fig:music_seqs}
\end{center}
\end{figure}

\begin{figure}[h]
    \begin{center}
    \includegraphics[width=0.48\textwidth]{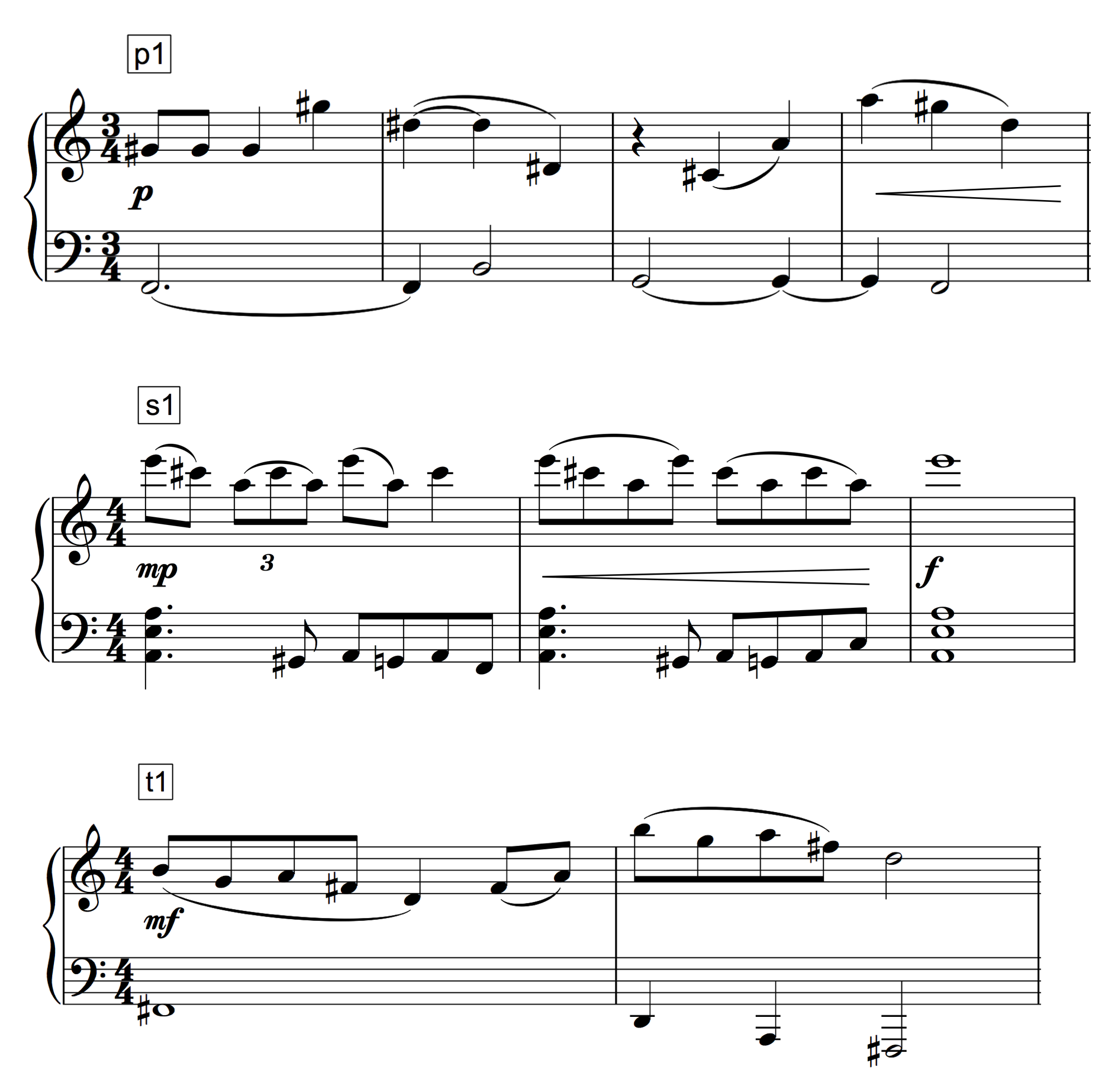}
    \caption{Examples of primary, secondary and tertiary snippets, respectively.}
    \label{fig:fig7}
    \end{center}
\end{figure}

Languages have different kinds of sentences; e.g., simple, compound, complex, and compound-complex sentence structures. By the same token, here we have three types of musical sequences: motif, basic (b\textunderscore seq) and composite (c\textunderscore seq) sequences , respectively (Fig. \ref{fig:music_seqs}).

In a nutshell, a motif is composed of four primary snippets, and a basic sequence is composed of a motif followed by a secondary snippet. A composite sequence will always start with a tertiary snippet, followed by two elements. Each of them can be one of three options: a tertiary snippet, a basic sequence, or another composite one.

\medskip

Unlike natural languages, the productions rules of this musical grammar are defined explicitly. Therefore it is possible to implement a parser to recover the derivation from the sequence. In fact, our CFG is parsable by a LL(1) parser, and hence is a LL(1) grammar \citep{ROSENKRANTZ1970226}. In contrast, parsing natural language requires contextual understanding of the words, so a statistical CCG parser \citep{yoshikawa:2017acl} combined with a conversion from CCG to DisCoCat \citep{yeung2021ccgbased} is necessary. By explicitly implementing a parser, we show that the grammatical structure is recoverable from a sentence and need not be explicitly provided to the DisCoCat model: other models are welcome to take advantage of it.

\medskip

Firstly, we generated a corpus of 100 compositions for piano. Then, we annotated the corpus according their meaning: \textit{rhythmic} or \textit{melodic}. As explained earlier, the overarching meaning of a composition describes an overall perceived property. Thus, compositions perceived as having prominent fast and loud rhythmic patterns were labelled as RIT (for rhythmic). Otherwise, those as having above all successions of notes forming melodies and harmonies were labelled as MEL (for melodic).

\medskip

The annotation of the corpus was carried out manually. The labellings were agreed upon by the authors after they independently listened to the 100 compositions and reached consensus. Eq. \ref{eq:corpus} shows how the corpus $\Sigma$ of annotated compositions $\sigma$ look like. The symbols $t3$, $g1$, $p3$, and so on, represent the snippets that form a respective composition.

\begin{equation}
\begin{aligned}
\Sigma = \lbrace { } & (1, \textrm{MEL}, \lbrack t3, \ g1, \ p3, \ p1, \ p3, \ p3, \ s3\rbrack), \\
                               & (2, \textrm{RIT}, \lbrack p4, \ p9, \ p7, \ p5, \ s1\rbrack), \\
                               & (3, \textrm{RIT}, \lbrack t3, \  g2, \ g2\rbrack), \\
                               & ... \rbrace
\end{aligned}
\label{eq:corpus}
\end{equation}

\medskip

The complete lexicon of musical snippets are provided in Appendix \ref{app:appendix2}. Although some snippets are more rhythmic than others, and vice-versa, there is considerable overlap of snippets between the two categories. This ensures that the machine learning task is realistic. After annotation, the the corpus is split into training, development, and test sets with a $50 + 25 + 25$ split, respectively. The compositions last for different durations, ranging for a few seconds to various minutes. A symbolic representation of the dataset is available in Appendix \ref{app:datasets}\footnote{Audio files are available at  [ \href{https://github.com/CQCL/Quanthoven/}{\color{blue}{\underline{https://github.com/CQCL/Quanthoven/}}} ]. Note, these were synthesised rather than played on a real instrument by a musician.}

\subsection{Pregroup Converter: from Context-Free Grammars to Pregroup Grammars}

In this section we give a functorial mapping from from musical CFGs to pregroups. The DisCoCat model is more naturally suitable for pregroup grammars as they are lexicalised grammars. In DisCoCat, most of the grammatical information is stored in the lexicon instead of the grammatical production rules. This is not the case for CFGs. Hence the need to convert to pregroups.

\medskip

In natural language, a transitive verb is a word that takes a noun on the left and a noun on the right to give a sentence, and so has the pregroup type $F(VP_{TV}) = n^r \cdot s \cdot n^l$. An adjective is a word that takes a noun on the right to give another noun, and so has the pregroup type $F(Adj) = n \cdot n^l$.

\medskip

Let us reason with analogies here: first we fix the types ground and primary elements, $s$ and $e_1$, to be the atomic types for our pregroup grammar. This is done by setting $F(s) = s$ and $F(e_1) = n$. Since a motif is made of four primary elements, we give it the pregroup type $F(m) = n \cdot n \cdot n \cdot n$. A secondary element takes a motif on the left to give a sentence, it has the pregroup type $F(e_2) = F(m^r \cdot s) = F(e_1 \cdot e_1 \cdot e_1 \cdot e_1)^r \cdot s = n^r \cdot n^r \cdot n^r \cdot n^r \cdot s$. Finally, a tertiary element takes two sentences on the right to give another sentence, and so it has the pregroup type $F(e_3) = s \cdot s^l \cdot s^l$.

\medskip

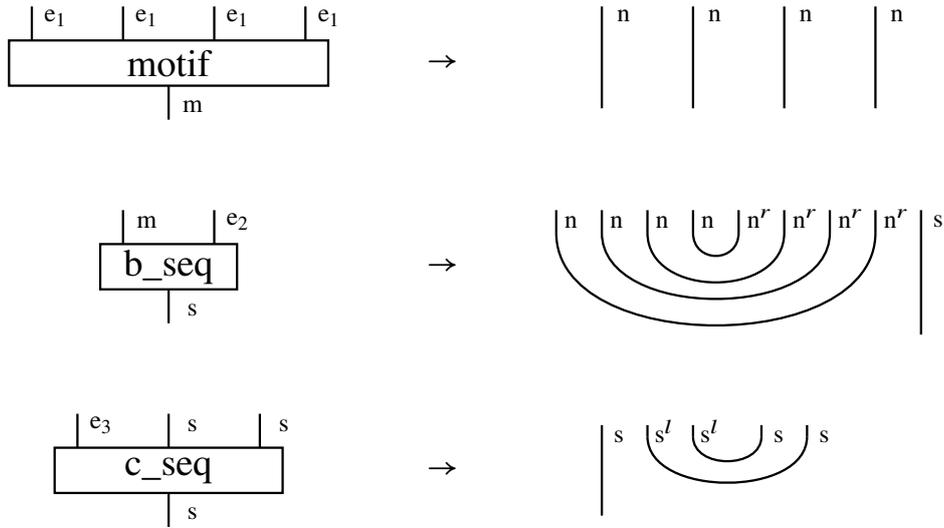
\begin{figure}[h]
	\begin{center}
	\scalebox{1.2}{\begin{tikzpicture}
	\begin{pgfonlayer}{nodelayer}
		\node [style=none] (0) at (0.75, 0.5) {};
		\node [style=none] (1) at (0.75, 1.125) {};
		\node [style=none] (2) at (0.75, 0.75) {};
		\node [style=none, right] (3) at (0.85, 1) {\scriptsize e$_1$};
		\node [style=none] (4) at (1.75, 1.125) {};
		\node [style=none] (5) at (1.75, 0.75) {};
		\node [style=none, right] (6) at (1.85, 1) {\scriptsize e$_1$};
		\node [style=none] (7) at (2.75, 1.125) {};
		\node [style=none] (8) at (2.75, 0.75) {};
		\node [style=none, right] (9) at (2.85, 1) {\scriptsize e$_1$};
		\node [style=none] (10) at (3.75, 1.125) {};
		\node [style=none] (11) at (3.75, 0.75) {};
		\node [style=none, right] (12) at (3.85, 1) {\scriptsize e$_1$};
		\node [style=none] (13) at (2.25, 0.25) {};
		\node [style=none] (14) at (2.25, -0.125) {};
		\node [style=none, right] (15) at (2.35, 0.025) {\scriptsize m};
		\node [style=none] (16) at (0.5, 0.25) {};
		\node [style=none] (17) at (4, 0.25) {};
		\node [style=none] (18) at (4, 0.75) {};
		\node [style=none] (19) at (0.5, 0.75) {};
		\node [style=none, fill=white] (20) at (2.25, 0.5) {motif};
		\node [style=none] (21) at (1.75, -1.75) {};
		\node [style=none] (22) at (1.75, -1.125) {};
		\node [style=none] (23) at (1.75, -1.5) {};
		\node [style=none, right] (24) at (1.85, -1.25) {\scriptsize m};
		\node [style=none] (25) at (2.75, -1.125) {};
		\node [style=none] (26) at (2.75, -1.5) {};
		\node [style=none, right] (27) at (2.85, -1.25) {\scriptsize e$_2$};
		\node [style=none] (28) at (2.25, -2) {};
		\node [style=none] (29) at (2.25, -2.375) {};
		\node [style=none, right] (30) at (2.35, -2.225) {\scriptsize s};
		\node [style=none] (31) at (1.5, -2) {};
		\node [style=none] (32) at (3, -2) {};
		\node [style=none] (33) at (3, -1.5) {};
		\node [style=none] (34) at (1.5, -1.5) {};
		\node [style=none, fill=white] (35) at (2.25, -1.75) {b\_seq};
		\node [style=none] (36) at (1.25, -4) {};
		\node [style=none] (37) at (1.25, -3.375) {};
		\node [style=none] (38) at (1.25, -3.75) {};
		\node [style=none, right] (39) at (1.35, -3.5) {\scriptsize e$_3$};
		\node [style=none] (40) at (2.25, -3.375) {};
		\node [style=none] (41) at (2.25, -3.75) {};
		\node [style=none, right] (42) at (2.35, -3.5) {\scriptsize s};
		\node [style=none] (43) at (3.25, -3.375) {};
		\node [style=none] (44) at (3.25, -3.75) {};
		\node [style=none, right] (45) at (3.35, -3.5) {\scriptsize s};
		\node [style=none] (46) at (2.25, -4.25) {};
		\node [style=none] (47) at (2.25, -4.625) {};
		\node [style=none, right] (48) at (2.35, -4.475) {\scriptsize s};
		\node [style=none] (49) at (1, -4.25) {};
		\node [style=none] (50) at (3.5, -4.25) {};
		\node [style=none] (51) at (3.5, -3.75) {};
		\node [style=none] (52) at (1, -3.75) {};
		\node [style=none] (53) at (2.25, -4) {c\_seq};
		\node [style=none] (54) at (7, 1.125) {};
		\node [style=none] (55) at (7, 0) {};
		\node [style=none] (56) at (8, 1.125) {};
		\node [style=none] (57) at (8, 0) {};
		\node [style=none] (58) at (9, 1.125) {};
		\node [style=none] (59) at (9, 0) {};
		\node [style=none] (60) at (10, 1.125) {};
		\node [style=none] (61) at (10, 0) {};
		\node [style=none] (62) at (6.5, -1.375) {};
		\node [style=none] (64) at (7, -1.375) {};
		\node [style=none] (66) at (7.5, -1.375) {};
		\node [style=none] (68) at (8, -1.375) {};
		\node [style=none] (70) at (8.5, -1.375) {};
		\node [style=none] (72) at (9, -1.375) {};
		\node [style=none] (74) at (9.5, -1.375) {};
		\node [style=none] (76) at (10, -1.375) {};
		\node [style=none] (77) at (10.5, -1.375) {};
		\node [style=none] (78) at (10.5, -2.5) {};
		\node [style=none] (79) at (7, -3.5) {};
		\node [style=none] (80) at (7.5, -3.625) {};
		\node [style=none] (81) at (8, -3.625) {};
		\node [style=none] (82) at (8.75, -3.625) {};
		\node [style=none] (83) at (9.25, -3.625) {};
		\node [style=none] (84) at (7, -4.5) {};
		\node [style=none, right] (85) at (10.075, 1.025) {\scriptsize n};
		\node [style=none, right] (86) at (9.075, 1.025) {\scriptsize n};
		\node [style=none, right] (87) at (8.075, 1.025) {\scriptsize n};
		\node [style=none, right] (88) at (7.075, 1.025) {\scriptsize n};
		\node [style=none, right] (89) at (10.525, -1.25) {\scriptsize s};
		\node [style=none, right] (90) at (10.075, -1.215) {\scriptsize n$^r$};
		\node [style=none, right] (91) at (9.575, -1.215) {\scriptsize n$^r$};
		\node [style=none, right] (92) at (9.075, -1.215) {\scriptsize n$^r$};
		\node [style=none, right] (93) at (8.575, -1.215) {\scriptsize n$^r$};
		\node [style=none] (94) at (6.5, -1.125) {};
		\node [style=none] (95) at (7, -1.125) {};
		\node [style=none] (96) at (7.5, -1.125) {};
		\node [style=none] (97) at (8, -1.125) {};
		\node [style=none] (98) at (8.5, -1.125) {};
		\node [style=none] (99) at (9, -1.125) {};
		\node [style=none] (100) at (9.5, -1.125) {};
		\node [style=none] (101) at (10, -1.125) {};
		\node [style=none] (102) at (10.5, -1.125) {};
		\node [style=none, right] (103) at (8, -1.25) {\scriptsize n};
		\node [style=none, right] (104) at (7.5, -1.25) {\scriptsize n};
		\node [style=none, right] (105) at (7, -1.25) {\scriptsize n};
		\node [style=none, right] (106) at (6.5, -1.25) {\scriptsize n};
		\node [style=none, right] (107) at (9.275, -3.625) {\scriptsize s};
		\node [style=none, right] (108) at (7.025, -3.625) {\scriptsize s};
		\node [style=none] (109) at (7.5, -3.5) {};
		\node [style=none] (110) at (8, -3.5) {};
		\node [style=none] (111) at (8.75, -3.5) {};
		\node [style=none] (112) at (9.25, -3.5) {};
		\node [style=none, right] (113) at (8.775, -3.625) {\scriptsize s};
		\node [style=none, right] (114) at (8.025, -3.565) {\scriptsize s$^l$};
		\node [style=none, right] (115) at (7.525, -3.565) {\scriptsize s$^l$};
		\node [style=none] (116) at (5.25, -4) {$\rightarrow$};
		\node [style=none] (117) at (5.25, -1.75) {$\rightarrow$};
		\node [style=none] (118) at (5.25, 0.5) {$\rightarrow$};
	\end{pgfonlayer}
	\begin{pgfonlayer}{edgelayer}
		\draw [in=90, out=-90] (1.center) to (2.center);
		\draw [in=90, out=-90] (4.center) to (5.center);
		\draw [in=90, out=-90] (7.center) to (8.center);
		\draw [in=90, out=-90] (10.center) to (11.center);
		\draw [in=90, out=-90] (13.center) to (14.center);
		\draw [-, fill=white] (16.center)
			 to (17.center)
			 to (18.center)
			 to (19.center)
			 to cycle;
		\draw [in=90, out=-90] (22.center) to (23.center);
		\draw [in=90, out=-90] (25.center) to (26.center);
		\draw [in=90, out=-90] (28.center) to (29.center);
		\draw [-, fill=white] (31.center)
			 to (32.center)
			 to (33.center)
			 to (34.center)
			 to cycle;
		\draw [in=90, out=-90] (37.center) to (38.center);
		\draw [in=90, out=-90] (40.center) to (41.center);
		\draw [in=90, out=-90] (43.center) to (44.center);
		\draw [in=90, out=-90] (46.center) to (47.center);
		\draw [-, fill=white] (49.center)
			 to (50.center)
			 to (51.center)
			 to (52.center)
			 to cycle;
		\draw (54.center) to (55.center);
		\draw (56.center) to (57.center);
		\draw (58.center) to (59.center);
		\draw (60.center) to (61.center);
		\draw [bend right=90, looseness=1.75] (68.center) to (70.center);
		\draw [bend right=90, looseness=1.25] (66.center) to (72.center);
		\draw [bend right=90] (64.center) to (74.center);
		\draw [bend right=90] (62.center) to (76.center);
		\draw (77.center) to (78.center);
		\draw (79.center) to (84.center);
		\draw [bend right=90, looseness=1.25] (81.center) to (82.center);
		\draw [bend right=90] (80.center) to (83.center);
		\draw (94.center) to (62.center);
		\draw (95.center) to (64.center);
		\draw (96.center) to (66.center);
		\draw (97.center) to (68.center);
		\draw (98.center) to (70.center);
		\draw (99.center) to (72.center);
		\draw (100.center) to (74.center);
		\draw (101.center) to (76.center);
		\draw (102.center) to (77.center);
		\draw (109.center) to (80.center);
		\draw (111.center) to (82.center);
		\draw (110.center) to (81.center);
		\draw (112.center) to (83.center);
	\end{pgfonlayer}
\end{tikzpicture}}
	\caption{Diagrammatic representation of pregroup contractions.}
	\label{fig:contractions}
	\end{center}
\end{figure}

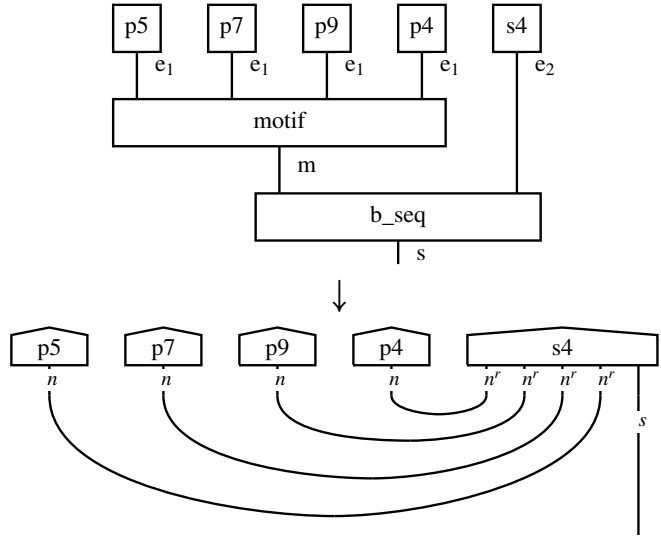
\begin{figure}[h]
    \begin{center}
    \scalebox{1.25}{\begin{tikzpicture}[baseline={(current bounding box).east}, node distance=5mm, text height=5pt, text depth=0pt, {every node/.style}={scale=0.6}]
	\begin{pgfonlayer}{nodelayer}
		\node [style=none] (1) at (4, 2.25) {};
		\node [style=none] (2) at (4, 0.75) {};
		\node [style=none, fill=white, right] (3) at (4.1, 2.1) {e$_2$};
		\node [style=none] (4) at (3, 2.25) {};
		\node [style=none] (5) at (3, 1.75) {};
		\node [style=none, fill=white, right] (6) at (3.1, 2.1) {e$_1$};
		\node [style=none] (7) at (2, 2.25) {};
		\node [style=none] (8) at (2, 1.75) {};
		\node [style=none, fill=white, right] (9) at (2.1, 2.1) {e$_1$};
		\node [style=none] (10) at (1, 2.25) {};
		\node [style=none] (11) at (1, 1.75) {};
		\node [style=none, fill=white, right] (12) at (1.1, 2.1) {e$_1$};
		\node [style=none] (13) at (0, 2.25) {};
		\node [style=none] (14) at (0, 1.75) {};
		\node [style=none, fill=white, right] (15) at (0.1, 2.1) {e$_1$};
		\node [style=none] (16) at (1.5, 1.25) {};
		\node [style=none] (17) at (1.5, 0.75) {};
		\node [style=none, fill=white, right] (18) at (1.6, 1.025) {m};
		\node [style=none] (19) at (2.75, 0.25) {};
		\node [style=none] (20) at (2.75, 0) {};
		\node [style=none, fill=white, right] (21) at (2.85, 0.1) {s};
		\node [style=none] (22) at (3.75, 2.25) {};
		\node [style=none] (23) at (4.25, 2.25) {};
		\node [style=none] (24) at (4.25, 2.75) {};
		\node [style=none] (25) at (3.75, 2.75) {};
		\node [style=none, fill=white] (26) at (4, 2.5) {s4};
		\node [style=none] (27) at (2.75, 2.25) {};
		\node [style=none] (28) at (3.25, 2.25) {};
		\node [style=none] (29) at (3.25, 2.75) {};
		\node [style=none] (30) at (2.75, 2.75) {};
		\node [style=none, fill=white] (31) at (3, 2.5) {p4};
		\node [style=none] (32) at (1.75, 2.25) {};
		\node [style=none] (33) at (2.25, 2.25) {};
		\node [style=none] (34) at (2.25, 2.75) {};
		\node [style=none] (35) at (1.75, 2.75) {};
		\node [style=none, fill=white] (36) at (2, 2.5) {p9};
		\node [style=none] (37) at (0.75, 2.25) {};
		\node [style=none] (38) at (1.25, 2.25) {};
		\node [style=none] (39) at (1.25, 2.75) {};
		\node [style=none] (40) at (0.75, 2.75) {};
		\node [style=none, fill=white] (41) at (1, 2.5) {p7};
		\node [style=none] (42) at (-0.25, 2.25) {};
		\node [style=none] (43) at (0.25, 2.25) {};
		\node [style=none] (44) at (0.25, 2.75) {};
		\node [style=none] (45) at (-0.25, 2.75) {};
		\node [style=none, fill=white] (46) at (0, 2.5) {p5};
		\node [style=none] (47) at (-0.25, 1.25) {};
		\node [style=none] (48) at (3.25, 1.25) {};
		\node [style=none] (49) at (3.25, 1.75) {};
		\node [style=none] (50) at (-0.25, 1.75) {};
		\node [style=none, fill=white] (51) at (1.5, 1.5) {motif};
		\node [style=none] (52) at (1.25, 0.25) {};
		\node [style=none] (53) at (4.25, 0.25) {};
		\node [style=none] (54) at (4.25, 0.75) {};
		\node [style=none] (55) at (1.25, 0.75) {};
		\node [style=none, fill=white] (56) at (2.75, 0.5) {b\_seq};
	\end{pgfonlayer}
	\begin{pgfonlayer}{edgelayer}
		\draw [in=90, out=-90] (1.center) to (2.center);
		\draw [in=90, out=-90] (4.center) to (5.center);
		\draw [in=90, out=-90] (7.center) to (8.center);
		\draw [in=90, out=-90] (10.center) to (11.center);
		\draw [in=90, out=-90] (13.center) to (14.center);
		\draw [in=90, out=-90] (16.center) to (17.center);
		\draw [in=90, out=-90] (19.center) to (20.center);
		\draw [-, fill=white] (22.center)
			 to (23.center)
			 to (24.center)
			 to (25.center)
			 to cycle;
		\draw [-, fill=white] (27.center)
			 to (28.center)
			 to (29.center)
			 to (30.center)
			 to cycle;
		\draw [-, fill=white] (32.center)
			 to (33.center)
			 to (34.center)
			 to (35.center)
			 to cycle;
		\draw [-, fill=white] (37.center)
			 to (38.center)
			 to (39.center)
			 to (40.center)
			 to cycle;
		\draw [-, fill=white] (42.center)
			 to (43.center)
			 to (44.center)
			 to (45.center)
			 to cycle;
		\draw [-, fill=white] (47.center)
			 to (48.center)
			 to (49.center)
			 to (50.center)
			 to cycle;
		\draw [-, fill=white] (52.center)
			 to (53.center)
			 to (54.center)
			 to (55.center)
			 to cycle;
	\end{pgfonlayer}
\end{tikzpicture}}\\
    \scalebox{1.25}{$\downarrow$} \\
    \scalebox{1.25}{\begin{tikzpicture}[scale=0.8, baseline={(current bounding box).east}, node distance=5mm, text height=5pt, text depth=0pt, {every node/.style}={scale=0.6}]
	\begin{pgfonlayer}{nodelayer}
		\node [style=none] (0) at (2.75, 0) {};
		\node [style=none, fill=white, anchor=west, scale=0.85] (1) at (3.125, -0.2) {$n$};
		\node [style=none] (2) at (3.25, 0) {};
		\node [style=none] (3) at (3.25, -0.4) {};
		\node [style=none] (4) at (2.75, 0) {};
		\node [style=none] (5) at (3.75, 0) {};
		\node [style=none] (6) at (3.75, 0.4) {};
		\node [style=none] (7) at (3.25, 0.5) {};
		\node [style=none] (8) at (2.75, 0.4) {};
		\node [style=none, fill=white, scale=1] (9) at (3.25, 0.2) {p4};
		\node [style=none, fill=white, anchor=west, scale=0.85] (10) at (4.375, -0.2) {$n^{r}$};
		\node [style=none] (11) at (4.5, 0) {};
		\node [style=none] (12) at (4.5, -0.4) {};
		\node [style=none] (13) at (5, 0) {};
		\node [style=none] (14) at (5, -0.4) {};
		\node [style=none, fill=white, anchor=west, scale=0.85] (15) at (5.375, -0.2) {$n^{r}$};
		\node [style=none] (16) at (5.5, 0) {};
		\node [style=none] (17) at (5.5, -0.4) {};
		\node [style=none] (18) at (4.25, 0) {};
		\node [style=none] (19) at (6.75, 0) {};
		\node [style=none] (20) at (6.75, 0.4) {};
		\node [style=none] (21) at (5.5, 0.5) {};
		\node [style=none] (22) at (4.25, 0.4) {};
		\node [style=none, fill=white, scale=1] (23) at (5.5, 0.2) {s4};
		\node [style=none] (33) at (3.875, -0.65) {};
		\node [style=none, fill=white, anchor=west, scale=0.85] (37) at (5.875, -0.2) {$n^{r}$};
		\node [style=none] (38) at (6, 0) {};
		\node [style=none] (39) at (6, -0.4) {};
		\node [style=none] (40) at (6.5, 0) {};
		\node [style=none] (41) at (6.5, -0.4) {};
		\node [style=none] (42) at (5.75, 0) {};
		\node [style=none] (43) at (6.5, -2.25) {};
		\node [style=none, fill=white, anchor=west, scale=0.85] (44) at (6.4, -0.75) {$s$};
		\node [style=none] (45) at (1.25, 0) {};
		\node [style=none, fill=white, anchor=west, scale=0.85] (46) at (1.625, -0.2) {$n$};
		\node [style=none] (47) at (1.75, 0) {};
		\node [style=none] (48) at (1.75, -0.4) {};
		\node [style=none] (49) at (1.25, 0) {};
		\node [style=none] (50) at (2.25, 0) {};
		\node [style=none] (51) at (2.25, 0.4) {};
		\node [style=none] (52) at (1.75, 0.5) {};
		\node [style=none] (53) at (1.25, 0.4) {};
		\node [style=none, fill=white, scale=1] (54) at (1.75, 0.2) {p9};
		\node [style=none] (55) at (-0.25, 0) {};
		\node [style=none, fill=white, anchor=west, scale=0.85] (56) at (0.125, -0.2) {$n$};
		\node [style=none] (57) at (0.25, 0) {};
		\node [style=none] (58) at (0.25, -0.4) {};
		\node [style=none] (59) at (-0.25, 0) {};
		\node [style=none] (60) at (0.75, 0) {};
		\node [style=none] (61) at (0.75, 0.4) {};
		\node [style=none] (62) at (0.25, 0.5) {};
		\node [style=none] (63) at (-0.25, 0.4) {};
		\node [style=none, fill=white, scale=1] (64) at (0.25, 0.2) {p7};
		\node [style=none] (65) at (-1.75, 0) {};
		\node [style=none, fill=white, anchor=west, scale=0.85] (66) at (-1.375, -0.2) {$n$};
		\node [style=none] (67) at (-1.25, 0) {};
		\node [style=none] (68) at (-1.25, -0.4) {};
		\node [style=none] (69) at (-1.75, 0) {};
		\node [style=none] (70) at (-0.75, 0) {};
		\node [style=none] (71) at (-0.75, 0.4) {};
		\node [style=none] (72) at (-1.25, 0.5) {};
		\node [style=none] (73) at (-1.75, 0.4) {};
		\node [style=none, fill=white, scale=1] (74) at (-1.25, 0.2) {p5};
		\node [style=none] (75) at (3.5, -1) {};
		\node [style=none] (76) at (3, -1.5) {};
		\node [style=none] (77) at (2.5, -2) {};
		\node [style=none, fill=white, anchor=west, scale=0.85] (78) at (4.875, -0.2) {$n^{r}$};
	\end{pgfonlayer}
	\begin{pgfonlayer}{edgelayer}
		\draw [in=90, out=-90] (2.center) to (3.center);
		\draw [-, fill=white] (4.center)
			 to (5.center)
			 to (6.center)
			 to (7.center)
			 to (8.center)
			 to cycle;
		\draw [in=90, out=-90] (11.center) to (12.center);
		\draw [in=90, out=-90] (13.center) to (14.center);
		\draw [in=90, out=-90] (16.center) to (17.center);
		\draw [-, fill=white] (18.center)
			 to (19.center)
			 to (20.center)
			 to (21.center)
			 to (22.center)
			 to cycle;
		\draw [in=180, out=-90, looseness=0.72] (3.center) to (33.center);
		\draw [in=0, out=-90, looseness=0.72] (12.center) to (33.center);
		\draw [in=90, out=-90] (38.center) to (39.center);
		\draw [in=90, out=-90] (40.center) to (41.center);
		\draw [in=90, out=-90] (41.center) to (43.center);
		\draw [in=90, out=-90] (47.center) to (48.center);
		\draw [-, fill=white] (49.center)
			 to (50.center)
			 to (51.center)
			 to (52.center)
			 to (53.center)
			 to cycle;
		\draw [in=90, out=-90] (57.center) to (58.center);
		\draw [-, fill=white] (59.center)
			 to (60.center)
			 to (61.center)
			 to (62.center)
			 to (63.center)
			 to cycle;
		\draw [in=90, out=-90] (67.center) to (68.center);
		\draw [-, fill=white] (69.center)
			 to (70.center)
			 to (71.center)
			 to (72.center)
			 to (73.center)
			 to cycle;
		\draw [in=-180, out=-90, looseness=0.75] (48.center) to (75.center);
		\draw [in=-90, out=0, looseness=0.75] (75.center) to (14.center);
		\draw [in=-180, out=-90, looseness=0.75] (58.center) to (76.center);
		\draw [in=-90, out=0, looseness=0.75] (76.center) to (17.center);
		\draw [in=-180, out=-90, looseness=0.75] (68.center) to (77.center);
		\draw [in=-90, out=0, looseness=0.75] (77.center) to (39.center);
	\end{pgfonlayer}
\end{tikzpicture}}
    \caption{Converting a CFG diagram into pregroup grammar.}
    \label{fig:conversion}
    \end{center}
\end{figure}

Once we have used our functor to convert the CFG types into the appropriate pregroup types, the CFG production rules become pregroup contractions, which are represented as `cups' (Fig. \ref{fig:contractions}). An example of this conversion on an actual musical grammar can be found in Fig. \ref{fig:conversion}.

\subsection{Optimiser: Diagrammatic Rewriting}

Once a CFG diagram is converted into a pregroup grammar diagram, a circuit functor can be applied to transfer the pregroup diagram into a quantum circuit; this will be explained below, in section \ref{subsec:preg2qcirc}. However, due the length of the compositions produced by our generation procedure, the resulting quantum circuit, after the functorial conversion has been applied, has too many qubits to be efficiently simulated before being executed on quantum hardware.

\medskip

Furthermore, to perform the highly entangled qubit measurements --- or Bell measurements --- required by pregroup grammars, we need to use postselections. To perform a postselection on a qubit, one measures the qubit and then discard the circuits that do not have the desired outcome on the respective qubit. As a rough estimate, postselection halves the number of usable shots\footnote{`Shots' is a term used to refer to repeated runs of a circuit.}. By performing $n$ postselections, we reduce the number of usable shots by a factor of $2^n$. For example, if a 12-qubit circuit requires 11 postselections, then we would expect approximately only 4 out of 8192 shots to be usable after postselection. This makes our results even more sensitive to the noise of a quantum computer.

\medskip

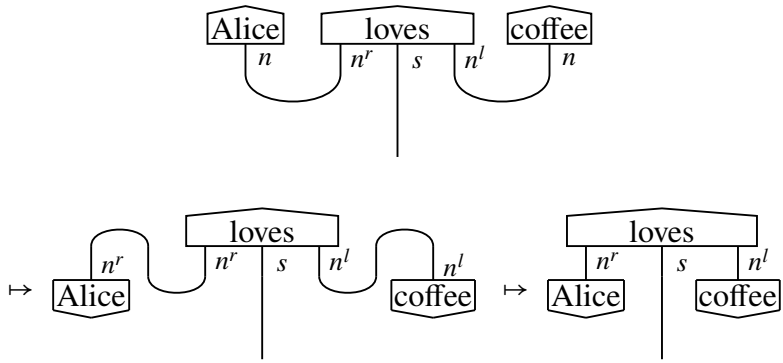
\begin{figure}[h]
\begin{center}
\ctikzfig{tikz/rewrite1}
\begin{center}
$\mapsto$ \begin{tikzpicture}[scale=1, baseline={([yshift=-1ex]current bounding box.center)}]
	\begin{pgfonlayer}{nodelayer}
		\node [style=none] (3) at (4, -0.4) {};
		\node [style=none, fill=white, anchor=west, scale=0.85] (10) at (4.875, -0.2) {$n^{r}$};
		\node [style=none] (11) at (4.75, 0) {};
		\node [style=none] (12) at (4.75, -0.4) {};
		\node [style=none, fill=white, anchor=west, scale=0.85] (13) at (5.625, -0.25) {$s$};
		\node [style=none] (14) at (5.5, 0) {};
		\node [style=none] (15) at (5.5, -0.4) {};
		\node [style=none, fill=white, anchor=west, scale=0.85] (16) at (6.375, -0.175) {$n^{l}$};
		\node [style=none] (17) at (6.25, 0) {};
		\node [style=none] (18) at (6.25, -0.4) {};
		\node [style=none] (19) at (4.5, 0) {};
		\node [style=none] (20) at (6.5, 0) {};
		\node [style=none] (21) at (6.5, 0.4) {};
		\node [style=none] (22) at (5.5, 0.5) {};
		\node [style=none] (23) at (4.5, 0.4) {};
		\node [style=none, fill=white, scale=1] (24) at (5.5, 0.2) {loves};
		\node [style=none] (27) at (7, -0.4) {};
		\node [style=none] (36) at (5.5, -1.5) {};
		\node [style=none] (37) at (2.75, -0.45) {};
		\node [style=none, fill=white, anchor=west, scale=0.85] (38) at (3.375, -0.25) {$n^r$};
		\node [style=none] (39) at (3.25, -0.45) {};
		\node [style=none] (40) at (2.75, -0.45) {};
		\node [style=none] (41) at (3.75, -0.45) {};
		\node [style=none] (42) at (3.75, -0.85) {};
		\node [style=none] (43) at (3.25, -0.95) {};
		\node [style=none] (44) at (2.75, -0.85) {};
		\node [style=none, fill=white, scale=1] (45) at (3.25, -0.65) {Alice};
		\node [style=none, fill=white, anchor=west, scale=0.85] (46) at (7.875, -0.25) {$n^l$};
		\node [style=none] (47) at (7.75, -0.45) {};
		\node [style=none] (48) at (7.2, -0.45) {};
		\node [style=none] (49) at (8.3, -0.45) {};
		\node [style=none] (50) at (8.3, -0.85) {};
		\node [style=none] (51) at (7.75, -0.95) {};
		\node [style=none] (52) at (7.2, -0.85) {};
		\node [style=none, fill=white, scale=1] (53) at (7.75, -0.65) {coffee};
		\node [style=none] (54) at (7, 0) {};
		\node [style=none] (55) at (7.75, 0) {};
		\node [style=none] (57) at (3.25, 0) {};
		\node [style=none] (58) at (4, 0) {};
	\end{pgfonlayer}
	\begin{pgfonlayer}{edgelayer}
		\draw [in=90, out=-90] (11.center) to (12.center);
		\draw [in=90, out=-90] (14.center) to (15.center);
		\draw [in=90, out=-90] (17.center) to (18.center);
		\draw [-, fill=white] (19.center)
			 to (20.center)
			 to (21.center)
			 to (22.center)
			 to (23.center)
			 to cycle;
		\draw [in=90, out=-90] (15.center) to (36.center);
		\draw [-, fill=white] (40.center) to (41.center);
		\draw [-, fill=white] (41.center) to (42.center);
		\draw [-, fill=white] (42.center) to (43.center);
		\draw [-, fill=white] (43.center) to (44.center);
		\draw [-, fill=white] (44.center) to (40.center);
		\draw [-, fill=white] (48.center) to (49.center);
		\draw [-, fill=white] (49.center) to (50.center);
		\draw [-, fill=white] (50.center) to (51.center);
		\draw [-, fill=white] (51.center) to (52.center);
		\draw [-, fill=white] (52.center) to (48.center);
		\draw (54.center) to (27.center);
		\draw (58.center) to (3.center);
		\draw [bend left=90] (57.center) to (58.center);
		\draw [bend right=90] (3.center) to (12.center);
		\draw [bend right=90] (18.center) to (27.center);
		\draw [bend left=90] (54.center) to (55.center);
		\draw (55.center) to (47.center);
		\draw (57.center) to (39.center);
	\end{pgfonlayer}
\end{tikzpicture} $\mapsto$ \begin{tikzpicture}[scale=1, baseline={([yshift=-1ex]current bounding box.center)}]
	\begin{pgfonlayer}{nodelayer}
		\node [style=none, fill=white, anchor=west, scale=0.85] (10) at (4.625, -0.2) {$n^{r}$};
		\node [style=none] (11) at (4.5, 0) {};
		\node [style=none, fill=white, anchor=west, scale=0.85] (13) at (5.625, -0.25) {$s$};
		\node [style=none] (14) at (5.5, 0) {};
		\node [style=none] (15) at (5.5, -0.4) {};
		\node [style=none, fill=white, anchor=west, scale=0.85] (16) at (6.625, -0.175) {$n^{l}$};
		\node [style=none] (17) at (6.5, 0) {};
		\node [style=none] (19) at (4.25, 0) {};
		\node [style=none] (20) at (6.75, 0) {};
		\node [style=none] (21) at (6.75, 0.4) {};
		\node [style=none] (22) at (5.5, 0.5) {};
		\node [style=none] (23) at (4.25, 0.4) {};
		\node [style=none, fill=white, scale=1] (24) at (5.5, 0.2) {loves};
		\node [style=none] (36) at (5.5, -1.5) {};
		\node [style=none] (37) at (4, -0.45) {};
		\node [style=none] (39) at (4.5, -0.45) {};
		\node [style=none] (40) at (4, -0.45) {};
		\node [style=none] (41) at (5, -0.45) {};
		\node [style=none] (42) at (5, -0.85) {};
		\node [style=none] (43) at (4.5, -0.95) {};
		\node [style=none] (44) at (4, -0.85) {};
		\node [style=none, fill=white, scale=1] (45) at (4.5, -0.65) {Alice};
		\node [style=none] (47) at (6.5, -0.45) {};
		\node [style=none] (48) at (5.95, -0.45) {};
		\node [style=none] (49) at (7.05, -0.45) {};
		\node [style=none] (50) at (7.05, -0.85) {};
		\node [style=none] (51) at (6.5, -0.95) {};
		\node [style=none] (52) at (5.95, -0.85) {};
		\node [style=none, fill=white, scale=1] (53) at (6.5, -0.65) {coffee};
	\end{pgfonlayer}
	\begin{pgfonlayer}{edgelayer}
		\draw [in=90, out=-90] (14.center) to (15.center);
		\draw [-, fill=white] (19.center)
			 to (20.center)
			 to (21.center)
			 to (22.center)
			 to (23.center)
			 to cycle;
		\draw [in=90, out=-90] (15.center) to (36.center);
		\draw [-, fill=white] (40.center) to (41.center);
		\draw [-, fill=white] (41.center) to (42.center);
		\draw [-, fill=white] (42.center) to (43.center);
		\draw [-, fill=white] (43.center) to (44.center);
		\draw [-, fill=white] (44.center) to (40.center);
		\draw [-, fill=white] (48.center) to (49.center);
		\draw [-, fill=white] (49.center) to (50.center);
		\draw [-, fill=white] (50.center) to (51.center);
		\draw [-, fill=white] (51.center) to (52.center);
		\draw [-, fill=white] (52.center) to (48.center);
		\draw (11.center) to (39.center);
		\draw (17.center) to (47.center);
	\end{pgfonlayer}
\end{tikzpicture}
\end{center}
\caption{Optimisation through rewriting.}
\label{fig:rewrite}
\end{center}
\end{figure}

One way to ameliorate both problems is to perform some diagrammatic rewriting. By rotating certain words in the pregroup diagram, we can eliminate the cups that correspond to Bell measurements using the `snake equation'. In terms of linear algebra, the rotation operation corresponds to taking the transpose. This rewriting technique, first used in \citep{Lorenz2021}, can also be applied to our musical pregroup diagrams. In Fig. \ref{fig:rewrite}, the number of postselections performed is halved from 8 to 4, so the number of useful shots has increase by a factor of 16. By applying this rewriting technique to our diagrams, we reduce the worse-case circuit size in our training set from 25 qubits to 13 qubits. This is a reduction of 12 postselections. Thus, the number of shots required is reduced by a factor of 4096.

\subsection{Circuit Converter: Translating Musical Compositions into Quantum Circuits}
\label{subsec:preg2qcirc}

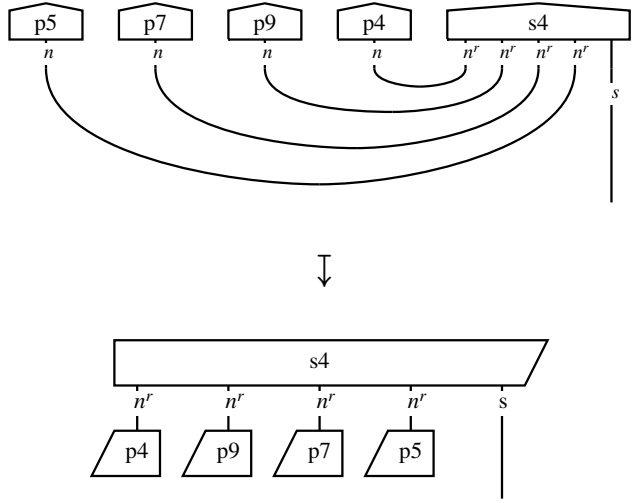
\begin{figure}[h]
    \begin{center}
    \scalebox{1.2}{\begin{tikzpicture}[scale=0.8, baseline={(current bounding box).east}, node distance=5mm, text height=5pt, text depth=0pt, {every node/.style}={scale=0.6}]
	\begin{pgfonlayer}{nodelayer}
		\node [style=none] (0) at (2.75, 0) {};
		\node [style=none, fill=white, anchor=west, scale=0.85] (1) at (3.125, -0.2) {$n$};
		\node [style=none] (2) at (3.25, 0) {};
		\node [style=none] (3) at (3.25, -0.4) {};
		\node [style=none] (4) at (2.75, 0) {};
		\node [style=none] (5) at (3.75, 0) {};
		\node [style=none] (6) at (3.75, 0.4) {};
		\node [style=none] (7) at (3.25, 0.5) {};
		\node [style=none] (8) at (2.75, 0.4) {};
		\node [style=none, fill=white, scale=1] (9) at (3.25, 0.2) {p4};
		\node [style=none, fill=white, anchor=west, scale=0.85] (10) at (4.375, -0.2) {$n^{r}$};
		\node [style=none] (11) at (4.5, 0) {};
		\node [style=none] (12) at (4.5, -0.4) {};
		\node [style=none] (13) at (5, 0) {};
		\node [style=none] (14) at (5, -0.4) {};
		\node [style=none, fill=white, anchor=west, scale=0.85] (15) at (5.375, -0.2) {$n^{r}$};
		\node [style=none] (16) at (5.5, 0) {};
		\node [style=none] (17) at (5.5, -0.4) {};
		\node [style=none] (18) at (4.25, 0) {};
		\node [style=none] (19) at (6.75, 0) {};
		\node [style=none] (20) at (6.75, 0.4) {};
		\node [style=none] (21) at (5.5, 0.5) {};
		\node [style=none] (22) at (4.25, 0.4) {};
		\node [style=none, fill=white, scale=1] (23) at (5.5, 0.2) {s4};
		\node [style=none] (33) at (3.875, -0.65) {};
		\node [style=none, fill=white, anchor=west, scale=0.85] (37) at (5.875, -0.2) {$n^{r}$};
		\node [style=none] (38) at (6, 0) {};
		\node [style=none] (39) at (6, -0.4) {};
		\node [style=none] (40) at (6.5, 0) {};
		\node [style=none] (41) at (6.5, -0.4) {};
		\node [style=none] (42) at (5.75, 0) {};
		\node [style=none] (43) at (6.5, -2.25) {};
		\node [style=none, fill=white, anchor=west, scale=0.85] (44) at (6.4, -0.75) {$s$};
		\node [style=none] (45) at (1.25, 0) {};
		\node [style=none, fill=white, anchor=west, scale=0.85] (46) at (1.625, -0.2) {$n$};
		\node [style=none] (47) at (1.75, 0) {};
		\node [style=none] (48) at (1.75, -0.4) {};
		\node [style=none] (49) at (1.25, 0) {};
		\node [style=none] (50) at (2.25, 0) {};
		\node [style=none] (51) at (2.25, 0.4) {};
		\node [style=none] (52) at (1.75, 0.5) {};
		\node [style=none] (53) at (1.25, 0.4) {};
		\node [style=none, fill=white, scale=1] (54) at (1.75, 0.2) {p9};
		\node [style=none] (55) at (-0.25, 0) {};
		\node [style=none, fill=white, anchor=west, scale=0.85] (56) at (0.125, -0.2) {$n$};
		\node [style=none] (57) at (0.25, 0) {};
		\node [style=none] (58) at (0.25, -0.4) {};
		\node [style=none] (59) at (-0.25, 0) {};
		\node [style=none] (60) at (0.75, 0) {};
		\node [style=none] (61) at (0.75, 0.4) {};
		\node [style=none] (62) at (0.25, 0.5) {};
		\node [style=none] (63) at (-0.25, 0.4) {};
		\node [style=none, fill=white, scale=1] (64) at (0.25, 0.2) {p7};
		\node [style=none] (65) at (-1.75, 0) {};
		\node [style=none, fill=white, anchor=west, scale=0.85] (66) at (-1.375, -0.2) {$n$};
		\node [style=none] (67) at (-1.25, 0) {};
		\node [style=none] (68) at (-1.25, -0.4) {};
		\node [style=none] (69) at (-1.75, 0) {};
		\node [style=none] (70) at (-0.75, 0) {};
		\node [style=none] (71) at (-0.75, 0.4) {};
		\node [style=none] (72) at (-1.25, 0.5) {};
		\node [style=none] (73) at (-1.75, 0.4) {};
		\node [style=none, fill=white, scale=1] (74) at (-1.25, 0.2) {p5};
		\node [style=none] (75) at (3.5, -1) {};
		\node [style=none] (76) at (3, -1.5) {};
		\node [style=none] (77) at (2.5, -2) {};
		\node [style=none, fill=white, anchor=west, scale=0.85] (78) at (4.875, -0.2) {$n^{r}$};
	\end{pgfonlayer}
	\begin{pgfonlayer}{edgelayer}
		\draw [in=90, out=-90] (2.center) to (3.center);
		\draw [-, fill=white] (4.center)
			 to (5.center)
			 to (6.center)
			 to (7.center)
			 to (8.center)
			 to cycle;
		\draw [in=90, out=-90] (11.center) to (12.center);
		\draw [in=90, out=-90] (13.center) to (14.center);
		\draw [in=90, out=-90] (16.center) to (17.center);
		\draw [-, fill=white] (18.center)
			 to (19.center)
			 to (20.center)
			 to (21.center)
			 to (22.center)
			 to cycle;
		\draw [in=180, out=-90, looseness=0.72] (3.center) to (33.center);
		\draw [in=0, out=-90, looseness=0.72] (12.center) to (33.center);
		\draw [in=90, out=-90] (38.center) to (39.center);
		\draw [in=90, out=-90] (40.center) to (41.center);
		\draw [in=90, out=-90] (41.center) to (43.center);
		\draw [in=90, out=-90] (47.center) to (48.center);
		\draw [-, fill=white] (49.center)
			 to (50.center)
			 to (51.center)
			 to (52.center)
			 to (53.center)
			 to cycle;
		\draw [in=90, out=-90] (57.center) to (58.center);
		\draw [-, fill=white] (59.center)
			 to (60.center)
			 to (61.center)
			 to (62.center)
			 to (63.center)
			 to cycle;
		\draw [in=90, out=-90] (67.center) to (68.center);
		\draw [-, fill=white] (69.center)
			 to (70.center)
			 to (71.center)
			 to (72.center)
			 to (73.center)
			 to cycle;
		\draw [in=-180, out=-90, looseness=0.75] (48.center) to (75.center);
		\draw [in=-90, out=0, looseness=0.75] (75.center) to (14.center);
		\draw [in=-180, out=-90, looseness=0.75] (58.center) to (76.center);
		\draw [in=-90, out=0, looseness=0.75] (76.center) to (17.center);
		\draw [in=-180, out=-90, looseness=0.75] (68.center) to (77.center);
		\draw [in=-90, out=0, looseness=0.75] (77.center) to (39.center);
	\end{pgfonlayer}
\end{tikzpicture}}\\
    \vspace{5mm}\scalebox{1.25}{$\downmapsto$}\vspace{5mm}\\
    \scalebox{1.2}{\begin{tikzpicture}[baseline={(current bounding box).east}, node distance=5mm, text height=5pt, text depth=0pt, {every node/.style}={scale=0.6}]
	\begin{pgfonlayer}{nodelayer}
		\node [style=none] (1) at (0, 4.25) {};
		\node [style=none] (2) at (0, 3.75) {};
		\node [style=none, fill=white, right] (3) at (-0.125, 4.05) {$n^r$};
		\node [style=none] (4) at (1, 4.25) {};
		\node [style=none] (5) at (1, 3.75) {};
		\node [style=none, fill=white, right] (6) at (0.875, 4.05) {$n^r$};
		\node [style=none] (7) at (2, 4.25) {};
		\node [style=none] (8) at (2, 3.75) {};
		\node [style=none, fill=white, right] (9) at (1.875, 4.05) {$n^r$};
		\node [style=none] (10) at (3, 4.25) {};
		\node [style=none] (11) at (3, 3.75) {};
		\node [style=none, fill=white, right] (12) at (2.875, 4.05) {$n^r$};
		\node [style=none] (13) at (4, 4.25) {};
		\node [style=none] (14) at (4, 3) {};
		\node [style=none, fill=white, right] (15) at (3.875, 4.05) {s};
		\node [style=none] (16) at (-0.25, 4.25) {};
		\node [style=none] (17) at (4.25, 4.25) {};
		\node [style=none] (18) at (4.5, 4.75) {};
		\node [style=none] (19) at (-0.25, 4.75) {};
		\node [style=none, fill=white] (20) at (2, 4.5) {s4};
		\node [style=none] (21) at (-0.5, 3.25) {};
		\node [style=none] (22) at (0.25, 3.25) {};
		\node [style=none] (23) at (0.25, 3.75) {};
		\node [style=none] (24) at (-0.25, 3.75) {};
		\node [style=none, fill=white] (25) at (0, 3.5) {p4};
		\node [style=none] (26) at (0.5, 3.25) {};
		\node [style=none] (27) at (1.25, 3.25) {};
		\node [style=none] (28) at (1.25, 3.75) {};
		\node [style=none] (29) at (0.75, 3.75) {};
		\node [style=none, fill=white] (30) at (1, 3.5) {p9};
		\node [style=none] (31) at (1.5, 3.25) {};
		\node [style=none] (32) at (2.25, 3.25) {};
		\node [style=none] (33) at (2.25, 3.75) {};
		\node [style=none] (34) at (1.75, 3.75) {};
		\node [style=none, fill=white] (35) at (2, 3.5) {p7};
		\node [style=none] (36) at (2.5, 3.25) {};
		\node [style=none] (37) at (3.25, 3.25) {};
		\node [style=none] (38) at (3.25, 3.75) {};
		\node [style=none] (39) at (2.75, 3.75) {};
		\node [style=none, fill=white] (40) at (3, 3.5) {p5};
	\end{pgfonlayer}
	\begin{pgfonlayer}{edgelayer}
		\draw [in=90, out=-90] (1.center) to (2.center);
		\draw [in=90, out=-90] (4.center) to (5.center);
		\draw [in=90, out=-90] (7.center) to (8.center);
		\draw [in=90, out=-90] (10.center) to (11.center);
		\draw [in=90, out=-90] (13.center) to (14.center);
		\draw [-, fill=white] (16.center)
			 to (17.center)
			 to (18.center)
			 to (19.center)
			 to cycle;
		\draw [-, fill=white] (21.center)
			 to (22.center)
			 to (23.center)
			 to (24.center)
			 to cycle;
		\draw [-, fill=white] (26.center)
			 to (27.center)
			 to (28.center)
			 to (29.center)
			 to cycle;
		\draw [-, fill=white] (31.center)
			 to (32.center)
			 to (33.center)
			 to (34.center)
			 to cycle;
		\draw [-, fill=white] (36.center)
			 to (37.center)
			 to (38.center)
			 to (39.center)
			 to cycle;
	\end{pgfonlayer}
\end{tikzpicture}}
    \caption{The pregroup diagram in Fig. \ref{fig:conversion} rewritten by removing the caps.}
    \label{fig:pg_rewrite}
    \end{center}
\end{figure}

With DisCoCat, a string diagram for a musical composition $\sigma$ (which is generated by a CFG and thus has a corresponding pregroup diagram) can be translated onto a parameterised quantum circuit $C_\sigma (\theta_\sigma)$ over a parameter set $\theta_\sigma$. The translation assigns a number of qubits $q_b$ to each wire carrying a type $b$. Therefore, Hilbert spaces are defined on the wires (Fig. \ref{fig:pg_rewrite}).

\begin{figure}
    \begin{center}
    \scalebox{1.2}{\begin{tikzpicture}[baseline={(current bounding box).east}, node distance=5mm, text height=5pt, text depth=0pt, {every node/.style}={scale=0.6}]
	\begin{pgfonlayer}{nodelayer}
		\node [style=none] (1) at (0, 4.25) {};
		\node [style=none] (2) at (0, 3.75) {};
		\node [style=none, fill=white, right] (3) at (-0.125, 4.05) {$n^r$};
		\node [style=none] (4) at (1, 4.25) {};
		\node [style=none] (5) at (1, 3.75) {};
		\node [style=none, fill=white, right] (6) at (0.875, 4.05) {$n^r$};
		\node [style=none] (7) at (2, 4.25) {};
		\node [style=none] (8) at (2, 3.75) {};
		\node [style=none, fill=white, right] (9) at (1.875, 4.05) {$n^r$};
		\node [style=none] (10) at (3, 4.25) {};
		\node [style=none] (11) at (3, 3.75) {};
		\node [style=none, fill=white, right] (12) at (2.875, 4.05) {$n^r$};
		\node [style=none] (13) at (4, 4.25) {};
		\node [style=none] (14) at (4, 3) {};
		\node [style=none, fill=white, right] (15) at (3.875, 4.05) {s};
		\node [style=none] (16) at (-0.25, 4.25) {};
		\node [style=none] (17) at (4.25, 4.25) {};
		\node [style=none] (18) at (4.5, 4.75) {};
		\node [style=none] (19) at (-0.25, 4.75) {};
		\node [style=none] (20) at (2, 4.5) {{s4}};
		\node [style=none] (21) at (-0.5, 3.25) {};
		\node [style=none] (22) at (0.25, 3.25) {};
		\node [style=none] (23) at (0.25, 3.75) {};
		\node [style=none] (24) at (-0.25, 3.75) {};
		\node [style=none] (25) at (0, 3.5) {{p4}};
		\node [style=none] (26) at (0.5, 3.25) {};
		\node [style=none] (27) at (1.25, 3.25) {};
		\node [style=none] (28) at (1.25, 3.75) {};
		\node [style=none] (29) at (0.75, 3.75) {};
		\node [style=none] (30) at (1, 3.5) {{p9}};
		\node [style=none] (31) at (1.5, 3.25) {};
		\node [style=none] (32) at (2.25, 3.25) {};
		\node [style=none] (33) at (2.25, 3.75) {};
		\node [style=none] (34) at (1.75, 3.75) {};
		\node [style=none] (35) at (2, 3.5) {{p7}};
		\node [style=none] (36) at (2.5, 3.25) {};
		\node [style=none] (37) at (3.25, 3.25) {};
		\node [style=none] (38) at (3.25, 3.75) {};
		\node [style=none] (39) at (2.75, 3.75) {};
		\node [style=none] (40) at (3, 3.5) {{p5}};
	\end{pgfonlayer}
	\begin{pgfonlayer}{edgelayer}
		\draw [in=90, out=-90] (1.center) to (2.center);
		\draw [in=90, out=-90] (4.center) to (5.center);
		\draw [in=90, out=-90] (7.center) to (8.center);
		\draw [in=90, out=-90] (10.center) to (11.center);
		\draw [in=90, out=-90] (13.center) to (14.center);
		\draw [-, fill=white, style=color1] (16.center)
			 to (17.center)
			 to (18.center)
			 to (19.center)
			 to cycle;
		\draw [-, fill=white, style=color2] (21.center)
			 to (22.center)
			 to (23.center)
			 to (24.center)
			 to cycle;
		\draw [-, fill=white, style=color3] (26.center)
			 to (27.center)
			 to (28.center)
			 to (29.center)
			 to cycle;
		\draw [-, fill=white, style=color4] (31.center)
			 to (32.center)
			 to (33.center)
			 to (34.center)
			 to cycle;
		\draw [-, fill=white, style=color5] (36.center)
			 to (37.center)
			 to (38.center)
			 to (39.center)
			 to cycle;
	\end{pgfonlayer}
\end{tikzpicture}} \\
    \vspace{5mm}\scalebox{1.25}{$\downmapsto$}\vspace{5mm} \\
    \scalebox{1.5}{\input{tikz/circ2.tikz}}
    \caption{The pregroup diagram in Fig. \ref{fig:pg_rewrite} converted to a quantum circuit, according to the IQP and Euler decomposition ans\"{a}tze. Regions of the diagram and circuit have been coloured to illustrate the transformation of each part of the diagram into a quantum circuit.}
    \label{fig:pg2circ}
    \end{center}
\end{figure}

\medskip

The snippet-state of the $\eta$-th snippet in a string diagram is reinterpreted as a pure quantum state prepared from a given reference product-state by a circuit $C_\eta$ (Eq. \ref{eq:purestate}).

\begin{equation}
C_\eta({\theta}_\eta)\ket{0}^{\otimes q_\eta }~,~~q_\eta = \sum_{i=1}^{|\eta|} {q_{b_i}}
\label{eq:purestate}
\end{equation}

\medskip

The width of a snippet-circuit depends on the number of qubits assigned to each of the $|\eta|$-many basic types $b \in B$ assigned to the snippet. Given a composition $\sigma$, the system firstly concatenates the snippet-states of the snippets as they appear in the composition. Concatenation corresponds to taking their tensor product (Eq. \ref{eq:tensorproduct}).

\begin{equation}
C_\sigma({\theta}_\sigma) \ket{0}^{\otimes q_{\sigma} } = {\bigotimes}_{\eta_j} C_{\eta_j}({\theta}_{\eta_j}) \ket{0}^{\otimes q_{\eta_j} }
\label{eq:tensorproduct}
\end{equation}

\medskip 

The circuit in Fig. \ref{fig:pg2circ} prepares the state $\ket {\sigma ({\theta}_\sigma)}$ from the all-zeroes basis states. A musical composition is parameterised by the concatenation of the parameters of its snippets: $\theta_\sigma =  \cup_{\eta \in \sigma}\theta_\eta$, where $\theta_\eta$ defines the snippet-embedding $\ket {\eta(\theta_\eta)}$.

\medskip

Then, Bell effects are applied, as determined by the cups representing pregroup reductions. Cups are interpreted as entangling circuits that implement a Bell effect by postselecting on a Bell state. This `effect of grammar', denoted $g$, acts on the state of the composition and returns the state $g({\ket {\sigma ({\theta}_\sigma)}})$.

\subsection{Training the System to Classify}
\label{sec:ml_model}

Here we describe how we trained the system with the aforementioned musical dataset to distinguish between two categories of music: rhythmic or melodic.

\medskip

As we are performing binary classification, we choose to represent the output of the system in one-hot encoding. This means that our two class labels are $[0,1]$ and $[1,0]$ corresponding to `melodic' and `rhythmic' music respectively. In the DisCoCat model, this can be conveniently achieved by setting the sentence dimension space (recall the whole circuit is represented by an open sentence wire) of our DisCoCat model to one qubit. The circuit $C_\sigma$ for each composition $\sigma$ is evaluated for the current set of parameters $\theta_\sigma$ on the quantum computer giving output state $\ket{C(\theta_\sigma)}$. The expected prediction $L^i_\text{pred}(\sigma, \theta)$ is given by the Born rule in Eq. \ref{eq:bornrule}, where $i \in \{0,1\}$ and $\theta =  \cup_{\sigma \in \Sigma}\theta_\sigma$.

\begin{equation}
L^i_\text{pred}(\sigma, \theta) \coloneqq  |\braket{i | C_\sigma(\theta_\sigma)}|^2
\label{eq:bornrule}
\end{equation}

\medskip

By running and measuring the outcome (either $[0, 1]$ or $[1, 0]$) of the circuit many times, the average outcome will converge towards this value. Then $L^i_\text{pred}(\sigma, \theta)$ is normalised to obtain a smooth probability distribution (Eq. \ref{eq:probdistrib}).

\begin{equation}
l^i_\text{pred}(\sigma, \theta) \coloneqq \frac{L^i_\text{pred} (\sigma, \theta) + \epsilon}{\sum\limits_{j} \Big(L^j_\text{pred}(\sigma, \theta) + \epsilon\Big)}
\label{eq:probdistrib}
\end{equation}

\medskip

The smoothing term is chosen to be $\epsilon = 10^{-9}$. The predicted label is obtained from the probability distribution by setting a decision threshold $t$ on $l^0_\text{pred}(\sigma, \theta)$: if $t < l^0_\text{pred}(\sigma, \theta)$ then the predicted label is $[0, 1]$. And if $t \geq l^0_\text{pred}(\sigma, \theta)$ then the predicted label is $[1, 0]$. This threshold can be adjusted depending on the desired sensitivity and specificity of the system. For this chapter, the threshold $t=0.5$ is selected.

\medskip

To predict the label for a given snippet well, we need to use optimisation methods to find optimal parameters for our model. We begin by comparing the predicted label with the training label using a loss function. Since we are performing a classification task, the binary cross-entropy loss function is chosen (Eq. \ref{eq:entropy}).

\begin{equation}
\mathrm{BCE}(\theta) = \sum_{\sigma}\sum_{i \in \{0, 1\}}{l^i_\text{label}(\sigma, \theta)\log(l^i_\text{pred}(\sigma, \theta))}
\label{eq:entropy}
\end{equation}

\medskip

A non-gradient based optimisation algorithm known as SPSA (Simultaneous Perturbation Stochastic Approximation) is used to minimise our loss function \citep{705889}. Alternatively, one could use gradient-based optimisation algorithms by differentiating the DisCoCat diagram \citep{Toumi_2021}. However, SPSA was found to be sufficient for our purposes.

\medskip

In order to minimise the loss function the system learns to classify the compositions by adjusting the parameters of the musical snippets. Philosophically, this is in contrast to typical connectionist machine learning models - that is, neural networks - where the weights of the network are adjusted rather than the weights of the word embeddings. This is because pregroup grammars are lexical. And in lexical grammars, whilst words have parameters that can be changed, the grammatical structure cannot be changed.

\begin{figure}[h]
\begin{center}
\includegraphics[width=0.6\textwidth]{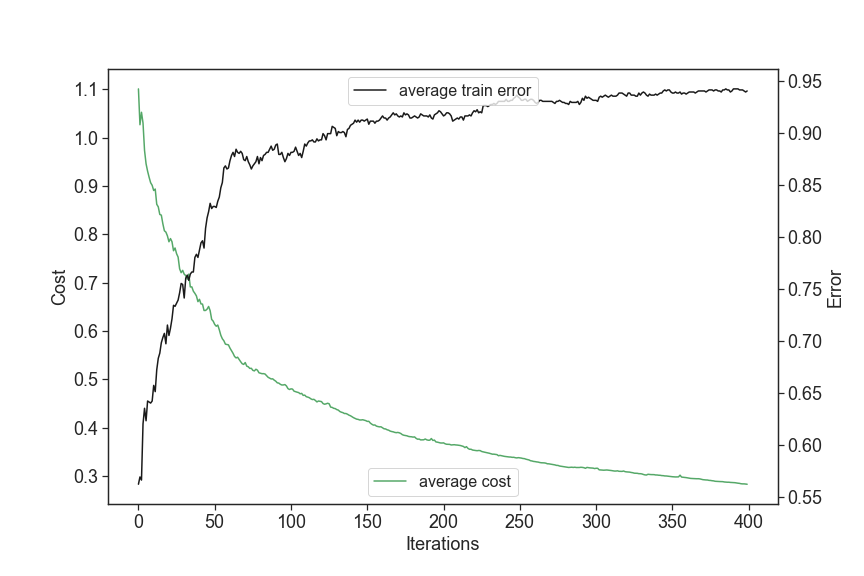}
\caption{Average over 20 different runs of a classical simulation of the training set showing the cost (green, left hand side axis) and the training error (black, right hand side axis).}
\label{fig:exact_sim}
\end{center}
\end{figure}

\medskip

Unfortunately, the training process cannot be performed entirely on a real quantum device at the time of writing. Access to quantum hardware still is limited, due to the demand for and the limited number of available quantum processors. This can be problematic for running variational tasks where circuits must be run for many iterations in order to get results.

\medskip

Therefore, we firstly pre-train the model parameters using exact tensor contraction of the quantum circuits on a classical computer. We use Python's JAX library\footnote{https://jax.readthedocs.io/en/latest/} for just-in-time compilation, which speeds up the pre-training. This speed up allows us to efficiently perform grid search and find the best hyper-parameter settings for the SPSA optimiser. The results for such a classical simulation are shown in Fig. \ref{fig:exact_sim} for the final settings chosen for the subsequent training.

\medskip

\begin{figure}[h]
\begin{center}
\includegraphics[width=0.6\textwidth]{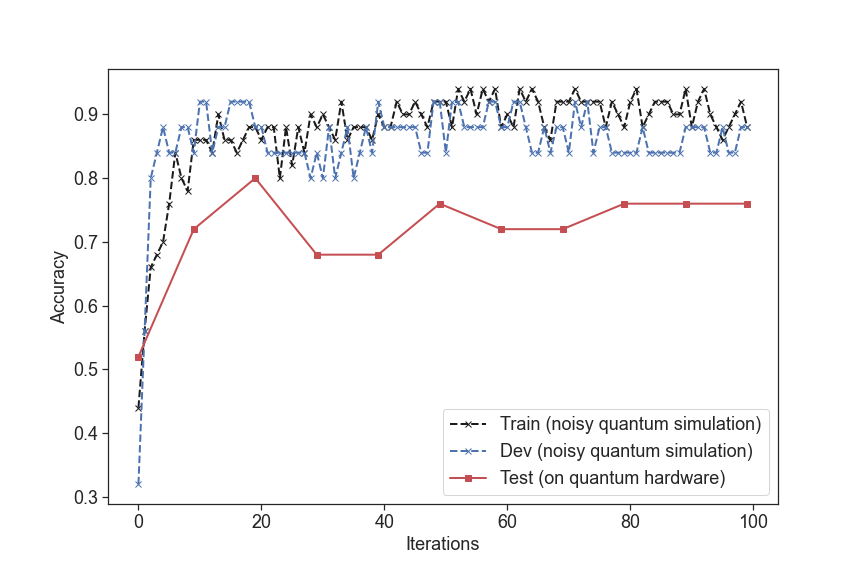}
\caption{Noisy quantum simulation to train the model (black), noisy quantum simulation of the development set (blue) and the test set run on on quantum hardware (red). The noise model used for the noisy simulation was that of the  \texttt{ibmq\_guadalupe} device, which is the hardware that we subsequently usedfor running the test set.}
\label{fig:noisy_sim}
\end{center}
\end{figure}

Then, we simulate the quantum process on a classical machine. In order to do this, we need to take into account the statistical nature of our experiments, which would involve running the same circuit thousands of times to obtain measurement statistics. And we also need to bear in mind the noise that is present in a real device. The parameters learnt from the exact pre-training is transferred to a model that closely resembles the real quantum device, which includes noise simulation. This will ultimately improve testing performance on the real quantum device. So, the same circuit is simulated 8,192 times with a noise model specific to the quantum device that will used for testing. This process is carried out for each composition in the training set, for a chosen number of iterations. The train and development set results from the quantum simulations are shown in Fig. \ref{fig:noisy_sim}.

\medskip

Once the pre-training phase is complete, and we are happy with the simulated results on the development set, then we take the parameters from the noisy quantum simulation and evaluate the test set on a real quantum computer. We used IBM Quantum's device \texttt{ibmq\_guadalupe}, which is a 16-qubit superconducting processor with a quantum volume of 32. Despite the expected differences between simulation and a quantum hardware processing, we can see that the model learns to classify compositions in the test set with an impressive final accuracy of 76\%, as shown in Fig. \ref{fig:noisy_sim}.

\subsubsection{Further Classification Tests}

The results from the testing with quantum hardware (Fig. \ref{fig:noisy_sim}) exceeded our expectations. We conducted additional tests to probe the system further, which confirmed the system's positive performance.

\medskip

For the new tests we ran our CFG to produce two additional datasets: one containing 60 compositions and another containing 90. As with the previous dataset, we auditioned and annotated the pieces manually. Again, both sets contain compositions of varying lengths. Then, we submitted the sets to the quantum classifier. Below we display the results from a different perspective than in Fig. \ref{fig:noisy_sim}.

\medskip

\begin{figure}[h]
\begin{center}
\includegraphics[width=0.7\textwidth]{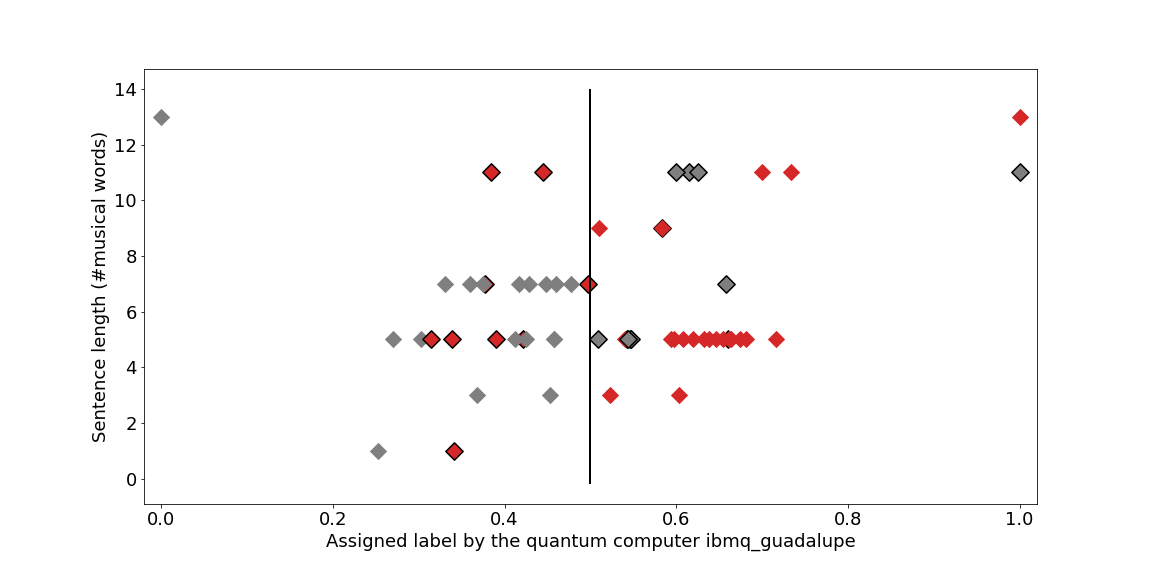}
\caption{Quantum computing classification of dataset with 60 compositions.}
\label{fig:results3}
\end{center}
\end{figure}

\begin{figure}[h]
\begin{center}
\includegraphics[width=0.7\textwidth]{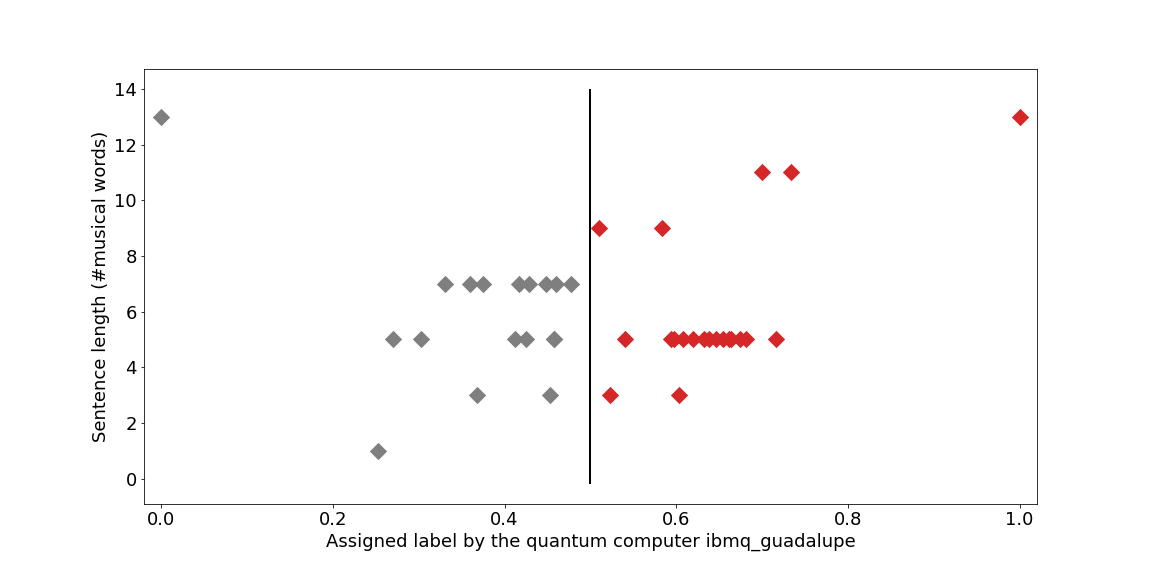}
\caption{A version of Fig. \ref{fig:results3} plotting only those pieces that were classified correctly.}
\label{fig:results1}
\end{center}
\end{figure}

Fig. \ref{fig:results3} shows that the system correctly classified 39 compositions from the set of 60. Melodic pieces corresponds to a value plotted on the first element in the tuple of $<$ 0.5 (grey), whereas the Rhythmic ones corresponds to a value $\geq$ 0.5 (red). Misclassified compositions have a black outline around them, as well as being the wrong colour for the side of the graph that they appear on. For clarity, Fig. \ref{fig:results1} plots only those compositions that were classified correctly.


\begin{figure}[h]
\begin{center}
\includegraphics[width=0.7\textwidth]{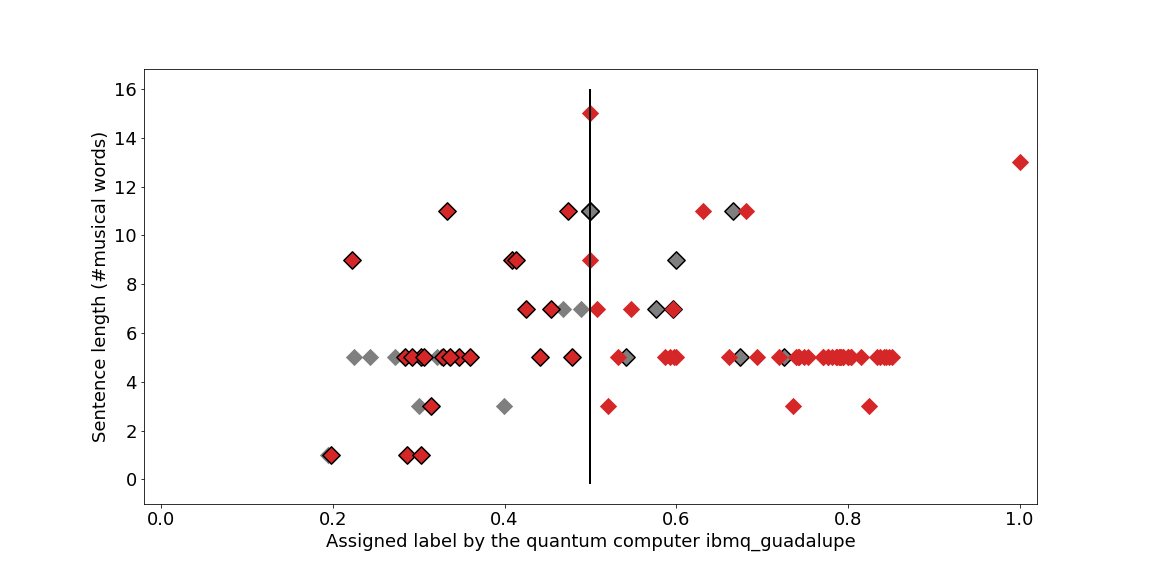}
\caption{Quantum computing classification of dataset with 90 compositions.}
\label{fig:results2}
\end{center}
\end{figure}

\begin{figure}[h]
\begin{center}
\includegraphics[width=0.7\textwidth]{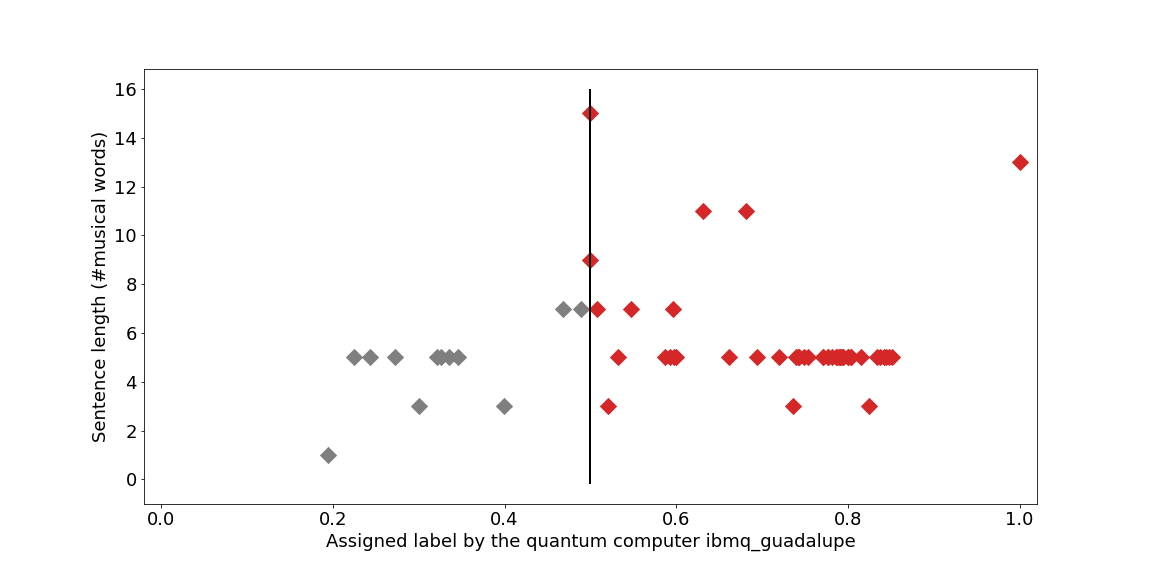}
\caption{A version of Fig. \ref{fig:results2} plotting only those pieces that were classified correctly.}
\label{fig:results4}
\end{center}
\end{figure}

Next, Fig. \ref{fig:results2} shows that the system correctly classified 59 compositions from the set of 90. Again, melodic pieces corresponds to a value plotted on the first element in the tuple of $<$ 0.5 (grey), whereas the Rhythmic ones corresponds to a value $\geq$ 0.5 (red). And Fig. \ref{fig:results4} shows the same results, but plots only those compositions that were classified correctly.

\section{\textit{Quanthoven}: Leveraging the Quantum Classifier to Compose}
\label{sec:generative}

\medskip

Composer Lejaren Hiller, who also was a chemistry teacher, allegedly is a pioneer of programming computers to generate music. In 1957, he teamed up with Leonard Isaacson, also a chemist and composer, to program the ILLIAC I\footnote{This was the first von Neumann architecture computer built and owned by a university in the USA. It was put into service on September 22, 1952.} (Illinois Automatic Computer) at the University of Illinois, USA, to compose music. The computer produced materials for their string quartet, entitled \textit{Illiac Suite}\footnote{Later retitled as ``String Quartet No. 4''.}.


\medskip

In their initial experiments, they programmed the computer to do two things: (a) generate musical data\footnote{That is, sequences of symbols representing notes, rhythms and expressions such as `pizzicato and `arco', and so on.} (pseudo-randomly), and  (b) test if the data satisfied musical rules. Musical data that did not satisfy the rules were discarded. Conversely, those that satisfied the rules were kept for the ongoing composition. For instance, for the first movement of \textit{Illiac Suite} they set the machine with rules from the famous treatise \textit{Gradus ad Parnassum}, written by Joseph Fux in the $17^{th}$ century \citep{Mann_1965}.

\medskip

\begin{figure}[h]
\begin{center}
\includegraphics[width=0.3\textwidth]{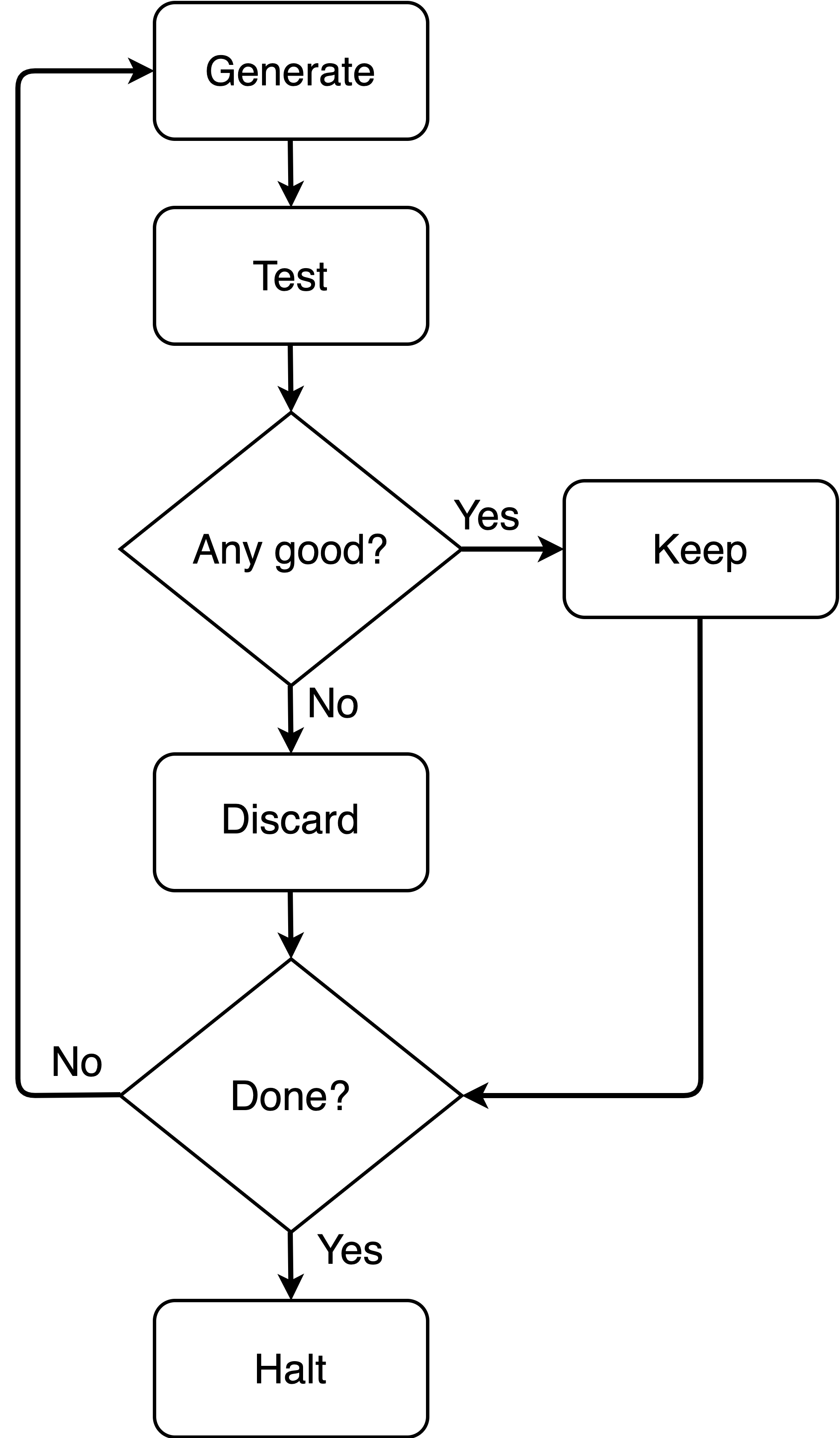}
\caption{Archetypal generate-and-test approach to algorithmic music composition.}
\label{fig:gentest}
\end{center}
\end{figure}

Hiller's approach became an archetype for algorithmic music composition with computers, which is often referred to as the \textit{generate-and-test} approach (Fig. \ref{fig:gentest}). Various variants have been implemented, with increasingly sophisticated methods for generating musical data and probing them \citep{Edwards_2011} \citep{Miranda2001}. Such systems might  generate notes that are checked one at a time, or phrases, larger sequences or even entire pieces. Human approval often takes place as well, whereby composers may further discard or amend computer-validated materials.

\medskip

The generate-and-test approach also informed the development of Artificial Intelligence (AI) techniques for musical composition \citep{Miranda2021} \citep{FernandezVico_2013}, including highly praised constraint-satisfaction methods \citep{AndersMiranda_2009}. Moreover, the generate-and-test method epitomises Evolutionary Computing techniques for music composition; e.g., using genetic algorithms \citep{Miranda_Biles_2007}. Such systems generate new sequences using biologically inspired processes (e.g., genetic mating, reproduction and mutation) and then employ `fitness measurements' to evaluate them \citep{Miranda_2011}.

\medskip

The quantum machine learning paradigm discussed above fits the generate-and-test approach to algorithmic music  composition very nicely. To probe this, we developed \textit{Quanthoven}, a system for composition with the classifier presented above acting as the evaluator. In a nutshell, the CFG (section \ref{sec:cfg}) generates compositions and the trained classifier (section \ref{sec:ml_model}) evaluates them (Fig. \ref{fig:gentest}).

\medskip

At the introduction, we mentioned that our ambition is to develop systems to aid musicians to compose `music for Alice's tea party', or a `tune to make Bob feel energetic'. In order to do this, the machine is required  to handle the meaning of music; that is, perceived properties of musical compositions, which holistically characterise their style. Our proof-of-concept gives a good hint about how this can be achieved.

\medskip

Effectively, our machine learning algorithm learned what we mean by `melodic' and `rhythmic' music. The CFG can generate pieces with snippets combined in ways that may render a piece more melodic than rhythmic, or vice-versa. Some might be neither to our ears, others might be both. But the quantum classifier is able to suggest whether a CFG-generated piece is one or the other.

\medskip

As a demonstration, we set ourselves the task of composing 4 pieces of music with \textit{Quanthoven}: two for relaxing at a tea party and two aimed at inducing excitement. We set \textit{Quanthoven} to generate two dozen pieces and save four, two for each category. The others were discarded on the spot; we do not even know how they sounded like.

\medskip

The saved pieces were encoded as MIDI files. Then, we uploaded these files into a music editor, formatted them, added titles and other details, and printed the scores. We hired a professional pianist to record them. The two pieces for relaxing at a tea party were entitled as \textit{Bob's Chamomile Slowdown} and \textit{Alice's Mushroom Trip}. The ones aimed at making one feel excited were entitled \textit{Bob's Cigar Buzz} and \textit{Alice's Caffeine Rush}.

\medskip

\begin{figure}[h]
\begin{center}
\includegraphics[width=0.64\textwidth]{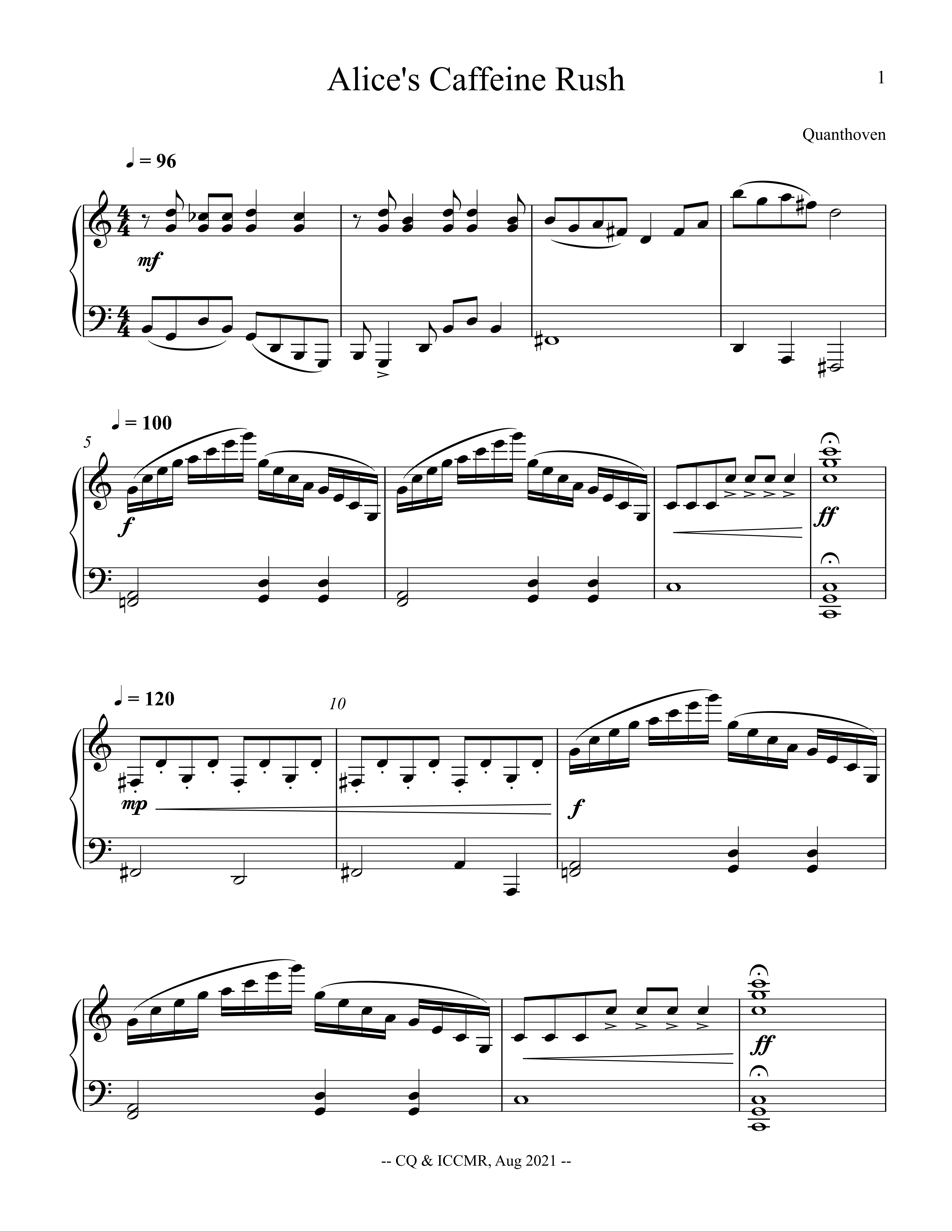}
\caption{First page of the composition \textit{Alice's Caffeine Rush}.}
\label{fig:alice_coffee_1}
\end{center}
\end{figure}

\begin{figure}[h]
\begin{center}
\includegraphics[width=0.64\textwidth]{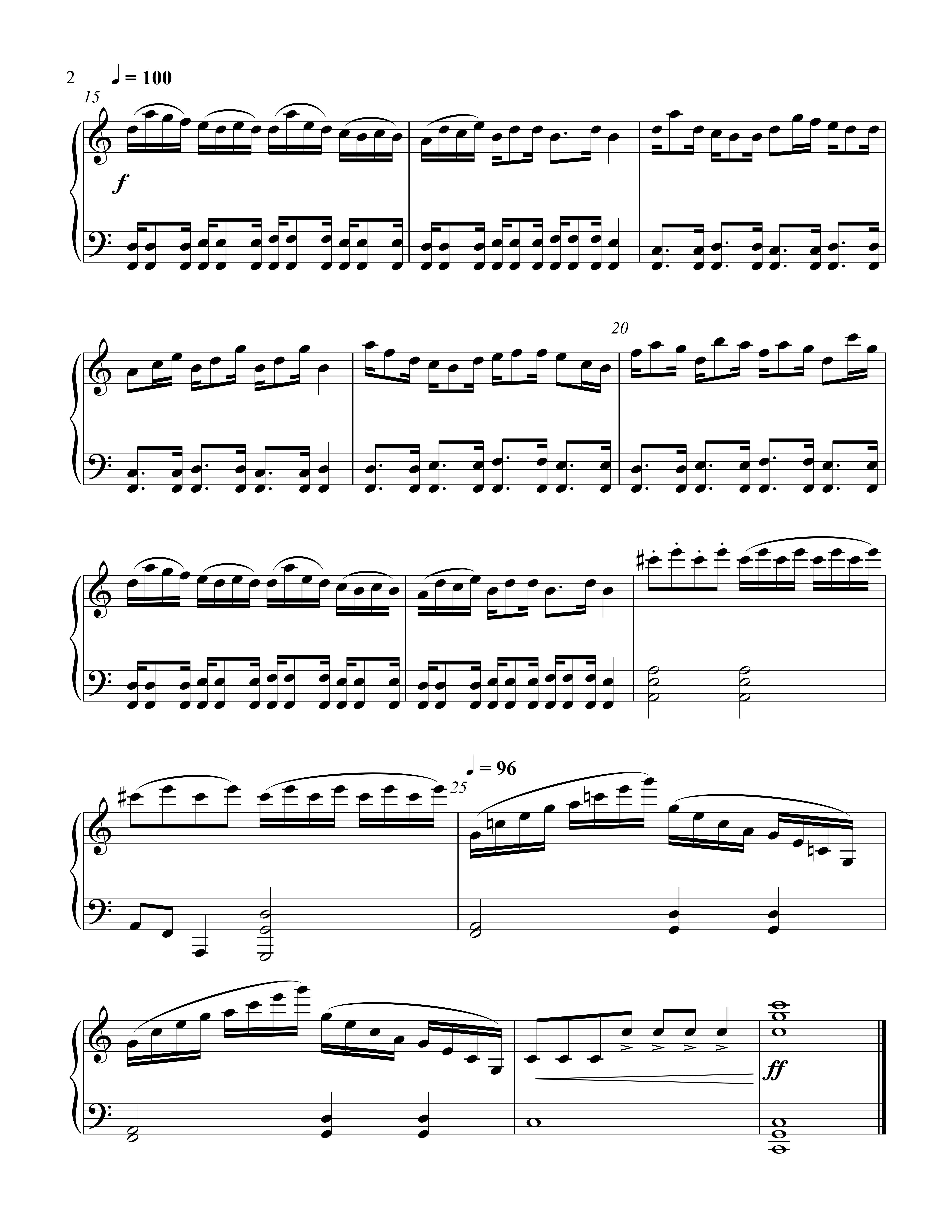}
\caption{Second page of the composition \textit{Alice's Caffeine Rush}.}
\label{fig:alice_coffee_2}
\end{center}
\end{figure}

The score for \textit{Alice's Caffeine Rush} is shown in Figs. \ref{fig:alice_coffee_1} and \ref{fig:alice_coffee_2}. The eyes of a trained musician would immediately see patterns that characterise this piece as rhythmic, in particular the section starting at bar 15. The scores for the other three pieces are provided in Appendix \ref{app:appendix3}.

\section{Final Remarks}

Recordings of all four pieces are released in \href{https://www.soundclick.com/artist/default.cfm?bandID=1491038}{\color{blue}{\underline{SoundClick}}} \citep{Quanthoven2021}. You are invited to listen to the compositions and make your own mind whether you agree with \textit{Quanthoven} or not. We are rather satisfied with the results. But of course, they are debatable.

\medskip

This chapter started by saying that people have different tastes and opinions about music. And it also noted that people perceive and react to music differently from each other. However, the five authors agreed of the `meanings' here. And the aim was to teach the machine to classify the pieces according to our opinions. Objectively, the system does what it says on the tin.

\medskip

We made the conscious decision of conducting this experiment with music generated from a CFG designed from the ground up, completely from scratch. Of course, we could have written a piece of software to extract a CFG from given corpora of musical scores; e.g., \citep{Bod_2001}. Ultimately, we would extract this information directly from audio recordings. However, we wanted to probe our approach with something other than existing styles of music. Moreover, we wanted to ensure that, at this stage of our research, the two categories in question share the same grammatical structure. Effective annotation methods for a much larger database should also be developed; for instance, based on information extracted from social media.
\medskip

As quantum computing hardware and error-correction technologies evolve, we will certainly continue pushing the boundaries in tandem, with compositions of increased sophistication, more instruments, longer durations, more categories (or `meanings'), and so on. These will probably require generative mechanisms other than CFG, something along the lines of transformational-generative grammars. Also, an important next step would be to design an actual quantum generative music engine drawing directly from the classifier, as opposed to using the classifier as a filter.

\section{Acknowledgements}

This project was developed during E. R. Miranda's research residency at Cambridge Quantum in 2021, which was partially funded by the QuTune Project\footnote{QuTune kick-started in the Spring of 2021 thanks to funding kindly provided by the UK National Quantum Technologies Programme's QCS Hub: [ https://iccmr-quantum.github.io/ ]}. The authors thank pianist, Lauryna Sableveciute, and recording engineers, John Lowndes and Manoli Moriaty, at Liverpool Hope University, for the recordings of \textit{Bob's Chamomile Slowdown}, \textit{Alice's Mushroom Trip}, \textit{Bob's Cigar Buzz}, and \textit{Alice's Caffeine Rush}.

\bibliographystyle{plainnat}
\bibliography{refs}

\newpage
\appendix
\section{Appendices}

\subsection{Context-Free Grammars and Monoidal Category Theory }
\label{app:appendix1}

A context-free grammar (CFG) is typically defined with a set of terminals words $T$, non-terminal symbols $N$, and a set of production rules of the form $ \text{A} \to \alpha $, where $\text{A} \in N$ is a non-terminal symbol and $\alpha \in (N \cup T)^*$ is a string of terminal words and non-terminal characters. Any sentence that can be produced by freely applying the production rules on the starting terminal symbol $S$ is considered to be a grammatical sentence.

\medskip

For example, below is a CFG which defines a collection of sentences:

$$\text{NP} \to \text{Adj NP}, \text{S} \to \text{NP V NP}, \text{NP} \to \text{N}$$
$$\text{N} \to n \in \{Alice, tea\}$$
$$\text{V} \to v \in \{drinks, likes, hates\}$$
$$\text{Adj} \to adj \in \{funny, silly, freaky\}$$

\medskip

The general framework of CFGs can be thought of as a freely generated monoidal category.  In diagrammatic notation, a monoidal category consists of boxes with input and output wires. The `types' of of the wires are the generating objects of the monoidal category. Wires of matching type can be connected by extending
the wire vertically.

\medskip

In a free monoidal category, any diagram produced by freely composing the boxes is a valid diagram in the category. This resembles how CFGs allow free applications of the production rules.

\medskip

\begin{figure}[h]
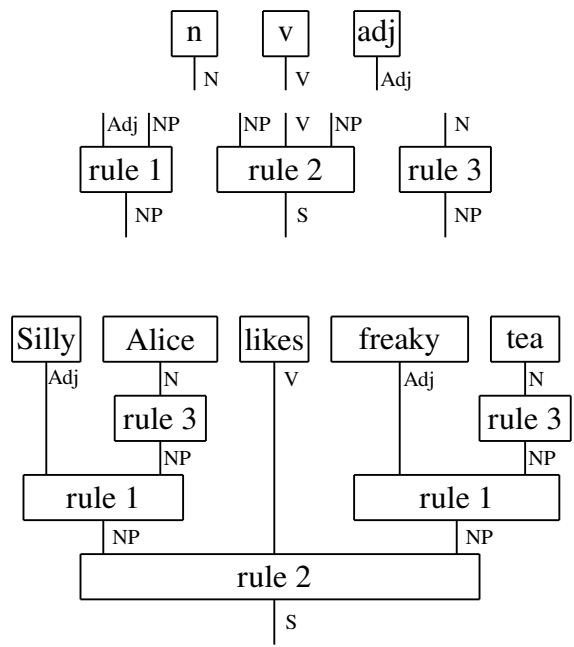

    \begin{center}
    \ctikzfig{tikz/cfg_example}
    \caption{A diagrammatic representation of the CFG example (top) and a sentence built using it (bottom).}
    \label{fig:cfgboxes}
    \end{center}
\end{figure}

Production rules in a CFG of the form $\text{A} \to \alpha$ are represented as a box with multiple input wires representing the string $\alpha$ and output wire representing the non-terminal $\text{A}$. Input wires representing terminal words in $\alpha$ are closed with boxes of type $I \to T$, where $I$ is an auxillary type for terminal types. For example, the CFG above can be described by freely composing the boxes in Fig. \ref{fig:cfgboxes}.

\newpage

\subsection{Lexicon of Musical Snippets}
\label{app:appendix2}

The lexicon for the music composition system presented in this chapter contains four types of snippets: ground (g), primary (p) secondary (s) and tertiary (t) snippets. Here, snippets are for music what words are for natural languages; languages have different types of words, such as nouns, verbs, and adjectives.

\begin{figure}[h]
\begin{center}
\includegraphics[width=0.48\textwidth]{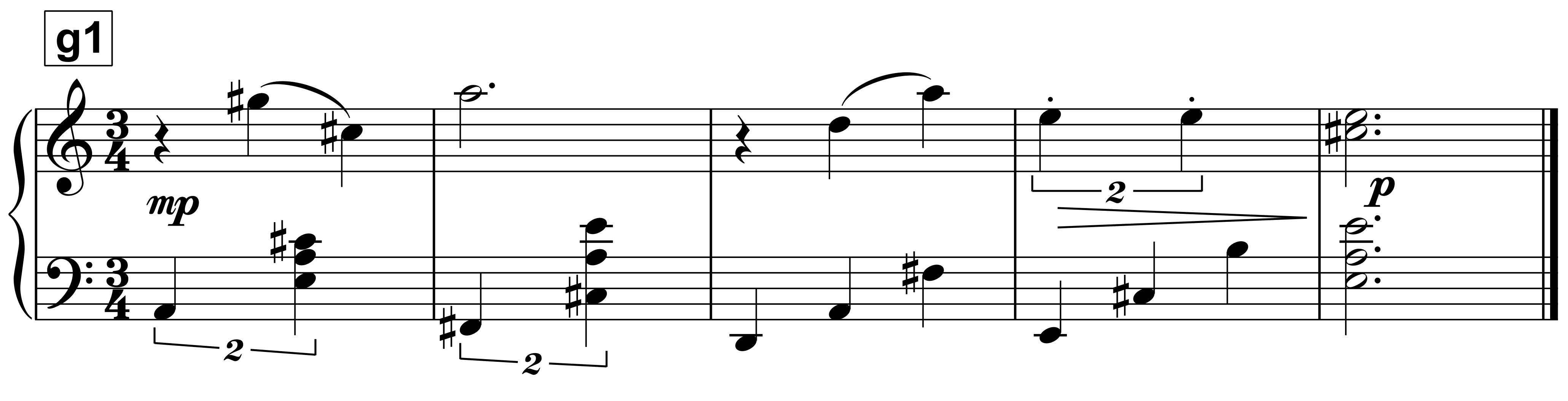}
\includegraphics[width=0.48\textwidth]{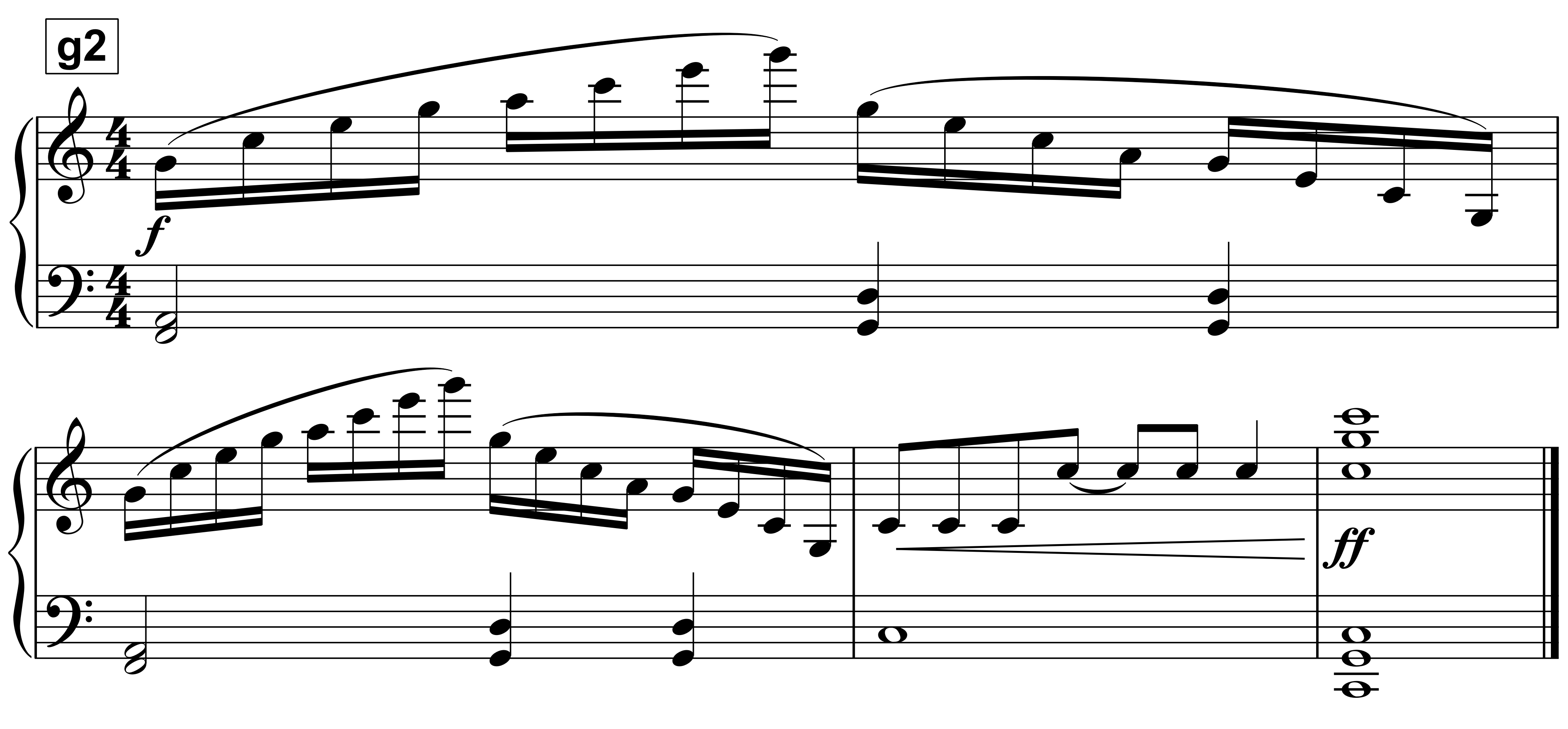}
\caption{Ground snippets g1 and g2.}
\label{fig:g1}
\end{center}
\end{figure}

\begin{figure}[h]
\begin{center}
\includegraphics[width=0.48\textwidth]{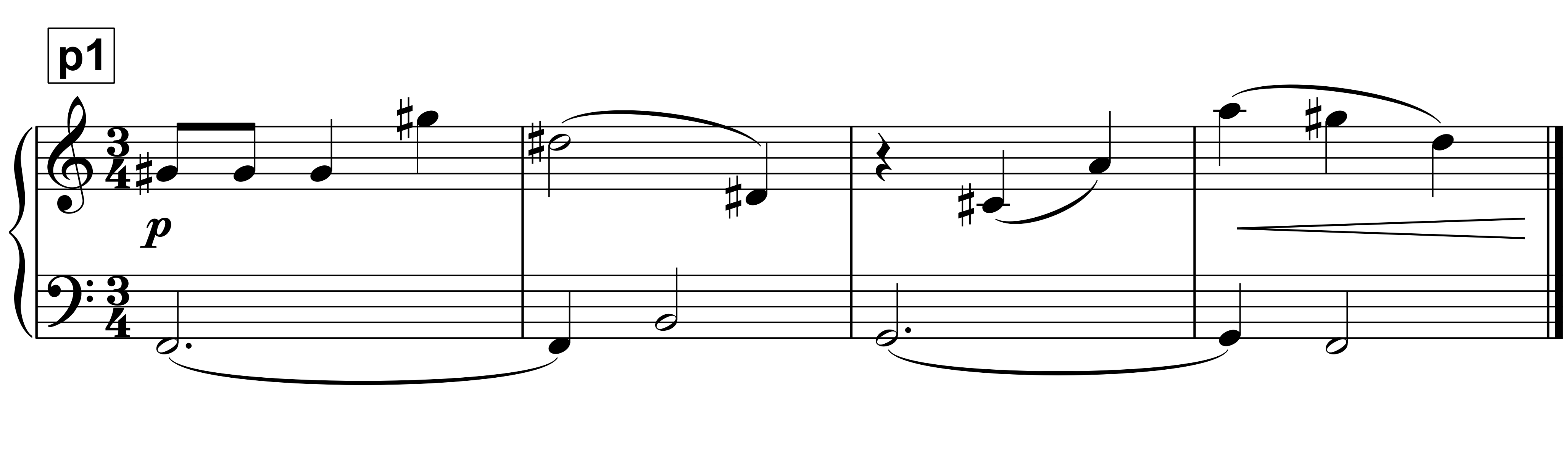}
\includegraphics[width=0.48\textwidth]{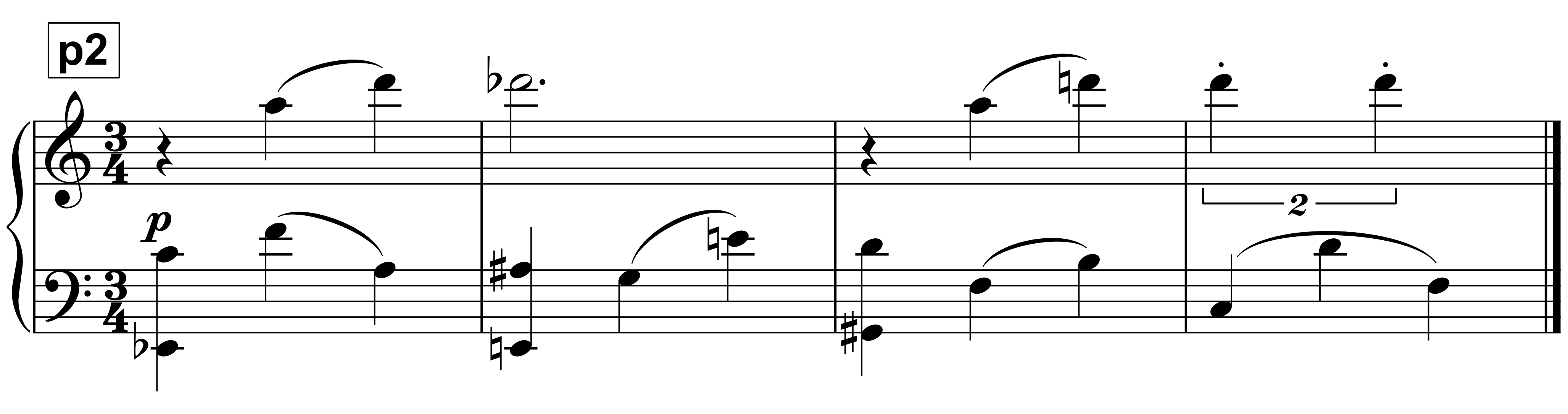} \\
\includegraphics[width=0.48\textwidth]{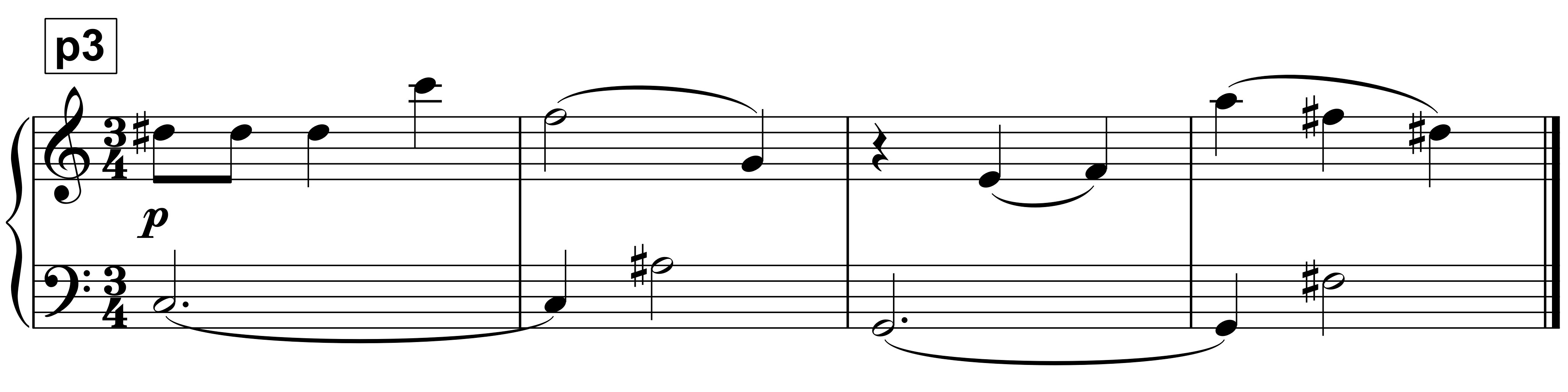}
\includegraphics[width=0.48\textwidth]{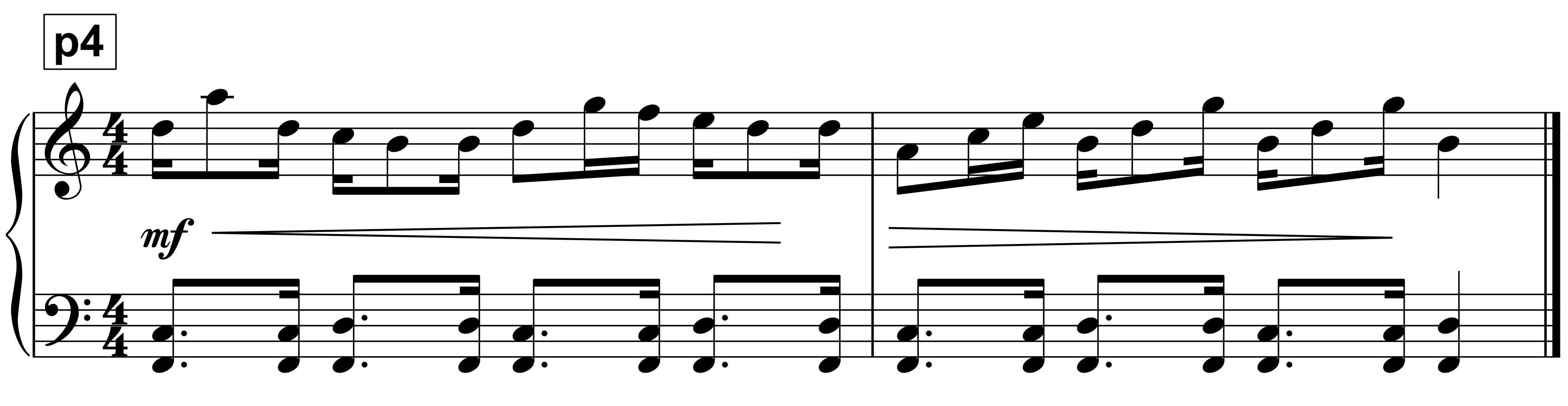} \\
\includegraphics[width=0.48\textwidth]{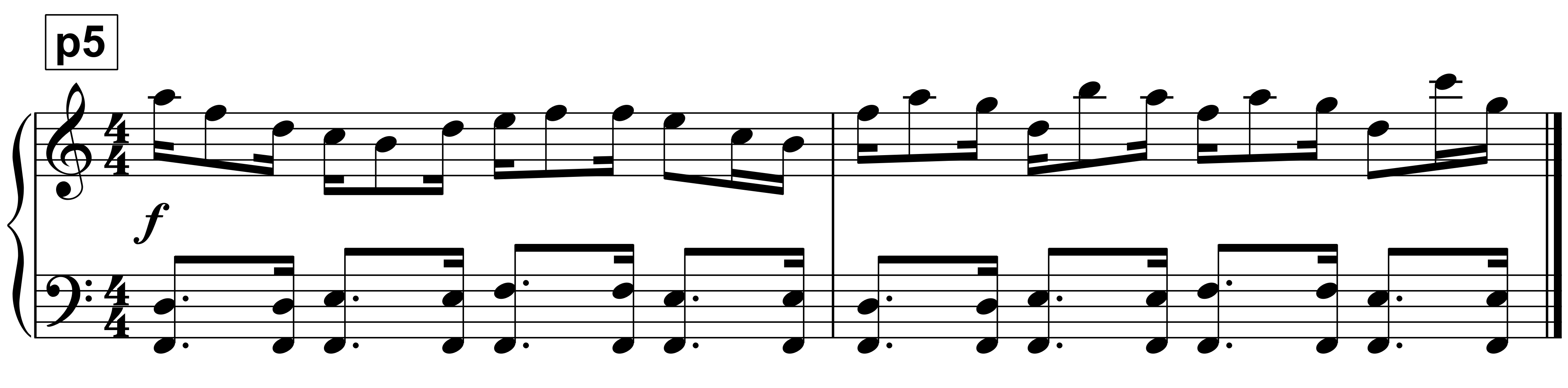}
\includegraphics[width=0.48\textwidth]{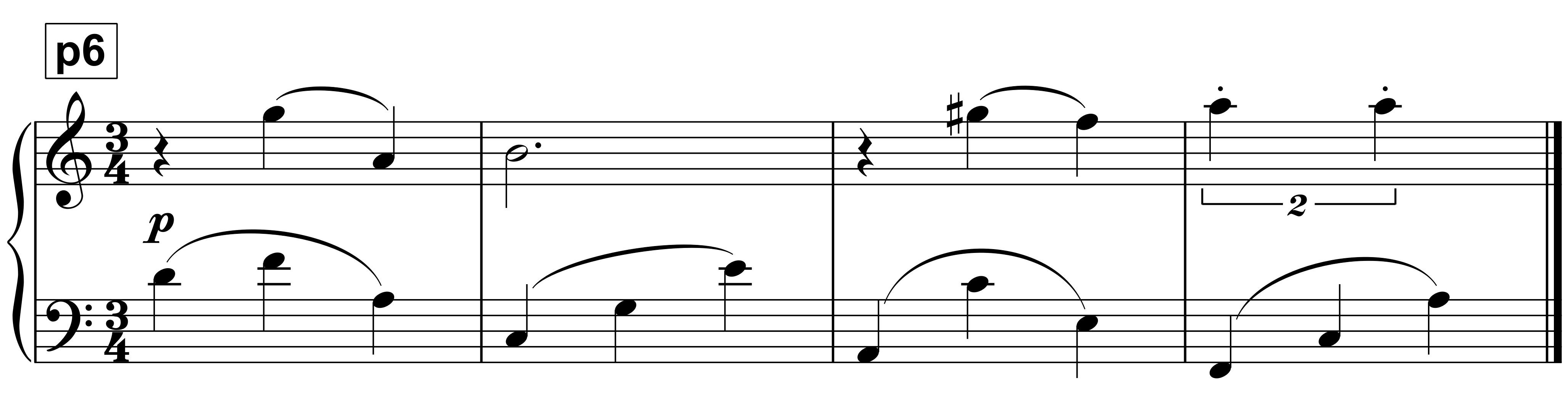} \\
\includegraphics[width=0.48\textwidth]{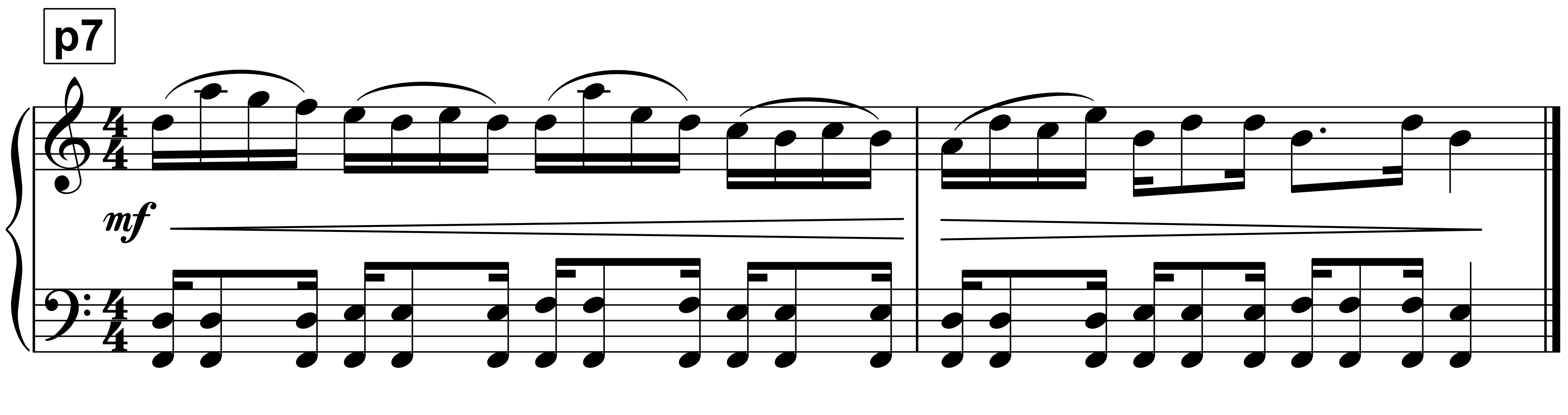}
\includegraphics[width=0.48\textwidth]{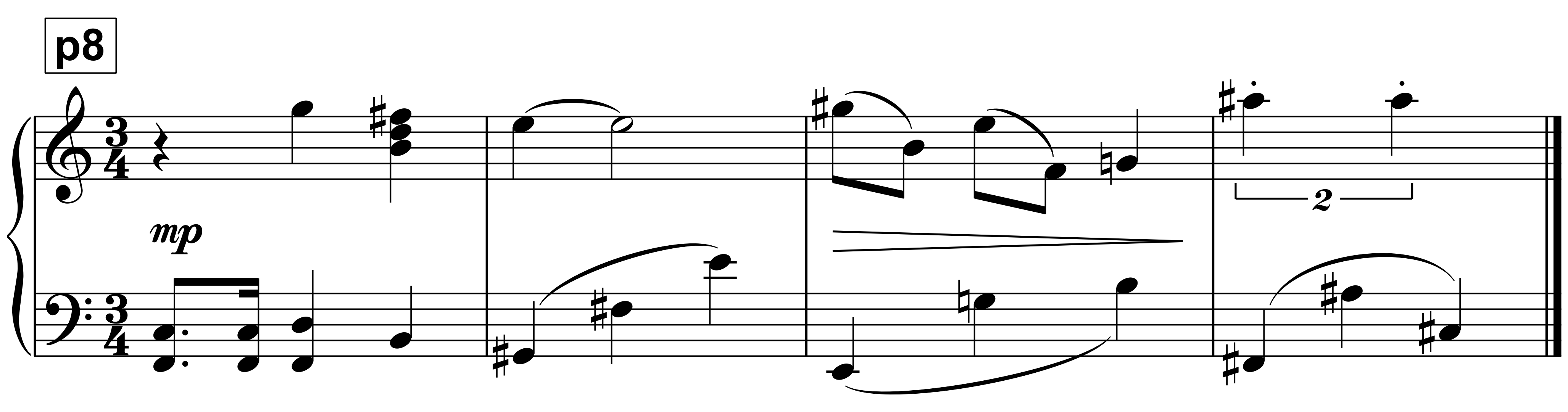} \\
\includegraphics[width=0.48\textwidth]{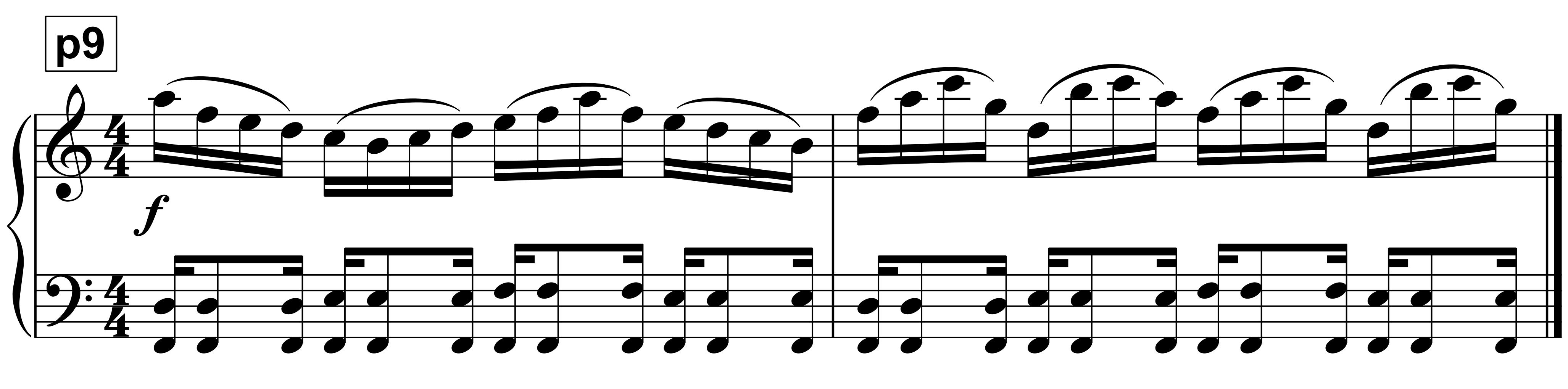}
\caption{Primary snippets p1, p2, p3, p4, p5, p6, p7, p8, and p9.}
\label{fig:p1}
\end{center}
\end{figure}

\begin{figure}[h]
\begin{center}
\includegraphics[width=0.48\textwidth]{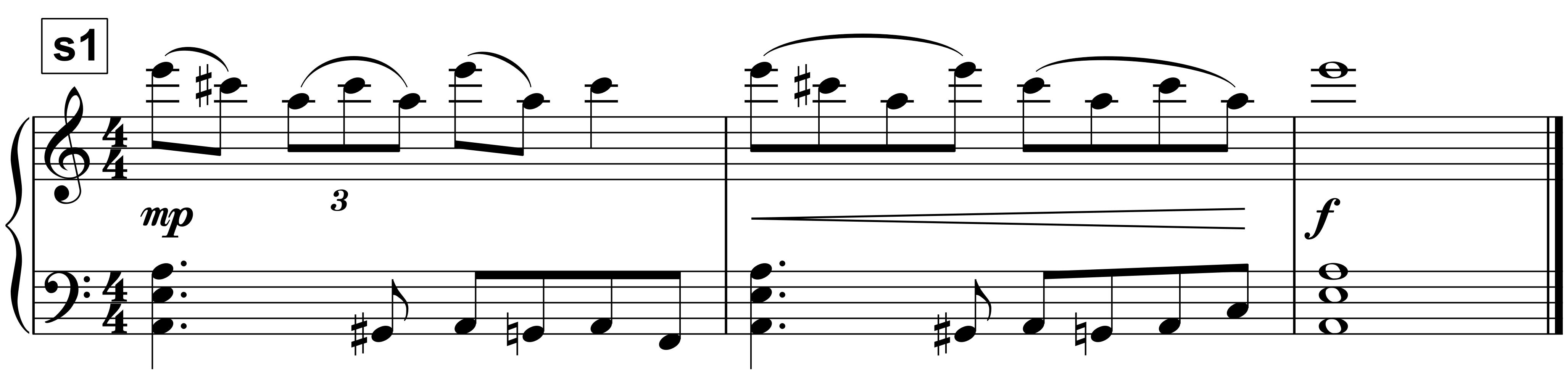} 
\includegraphics[width=0.48\textwidth]{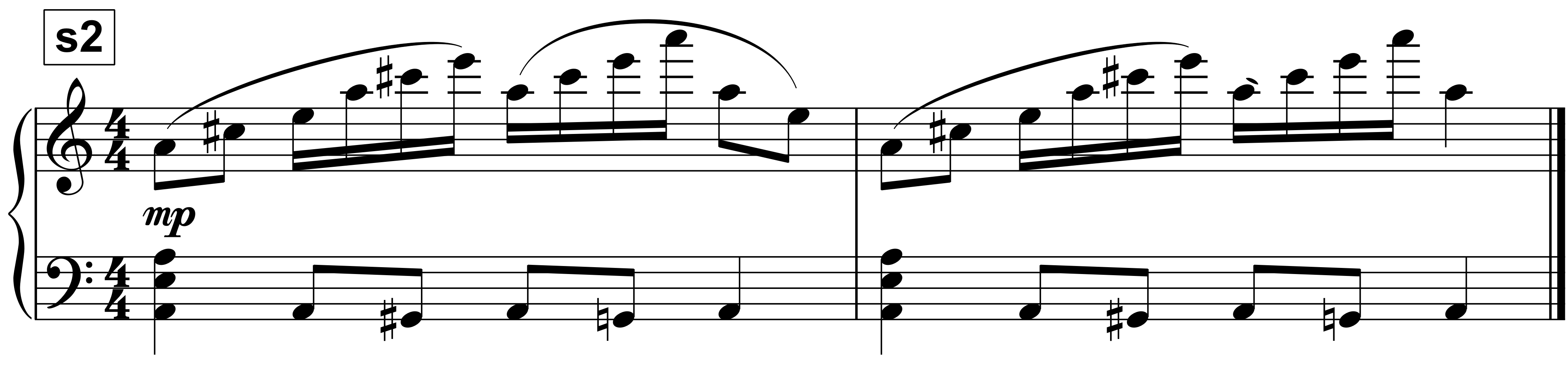} \\
\includegraphics[width=0.48\textwidth]{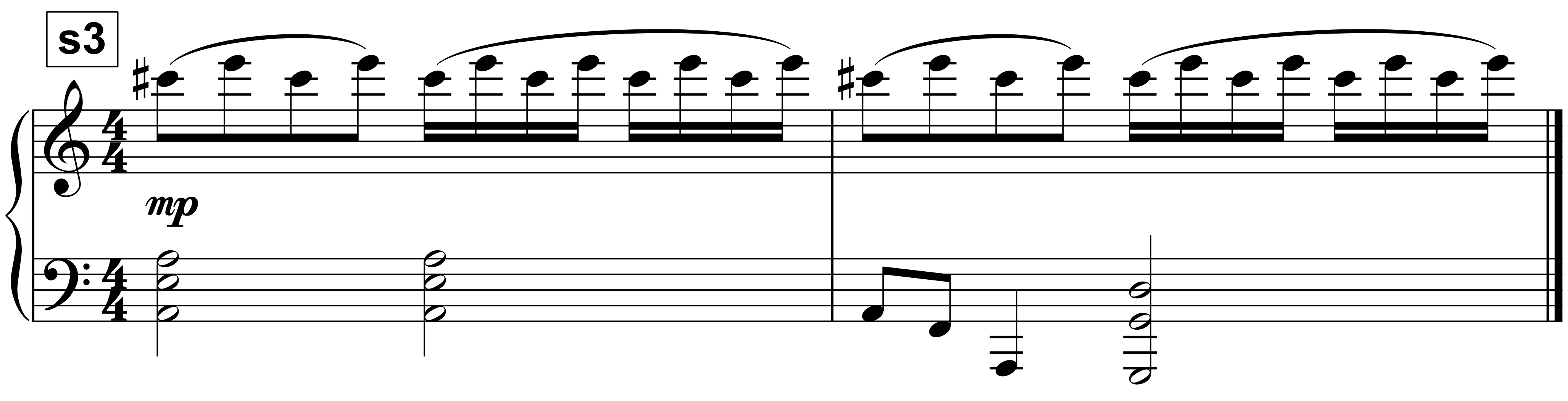} 
\includegraphics[width=0.48\textwidth]{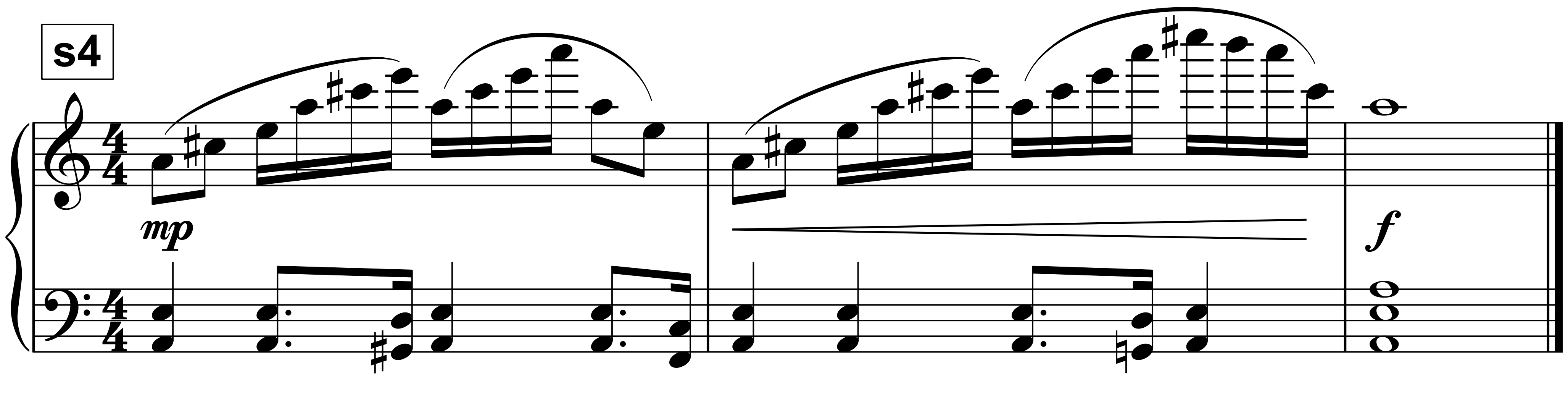}
\caption{Secondary snippets s1, s2, s3, and s4.}
\label{fig:s1}
\end{center}
\end{figure}

\newpage

\begin{figure}[h]
\begin{center}
\includegraphics[width=0.48\textwidth]{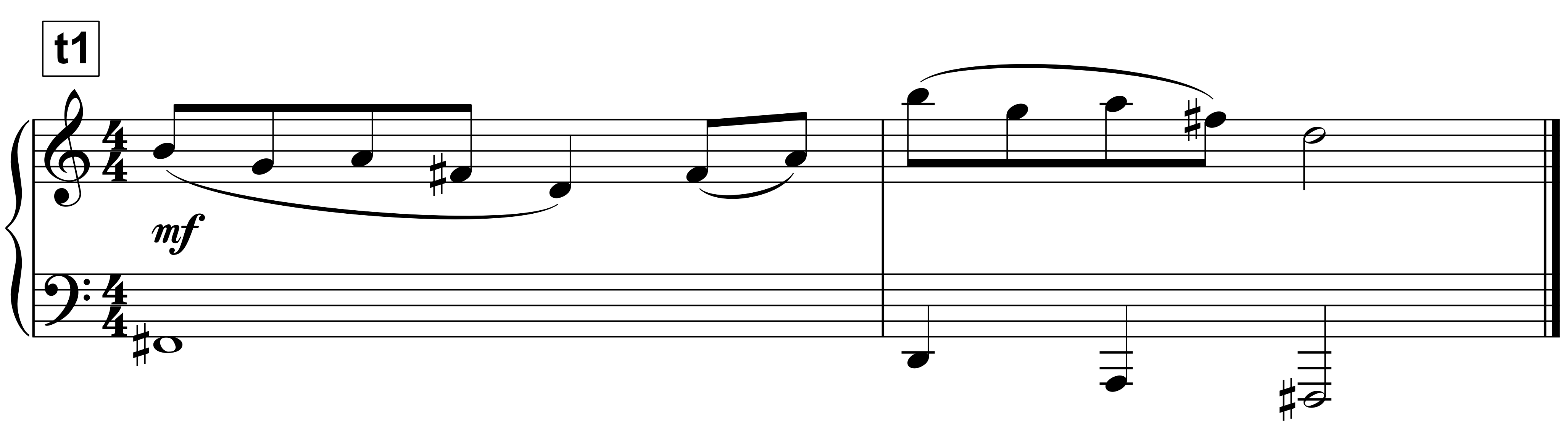}
\includegraphics[width=0.48\textwidth]{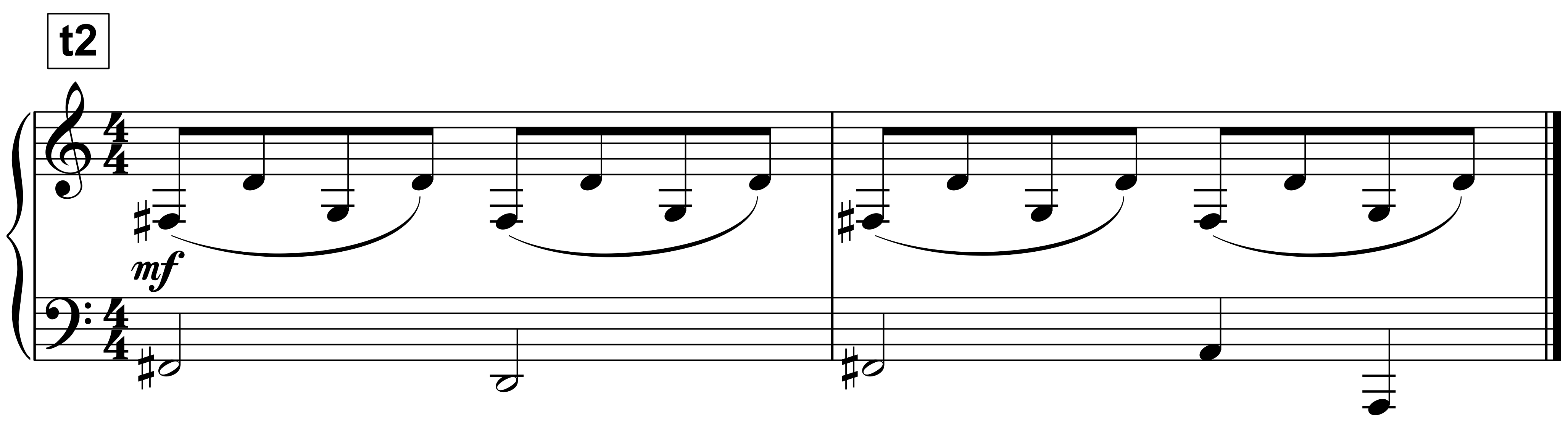}\\
\includegraphics[width=0.48\textwidth]{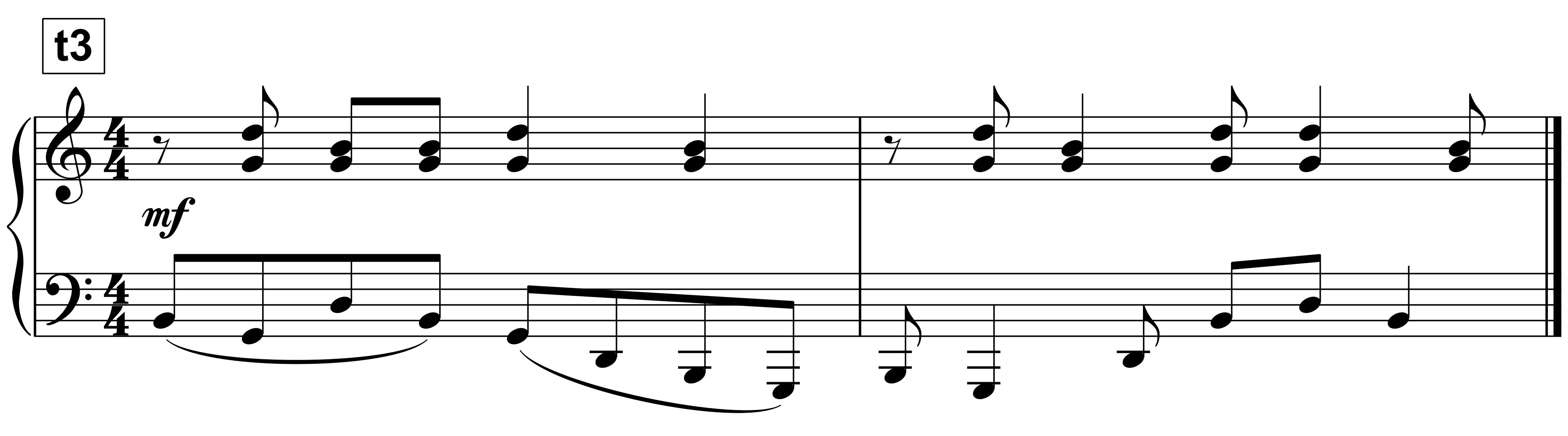}
\caption{Tertiary snippets t1, t2 and t3.}
\label{fig:t1}
\end{center}
\end{figure}

\newpage

\subsection{Musical Datasets}
\label{app:datasets}

\subsubsection{Training Data}

$\Sigma_{\alpha} = \lbrace (1, \textrm{MEL, \lbrack t3, g1, g1\rbrack)},$ \\
$(2, \textrm{MEL, \lbrack t3, p8, p1, p8, p1, s4, g1\rbrack)},$ \\
$(3, \textrm{RIT, \lbrack t3, p9, p9, p5, p9, s4, g2\rbrack)},$ \\
$(4, \textrm{RIT, \lbrack p9, p4, p4, p4, s3\rbrack)},$ \\
$(5, \textrm{RIT, \lbrack p4, p9, p9, p9, s4\rbrack)},$ \\
$(6, \textrm{RIT, \lbrack p5, p9, p4, p5, s1\rbrack)},$ \\
$(7, \textrm{RIT, \lbrack t3, p9, p5, p7, p7, s1, g2\rbrack)},$ \\
$(8, \textrm{MEL, \lbrack p1, p8, p2, p3, s4\rbrack)},$ \\
$(9, \textrm{MEL, \lbrack p6, p1, p8, p3, s4\rbrack)},$ \\
$(10, \textrm{MEL, \lbrack t2, g1, t1, g1, g1\rbrack)},$ \\
$(11, \textrm{RIT, \lbrack p4, p5, p5, p7, s4\rbrack)},$ \\
$(12, \textrm{MEL, \lbrack t2, t3, t2, g1, g1, g1, g1\rbrack)},$ \\
$(13, \textrm{MEL, \lbrack p8, p1, p6, p2, s4\rbrack)},$ \\
$(14, \textrm{RIT, \lbrack p5, p7, p7, p9, s2\rbrack)},$ \\
$(15, \textrm{RIT, \lbrack p5, p7, p5, p9, s4\rbrack)},$ \\
$(16, \textrm{RIT, \lbrack t3, g2, p9, p9, p7, p7, s1\rbrack)},$ \\
$(17, \textrm{MEL, \lbrack p6, p2, p3, p2, s1\rbrack)},$ \\
$(18, \textrm{MEL, \lbrack t1, g1, p3, p1, p1, p1, s2\rbrack)},$ \\
$(19, \textrm{RIT, \lbrack p7, p7, p5, p7, s1\rbrack)},$ \\
$(20, \textrm{RIT, \lbrack t3, p5, p9, p5, p7, s2, g2\rbrack)},$ \\
$(21, \textrm{RIT, \lbrack p7, p4, p4, p9, s2\rbrack)},$ \\
$(22, \textrm{RIT, \lbrack p5, p9, p5, p4, s1\rbrack)},$ \\
$(23, \textrm{RIT, \lbrack p9, p9, p7, p9, s1\rbrack)},$ \\
$(24, \textrm{RIT, \lbrack p7, p7, p9, p7, s1\rbrack)},$ \\
$(25, \textrm{RIT, \lbrack p9, p4, p5, p7, s4\rbrack)},$ \\
$(26, \textrm{MEL, \lbrack t3, p8, p6, p1, p2, s2, g1\rbrack)},$ \\
$(27, \textrm{MEL, \lbrack t2, p6, p8, p8, p8, s1, g1\rbrack)},$ \\
$(28, \textrm{MEL, \lbrack p6, p3, p6, p6, s2\rbrack)},$ \\
$(29, \textrm{MEL, \lbrack p6, p8, p3, p1, s2\rbrack)},$ \\
$(30, \textrm{MEL, \lbrack p1, p2, p6, p6, s4\rbrack)},$ \\
$(31, \textrm{MEL, \lbrack p6, p1, p2, p8, s3\rbrack)},$ \\
$(32, \textrm{MEL, \lbrack p1, p1, p6, p3, s2\rbrack)},$ \\
$(33, \textrm{RIT, \lbrack p4, p9, p7, p5, s1\rbrack)},$ \\
$(34, \textrm{RIT, \lbrack t2, g2, p5, p4, p5, p4, s1\rbrack)},$ \\
$(35, \textrm{RIT, \lbrack p9, p9, p9, p5, s3\rbrack)},$ \\
$(36, \textrm{RIT, \lbrack p9, p9, p5, p5, s4\rbrack)},$ \\
$(37, \textrm{MEL, \lbrack p6, p2, p6, p2, s1\rbrack)},$ \\
$(38, \textrm{RIT, \lbrack p5, p5, p5, p5, s4\rbrack)},$ \\
$(39, \textrm{MEL, \lbrack p6, p1, p1, p1, s1\rbrack)},$ \\
$(40, \textrm{RIT, \lbrack p5, p7, p5, p4, s2\rbrack)},$ \\
$(41, \textrm{MEL, \lbrack t2, g1, p1, p6, p2, p6, s2\rbrack)},$ \\
$(42, \textrm{MEL, \lbrack p1, p6, p3, p8, s2\rbrack)},$ \\
$(43, \textrm{RIT, \lbrack p4, p5, p7, p7, s1\rbrack)},$ \\
$(44, \textrm{MEL, \lbrack p1, p6, p8, p1, s2\rbrack)},$ \\
$(45, \textrm{MEL, \lbrack p6, p1, p2, p3, s1\rbrack)},$ \\
$(46, \textrm{RIT, \lbrack t2, g2, t1, g2, p9, p5, p4, p9, s2\rbrack)},$ \\
$(47, \textrm{MEL, \lbrack t3, t1, g1, g1, p8, p6, p6, p8, s4\rbrack)},$ \\
$(48, \textrm{MEL, \lbrack p2, p3, p8, p6, s2\rbrack)},$ \\
$(49, \textrm{RIT, \lbrack t1, g2, t1, p7, p4, p7, p7, s4, g2\rbrack)},$ \\
$(50, \textrm{RIT, \lbrack t1, t2, g2, p4, p9, p4, p7, s4, g2\rbrack)} \rbrace$
\subsubsection{Development Data}

$\Sigma_{\beta}= \lbrace (51, \textrm{MEL, \lbrack t1, g1, p8, p6, p8, p2, s4\rbrack)},$ \\
$(52, \textrm{MEL, \lbrack t1, p6, p8, p2, p8, s4, g1\rbrack)},$ \\
$(53, \textrm{RIT, \lbrack p4, p5, p7, p4, s2\rbrack)},$ \\
$(54, \textrm{MEL, \lbrack t2, p2, p8, p1, p6, s2, g1\rbrack)},$ \\
$(55, \textrm{MEL, \lbrack p8, p2, p3, p1, s2\rbrack)},$ \\
$(56, \textrm{MEL, \lbrack t3, p6, p2, p8, p2, s3, g1\rbrack)},$ \\
$(57, \textrm{RIT, \lbrack p4, p9, p4, p9, s1\rbrack)},$ \\
$(58, \textrm{RIT, \lbrack p7, p9, p5, p4, s4\rbrack)},$ \\
$(59, \textrm{RIT, \lbrack t2, p7, p7, p4, p9, s1, g2\rbrack)},$ \\
$(60, \textrm{RIT, \lbrack p9, p9, p7, p5, s4\rbrack)},$ \\
$(61, \textrm{RIT, \lbrack p5, p9, p5, p7, s1\rbrack)},$ \\
$(62, \textrm{RIT, \lbrack p4, p5, p4, p9, s2\rbrack)},$ \\
$(63, \textrm{RIT, \lbrack p9, p4, p9, p9, s4\rbrack)},$ \\
$(64, \textrm{RIT, \lbrack p4, p7, p5, p5, s3\rbrack)},$ \\
$(65, \textrm{MEL, \lbrack p2, p1, p2, p2, s3\rbrack)},$ \\
$(66, \textrm{RIT, \lbrack p7, p7, p9, p7, s4\rbrack)},$ \\
$(67, \textrm{MEL, \lbrack p6, p3, p6, p6, s4\rbrack)},$ \\
$(68, \textrm{RIT, \lbrack p9, p7, p9, p4, s3\rbrack)},$ \\
$(69, \textrm{RIT, \lbrack p9, p9, p4, p5, s2\rbrack)},$ \\
$(70, \textrm{MEL, \lbrack p8, p3, p2, p6, s2\rbrack)},$ \\
$(71, \textrm{RIT, \lbrack p9, p4, p5, p4, s4\rbrack)},$ \\
$(72, \textrm{MEL, \lbrack p6, p6, p2, p3, s4\rbrack)},$ \\
$(73, \textrm{RIT, \lbrack p7, p7, p4, p5, s2\rbrack)},$ \\
$(74, \textrm{MEL, \lbrack t3, g1, t3, g1, p1, p2, p2, p3, s2\rbrack)},$ \\
$(75, \textrm{MEL, \lbrack t1, g1, g1\rbrack)} \rbrace$

\subsubsection{Testing Data}

$\Sigma_{\gamma}= \lbrace (76, \textrm{RIT, \lbrack p9, p9, p9, p9, s1\rbrack)},$ \\
$(77, \textrm{MEL, \lbrack p3, p6, p8, p2, s4\rbrack)},$ \\
$(78, \textrm{RIT, \lbrack p4, p7, p5, p9, s1\rbrack)},$ \\
$(79, \textrm{RIT, \lbrack p9, p9, p5, p5, s3\rbrack)},$ \\
$(80, \textrm{RIT, \lbrack p7, p9, p9, p7, s2\rbrack)},$ \\
$(81, \textrm{RIT, \lbrack t3, g2, p4, p4, p9, p9, s4\rbrack)},$ \\
$(82, \textrm{RIT, \lbrack p4, p4, p7, p9, s2\rbrack)},$ \\
$(83, \textrm{RIT, \lbrack p4, p5, p9, p4, s4\rbrack)},$ \\
$(84, \textrm{MEL, \lbrack t3, p6, p6, p8, p3, s4, g1\rbrack)},$ \\
$(85, \textrm{MEL, \lbrack p2, p6, p1, p8, s3\rbrack)},$ \\
$(86, \textrm{MEL, \lbrack p6, p2, p8, p3, s4\rbrack)},$ \\
$(87, \textrm{MEL, \lbrack p8, p3, p1, p2, s2\rbrack)},$ \\
$(88, \textrm{MEL, \lbrack t1, g1, t3, g1, g1\rbrack)},$ \\
$(89, \textrm{RIT, \lbrack p9, p7, p9, p9, s2\rbrack)},$ \\
$(90, \textrm{RIT, \lbrack p9, p7, p7, p9, s3\rbrack)},$ \\
$(91, \textrm{MEL, \lbrack t3, g1, p3, p1, p3, p3, s3\rbrack)},$ \\
$(92, \textrm{MEL, \lbrack p3, p3, p1, p8, s3\rbrack)},$ \\
$(93, \textrm{MEL, \lbrack t2, p2, p1, p3, p8, s3, g1\rbrack)},$ \\
$(94, \textrm{MEL, \lbrack t3, g1, p3, p8, p3, p3, s2\rbrack)},$ \\
$(95, \textrm{MEL, \lbrack p2, p1, p6, p6, s3\rbrack)},$ \\
$(96, \textrm{RIT, \lbrack t1, p5, p9, p4, p9, s3, g2\rbrack)},$ \\
$(97, \textrm{RIT, \lbrack t2, p9, p7, p5, p5, s4, g2\rbrack)},$ \\
$(98, \textrm{MEL, \lbrack t2, p2, p1, p6, p1, s4, g1\rbrack)},$ \\
$(99, \textrm{RIT, \lbrack t3, p9, p5, p9, p9, s1, t3, g2, g2\rbrack)},$ \\
$(100, \textrm{MEL, \lbrack t2, g1, g1\rbrack)} \rbrace$

\newpage

\subsection{Compositions}
\label{app:appendix3}

\begin{figure}[h]
\begin{center}
\includegraphics[width=0.76\textwidth]{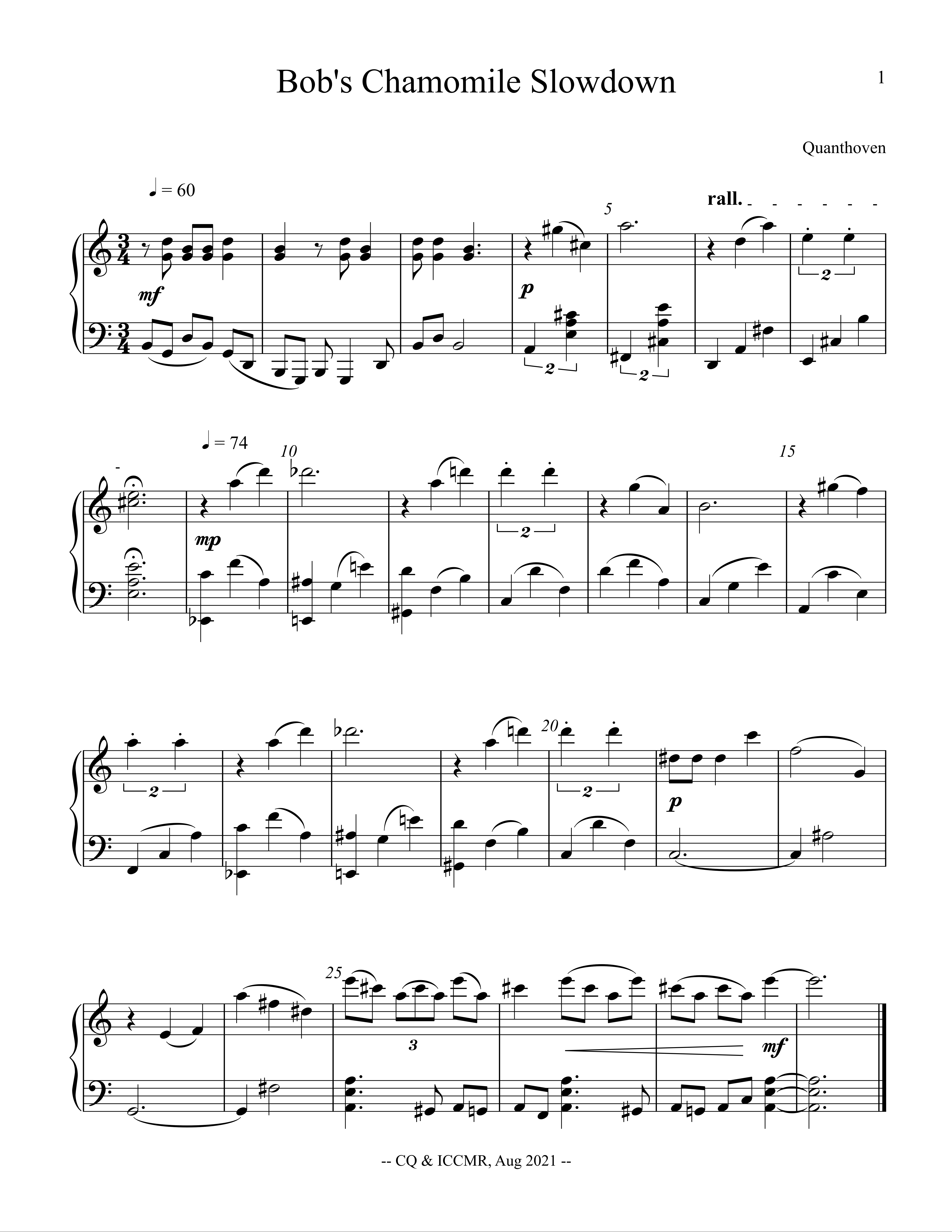}
\caption{The composition \textit{Bob's Chamomile Slowdown}.}
\label{fig:bob_chamo}
\end{center}
\end{figure}

\begin{figure}[h]
\begin{center}
\includegraphics[width=0.78\textwidth]{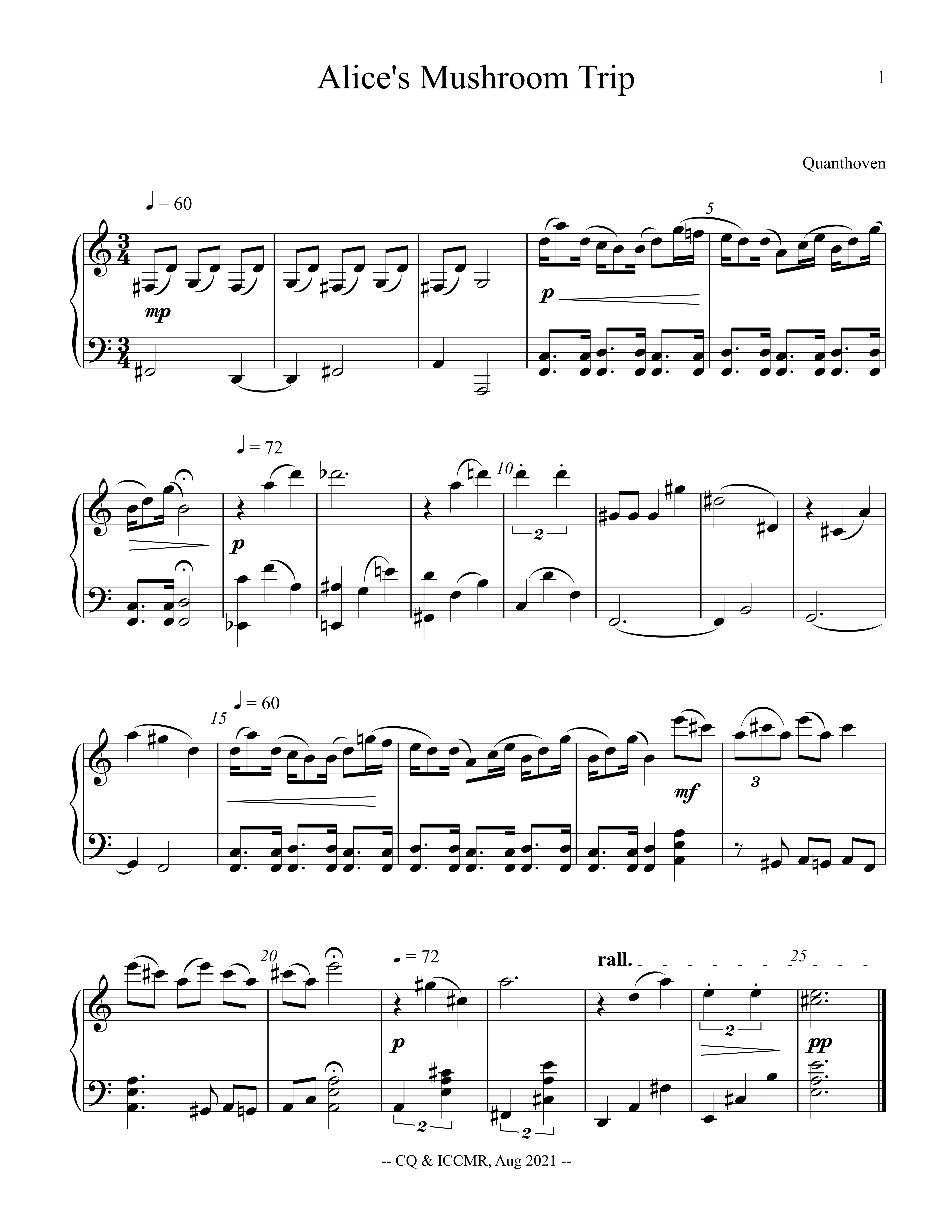}
\caption{The composition \textit{Alice's Mushroom Trip}.}
\label{fig:alice_muschroom}
\end{center}
\end{figure}

\begin{figure}[h]
\begin{center}
\includegraphics[width=0.78\textwidth]{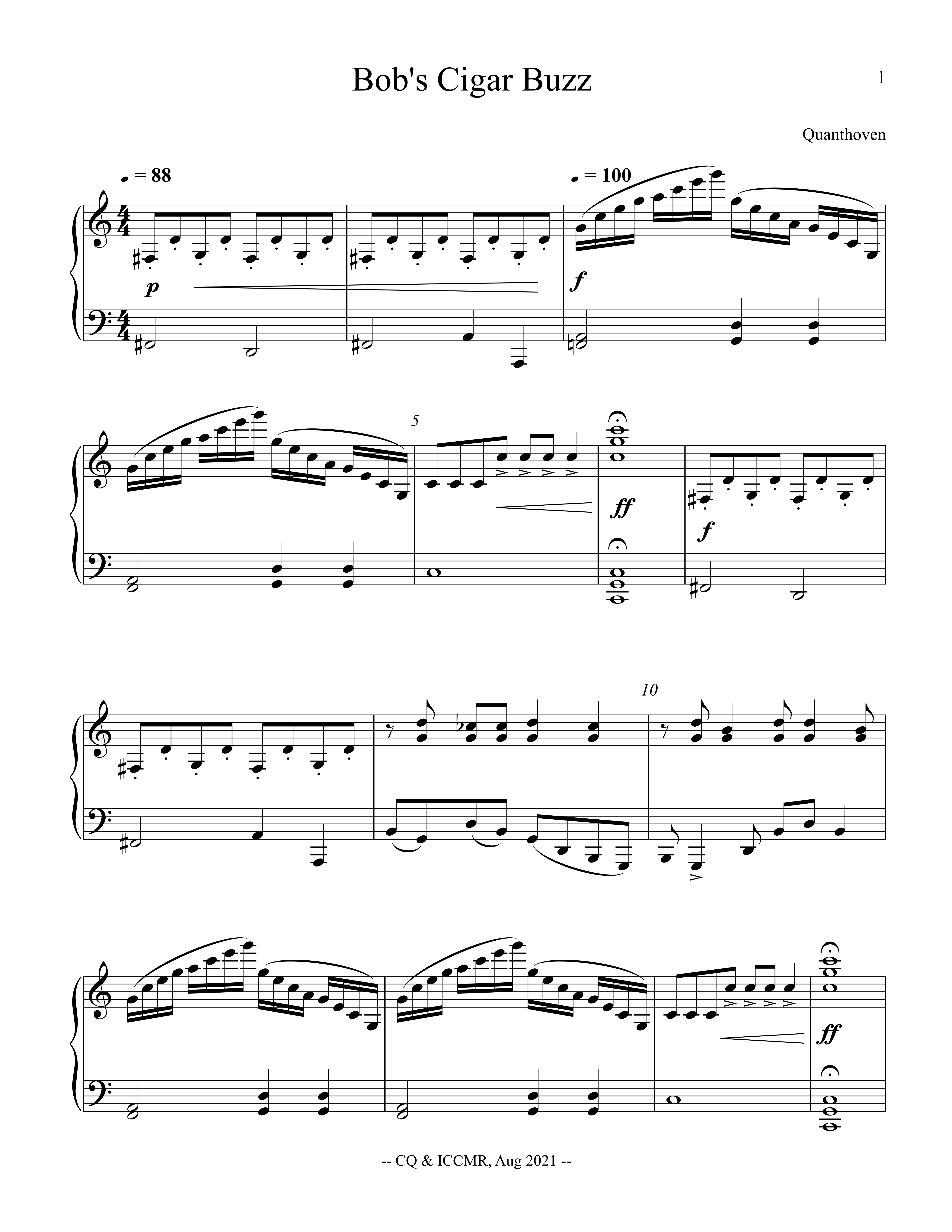}
\caption{First page of the composition \textit{Bob's Cigar Rush}.}
\label{fig:alice_muschroom}
\end{center}
\end{figure}

\begin{figure}[h]
\begin{center}
\includegraphics[width=0.78\textwidth]{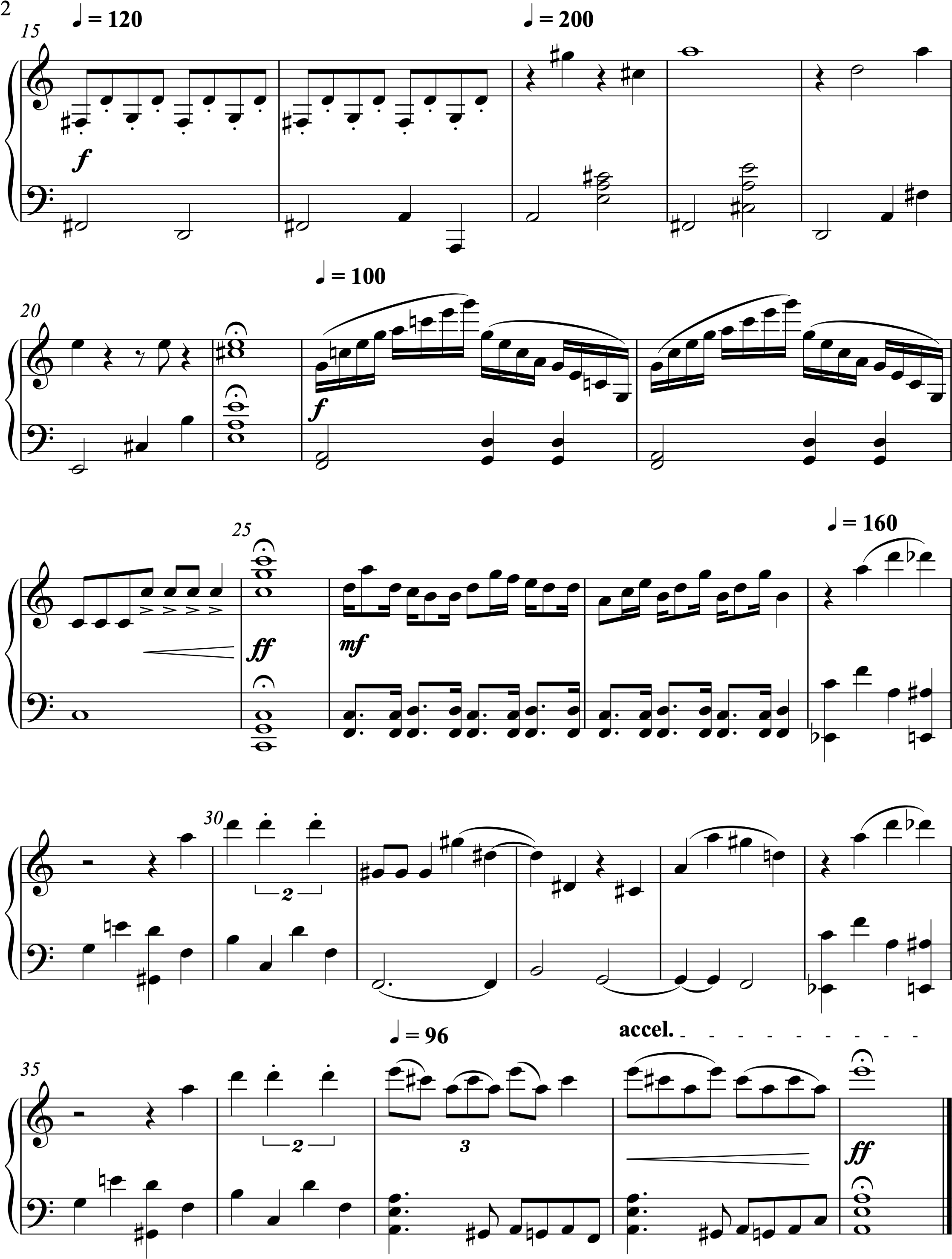}
\caption{Second page of the composition \textit{Bob's Cigar Rush}.}
\label{fig:alice_muschroom}
\end{center}
\end{figure}

\end{document}